\documentclass{aa}

\usepackage{txfonts}
\usepackage{color, colortbl}
\usepackage[colorlinks, citecolor=blue, linkcolor=blue, urlcolor=blue]{hyperref}

\usepackage{graphicx}
\usepackage{txfonts}
\usepackage{pdflscape}	
\usepackage{color, colortbl}

\usepackage{newtxtext,newtxmath}

\usepackage[T1]{fontenc}
\usepackage{ae,aecompl}

\usepackage{pdflscape}	

\usepackage{listings}
\usepackage{pdflscape}
\usepackage{longtable}
\usepackage{wasysym}

\usepackage{graphicx}	
\usepackage{amsmath}	
\usepackage{amssymb}	
\usepackage{multirow}

\newcommand{\gaia}{{\it Gaia}}
\newcommand{\numberR}{nine}
\newcommand{\sample}{4847}
\newcommand{\kms}{\,km\,s$^{-1}$}	

\usepackage{soul}
\usepackage{blindtext}
\begin{document} 


\title{Milky Way archaeology using RR Lyrae and type II Cepheids II. High velocity RR Lyrae stars, and mass of the Milky Way}

\author{Z.~Prudil\inst{1}, A.~J~Koch-Hansen\inst{1}, B.~Lemasle\inst{1}, E.~K.~Grebel\inst{1}, T.~Marchetti\inst{2}, C.~J.~Hansen\inst{3,4}, J.~Crestani\inst{5,6}, V.~F.~Braga\inst{6,7}, G.~Bono\inst{5,6}, B.~Chaboyer\inst{8} , M.~Fabrizio\inst{6,7}, M. Dall'Ora\inst{9} \and C.~E.~Mart\'inez-V\'azquez\inst{10,11} }

\institute{
Astronomisches Rechen-Institut, Zentrum f{\"u}r Astronomie der Universit{\"a}t Heidelberg, M{\"o}nchhofstr. 12-14, D-69120 Heidelberg, Germany \\ \href{mailto:prudilz@ari.uni-heidelberg.de}{prudilz@ari.uni-heidelberg.de}
\and European Southern Observatory, Karl-Schwarzschild-Strasse 2, 85748 Garching bei M{\"u}nchen, Germany
\and TU Darmstadt, IKP, Schlossgartenstr. 2 (S2|11), 64289 Darmstadt, Germany
\and Goethe University Frankfurt, IAP, Max-von-Laue-Str. 12, Frankfurt am Main, 60438 Frankfurt, Germany
\and Dipartimento di Fisica, Universit\`a di Roma Tor Vergata, via della Ricerca Scientifica 1, 00133 Roma, Italy 
\and INAF -- Osservatorio Astronomico di Roma, via Frascati 33, 00078 Monte Porzio Catone, Italy
\and Space Science Data Center -- ASI, via del Politecnico snc, 00133 Roma, Italy
\and Department of Physics and Astronomy, Dartmouth College, Hanover, NH 03755, USA	
\and INAF -- Osservatorio Astronomico di Capodimonte, Salita Moiariello 16, 80131 Napoli, Italy
\and Gemini Observatory/NSF's NOIRLab, 670 N. A'ohoku Place, Hilo, HI 96720, USA
\and Cerro Tololo Inter-American Observatory, NSF's National Optical-Infrared Astronomy Research Laboratory, Casilla 603, La Serena, Chile	}

\date{\today}

\abstract
{We report the discovery of high velocity candidates among RR~Lyrae stars found in the Milky Way halo. We identified \numberR~RR~Lyrae stars with Galactocentric velocities exceeding the local escape velocity based on the assumed Galaxy potential. Based on close examination of their orbits', we ruled out their ejection location in the Milky Way disk and bulge. The spatial distribution revealed that seven out of \numberR~pulsators overlap with the position of the Sagittarius stellar stream. Two out of these seven RR~Lyrae stars can be tentatively linked to the Sagittarius dwarf spheroidal galaxy on the basis of their orbits. Focusing on the high-velocity tail of the RR~Lyrae velocity distribution we estimate the escape velocity in the Solar neighborhood to be $v_{\rm esc} = 512^{+94}_{-37}$\kms~($4$ to $12$\,kpc), and beyond the Solar neighborhood as $v_{\rm esc} = 436^{+44}_{-22}$\kms~and $v_{\rm esc} = 393^{+53}_{-26}$\kms~(for distances between $12$ to $20$\,kpc and $20$ to $28$\,kpc), respectively. We utilized three escape velocity estimates together with the local circular velocity to estimate the Milky Way mass. The resulting measurement $M_{\rm 200} = 0.83^{+0.29}_{-0.16} \cdot 10^{12}$\,M$_{\odot}$ falls on the lower end of the current Milky Way mass estimates, but once corrected for the likely bias in the escape velocity (approximately $10$ percent increase of the escape velocity), our mass estimate yields $M_{\rm 200} = 1.26^{+0.40}_{-0.22} \cdot 10^{12}$\,M$_{\odot}$, which is in agreement with estimates based on different diagnostics of the Milky Way mass. The MW mass within $20$\,kpc then corresponds to $M_{\rm MW} \left(r < 20\,\text{kpc} \right) = 1.9^{+0.2}_{-0.1} \times 10^{11}$\,M$_{\odot}$ without correction for bias, and $M_{\rm MW} \left(r < 20\,\text{kpc} \right) = 2.1^{+0.2}_{-0.1} \times 10^{11}$\,M$_{\odot}$ corrected for a likely offset in escape velocities.}

\titlerunning{Unbound RR Lyrae stars, and mass of the Milky Way}
\authorrunning{Prudil et al.}
\maketitle

\section{Introduction} \label{sec:Intro}

Our Galaxy, the Milky Way (MW), harbors stellar populations with a variety of dynamical properties that reflect the underlying mass distribution. Particularly, stars on the edge (and beyond) of the MW escape velocity distribution provide an important perspective both on the small and large dynamical scale of the MW environment. 

Focusing on the small scale, stars in a galaxy can be accelerated by dynamical interactions with their close surroundings. Principally two main production classes for high-velocity stars have been introduced. The \textit{runaway stars} are mainly young (O-B type) stars found far away from a star-forming galaxy disk, deep in the stellar halo \citep{Blaauw1961}. The mechanism responsible for their ejection involves two-body \citep[supernova explosion in a binary, e.g.,][]{Blaauw1961,Portegies2000,Evans2020} and multi-body \citep[close encounters in stellar systems like young massive star clusters, e.g.,][]{Poveda1967,Leonard1991,Gvaramadze2009} systems. Typically, the runaway stars from the aforementioned production channels can attain velocities from several dozens up to low-hundreds of km\,s$^{-1}$ \citep{Portegies2000,Perets2012}, with very few exceeding the Galactic escape velocity \citep[the so called hyper-runaway stars, e.g.,][]{Perets2012,Brown2015,Evans2020}.

The second type of extreme velocity stars is the \textit{hypervelocity stars}, which are the fastest moving stars in a galaxy. The ejection mechanism of hypervelocity stars is linked to a supermassive black hole (SMBH) in the Galactic center \citep{Hills1988} where during a close encounter, a binary system can be disrupted. Then one of the components can be ejected with an extreme velocity \citep[up to thousands of km\,s$^{-1}$, e.g.,][]{Hills1988,Yu2003,Bromley2006} exceeding the galaxy escape velocity. The first candidate hypervelocity stars were found by \citet{Brown2005}, and since then, hundreds of more candidates have been discovered \citep[e.g.,][]{Edelmann2005,Kollmeier2009,Brown2014,Brown2015,Vickers2015}. Especially with the advent of the \gaia~astrometric mission, the number of hypervelocity candidates has increased \citep[e.g.,][]{Hattori2018,Marchetti2019,Marchetti2021,Li2021High591} with some of the objects reaching velocities beyond $>1500$\,\kms \citep{Koposov2020HighVel}. For a full review on high-velocity and particularly hypervelocity stars, see \citet{Brown2015Rev}.

The outlined ejection mechanisms are not intrinsic only to the MW, but to other star-forming galaxies. For example, star-forming dwarf galaxies like the Large and Small Magellanic Clouds (LMC and SMC) can be a source of stars with extreme velocities that were ejected (runaway stars), and are now passing through the MW halo \citep{Boubert2017,Hattori2018metPoor_LMC,Erkal2019HypLMC}. The infall of globular clusters and dwarf galaxies toward the Galactic center can lead to the ejection of stars with extreme velocities \citep{Abadi2009,Capuzzo2015,Fragione2016} during the interaction with the SMBH. Theoretical predictions also suggest a sub-population of hypervelocity stars ejected from the Andromeda galaxy by its SMBH \citep{Sherwin2008}.

Not all stars accelerated by the mechanisms described above surpass the Galaxy's escape velocity and become unbound. Some objects populate the high-velocity tail of the Galaxy's velocity distribution. The escape velocity boundary at the high-velocity tail decreases with increasing Galactocentric radius, and is set by the Galaxy's potential and thus mass distribution. Therefore, weakly bound halo stars can be utilized to estimate the escape velocity at a given radius, and in turn, for a selected gravitational potential, they can provide a constraint on the MW mass. The mass of our Galaxy is one of the large-scale fundamental parameters that still remain uncertain, and current estimates center around 10$^{12}$\,M$_{\odot}$ \citep[for an illustration of the MW mass estimates see Fig.~7 in][]{Callingham2019}. The extensive dynamical studies of the MW satellite galaxies \citep[e.g.,][]{Watkins2010,Lepine2011,Koch2012,Callingham2019}, globular clusters \citep[e.g.,][]{Eadie2016,Sohn2018}, stellar streams \citep[e.g.,][]{Pearson2017,Malhan2018,Erkal2019}, and halo stars \citep[e.g.,][]{Shen2022} yield numerous mass estimates for the MW. 

The study by \citet{Leonard1990} put forward an approach to model the velocity distribution by a power-law. This technique has been often utilized to estimate the Galactic escape velocity, and subsequently, the mass of the MW \citep[e.g.,][]{Smith2007,Piffl2014}. Particularly with the availability of astrometric data from the \gaia~mission~\citep{Gaia2016}, many studies have been focusing on estimating the Galactic escape velocity in the Solar neighborhood \citep[i.e.,][]{Monari2018,Deason2019,Grand2019,Koppelman2021,Necib2021II}. In the studies mentioned above, the estimates of the Galactic escape velocity fall in the range from $480$ to $640$\,\kms~where a subtle difference in the analysis (prior assumptions of model parameters, minimization technique) can play a role in the resulting escape velocity. The treatment of substructures, like the Gaia-Sausage-Enceladus \citep{Belokurov2018,Helmi2018Nature} and stellar streams \citep[e.g.,][]{Ibata2019} around the Solar neighborhood can also lead to discrepancies in estimates of the Galactic escape velocity \citep{Grand2019,Necib2021I}. These differences can lead to underestimating the escape velocity by approximately $10$\, percent \citep[e.g.,][]{Smith2007,Grand2019}; thus, caution is needed in interpreting the derived escape velocity values.

In what follows, we explore both the large- and small-scale dynamical aspects of the MW environment by utilizing RR~Lyrae variables. RR~Lyrae pulsators are old \citep[above $10$\,Gyr, see, e.g.,][]{Savino2020} helium-burning giants, located on the horizontal branch inside the instability strip (IS), exhibiting periodic variations on scales from a few hours up to one day. RR~Lyrae stars can be divided into three main groups based on the pulsation mode, where fundamental-mode pulsators are denoted as RRab, first-overtone variables are referred to as RRc stars, and double-mode pulsators are marked as RRd. The periodicity in brightness changes is driven by the interplay between pressure and temperature and can be utilized together with metallicity to estimate the stars' luminosity (PLZ relations), and subsequently, their distance \citep{Bono2019}. All these properties make them invaluable distance indicators within the Local Group, and probes of the Galactic structure and dynamics \citep[e.g,][]{Koposov2019Orphan,Price-Whelan2019,Prudil2020Disk}.

RR~Lyrae stars with high velocities have been identified in the Galactic bulge by the Bulge Radial Velocity Assay for RR~Lyrae stars \citep[BRAVA-RR,][]{Kunder2015,Prudil2019Kin,Kunder2020}, and one of them was subsequently followed up by a spectroscopic study \citep{Hansen2016}. All of these appear to be bound and kinematically associated with the MW halo (although currently passing through the Galactic bulge), nonetheless, they provide important insight into the high-velocity distribution of the old population in the Galactic center. Unlike the majority of known unbound stars in the MW, RR~Lyrae variables are typically metal-poor stars \citep[$\text{[Fe/H]} \approx -1.55$\,dex,][]{Hansen2011,Crestani2021} associated with the old MW population ($>10$\,Gyr). It is important to mention that some \textit{old}, metal-poor stars with extreme velocities have been detected in the Solar neighborhood \citep[e.g.,][]{Hattori2018metPoor_LMC,Du2018,Huang2021}, with ages above 1\,Gyr, where one of the objects is a known RR~Lyrae star \citep[Gaia-T-ES14 in][]{Hattori2018metPoor_LMC}.

We present the second paper of our series focused on the old population pulsators as a tool to study the MW's formation history. The present study aims at identifying high-velocity RR~Lyrae stars in the MW halo that appear to be unbound and at modeling the velocity distribution at the Solar radius and beyond. The paper is organized as follows: Section~\ref{sec:RR-Setup} describes the dataset, criteria imposed on the spatial and kinematical parameters, and outlines the setup for the dynamical analysis. In Section~\ref{sec:WeShallSee} we examine the properties of unbound RR~Lyrae candidates. In Section~\ref{sec:HighTail} we attempt to model the high-velocity tail the of the RR~Lyrae distribution and estimate the Galactic escape velocity at different radii. Section~\ref{sec:MassOfMW} focuses on estimating the mass of the MW using the calculated escape velocities. Finally, Section~\ref{sec:Summary} summarizes our findings.

\section{Dataset and analysis} \label{sec:RR-Setup}

As an initial dataset, we utilized our sample of 4247 RR~Lyrae from the first paper of our series \citep{Prudil2021Orphan}, for which we have the complete 6D information on their position and motion (equatorial coordinates $\alpha$, $\delta$, proper motions $\mu_{\alpha^{\ast}}$, $\mu_{\delta}$, distances $d$, systemic velocities $v_{\rm sys}$) together with metallicities derived using the calibration by \citet{Crestani2021}, and used also in \citet{Fabrizio2021}. Our dataset draws from three sources; proper motions from the third early data release of the \gaia~astrometric mission \citep[EDR3,][]{GaiaEDR3Summary2020}, distances derived based on the $i$-band mean magnitudes obtained by the Panoramic Survey Telescope and Rapid Response System \citep[PanSTARRS-1,][]{Chambers2016,Sesar2017a}, and systemic velocities corrected for radius variation, determined from the low-resolution spectra collected by the Sloan Digital Sky Survey in its fifteenth data release \citep[SDSS DR15,][]{York2000,Aguado2019}. Details on the dataset, including estimating uncertainties, can be found in \citet{Prudil2021Orphan}. 

To increase our chances of finding RR~Lyrae variables with high velocities, we expand our initial dataset with the low-resolution spectra from the LAMOST\footnote{The LAMOST acronym stands for the Large Sky Area Multi-Object Fiber Spectroscopic Telescope, which refers to the Guoshoujing telescope.} Experiment for Galactic Understanding and Exploration \citep[LEGUE,][]{Deng2012,Liu2014} survey, specifically from its sixth data release (DR6, version 2). Similarly to the SDSS, LAMOST uses a multi-object fiber-fed spectrograph providing low-resolution spectra ($R\sim1800$) with a wavelength coverage between 3700\,\AA~and 9100\,\AA~\citep{Zhao2012Lamost}. Each spectrum contains flux, inverse variance of the flux, and vacuum wavelengths in the heliocentric reference frame. An object observed by the LAMOST survey has at least one co-added spectrum composed of 1 or up to 6 individual exposures (typically 2 or 3), where the exposures range from a few minutes up to 45 minutes. Due to the overlapping fields of view for better coverage, some objects have been observed several times with different plates and fibers. 

Using our RR~Lyrae catalog based on the variability sample from \gaia~data release 2 \citep[DR2][]{Clementini2019} and the Catalina sky survey \citep[CSS,][]{Drake2009,Drake2013,Drake2013stream,Drake2014CatVari,Abbas2014} we found matches for 3003 objects with a total of 4754 spectra, which also have $i$-band mean magnitudes in the \citet{Sesar2017a} catalog of RR~Lyrae stars. To estimate systemic velocities and distances for the matched stars, we proceeded similarly as in \citet[][see appendix A and B]{Prudil2021Orphan}:
\begin{itemize} 
\itemsep0em
\item{Using the \texttt{iSpec} package \citep{Blanco2014,Blanco2019iSpec} we synthesized a spectrum with a characteristic set of stellar properties for a halo type RR~Lyrae star \citep[see, e.g.,][]{For2011chem,Sneden2017,Crestani2021}. We used the following stellar parameters: [Fe/H]$ = -1.5$\,dex, $T_{\rm eff} = 6600$\,K, log\,$g = 2.25$\,dex, and microturbulence velocity $\xi_{\rm turb} = 3.5$\,km\,s$^{-1}$. For the spectral synthesis, we used a python wrapper implemented in the \texttt{iSpec} module for the MOOG radiative transfer code \citep[February 2017 version,][]{Sneden1973}, a line list from VALD\footnote{\url{http://vald.astro.uu.se/}}, the solar reference scale of \citet{Asplund2009}, and ATLAS9 model atmospheres \citep{Castelli2003}.}
\item{In the spectral synthesis, we focused on the four most prominent features in the low-resolution spectra of RR~Lyrae stars, the Balmer lines (H$\alpha$, H$\beta$, H$\gamma$, H$\delta$) and their close neighborhood (defined as the region around each Balmer profile $\pm$100\,\AA).}
\item{Wavelengths for each spectrum were transformed from the vacuum to air wavelengths using the same prescription as for the SDSS \citep{Ciddor1996}.}
\item{Individual line-of-sight velocities were obtained using  \texttt{iSpec} by cross-correlating each Balmer line with its synthetic counterpart. Uncertainties on the respective line-of-sight measurements were calculated through a Monte-Carlo error simulation, in which we varied fluxes using their uncertainties (under the assumption that they follow a Gaussian distribution).}
\item{The systemic velocities were determined by the minimalization process that utilizes spectroscopic products (line-of-sight velocities), photometry (pulsation type and ephemerides), and line-of-sight velocity templates \citep{Braga2021}.}
\item{We slightly modified our approach of estimating the systemic velocities due to little to no time overlap between the spectroscopic and photometric observations. Time offsets between photometric and spectroscopic observation can cause an issue since RR~Lyrae stars experience changes in their pulsation periods as they evolve off the zero-age horizontal branch. These variations can lead to shifts in the time of mean brightness, $M_\mathrm{0}$, which potentially results in a phase offset and subsequently affects the determination of systemic velocities. The accurate estimate of the pulsation phase during the spectroscopic observation plays a crucial role. Thus, for the case of the LAMOST spectra, we increased the boundaries for our uniform ($\mathcal{U}$) prior on $M_\mathrm{0}$ from 0.1 \citep[as for SDSS spectra see appendix B in][]{Prudil2021Orphan} to 0.2 to better sample the uncertainties on the posterior:
\begin{equation}
p(\boldsymbol{\theta}_{n}) = \mathcal{U}(-0.2 < \Delta M_{0,n} < 0.2)\\. 
\end{equation}}
\item{The systemic velocity, $v_\mathrm{sys}$, of a given star was estimated as a weighted average between systemic velocities determined from single Balmer profiles.}
\item{In the case of distances, we used $i$-band mean magnitudes from \citet{Sesar2017a}, together with the information on metallicity from \citet{Crestani2021} for the crossmatched LAMOST portion of our dataset. To account for reddening, we used recalibrated extinction maps from \citet{Schlafly2011}. All of the aforementioned quantities entered the PLZ relation from \citet[see Table 1,][]{Sesar2017a} and subsequently led to distances for individual objects from our LAMOST dataset.}
\end{itemize}
From the 3003 initial LAMOST RR~Lyrae cross-match (within 1 arcsec), we removed 242 objects due to their problematic spectra (low signal-to-noise ratio, unphysical behavior in the spectrum, etc.). Additionally, 821 variables we previously analyzed in the SDSS dataset; for those stars we used the values provided by the SDSS data products\footnote{For comparison, the difference between systemic velocities determined on the SDSS and LAMOST spectra for overlapping dataset centers at $-9$\kms~(lower than usual one $\sigma$ uncertainty of a given systemic velocity) with a dispersion of $28$\kms.}. The remaining 1940 objects were added to our dataset of 4247 RR~Lyrae stars from the SDSS sample, increasing our original sample to 6187 RR~Lyrae variables with 7D information on their spatial, dynamical, and metallicity distribution. 

\subsection{Astrometric cuts} \label{subsec:AstroCut}

To ensure the purity of our data sample, we employed a cut on the significance of proper motions similar to the one used in \citet{Prudil2021Orphan}. Unlike \citet{Prudil2021Orphan}, in our study, we required higher proper motion significance, and we introduced four additional metrics on the quality of the astrometric solution based on \citet{Fabricius2020EDR3}, \citet{Lindegren2020GaiaAstrometry}, and \citet{Smart2020CloseEDR3}:
\begin{gather} \label{eq:PMcut}
\sqrt{ \sum \mathbf{V}^{2} / \text{tr}(\mathbf{\Sigma^{\ast}_{\rm pm}}) } > 4.0 \\
\text{\texttt{ipd\_gof\_harmonic\_amplitude}} < 0.4 \\
\text{\texttt{ipd\_frac\_multi\_peak}} < 2. \\
\text{\texttt{astrometric\_excess\_noise\_sig}} \leq 2.0 \\
\text{\texttt{visibility\_periods\_used}} \geq 13 \hspace{1cm}. \label{eq:LastEq}
\end{gather}
Here, $\mathbf{V}$ and $\mathbf{\Sigma^{\ast}_{\rm pm}}$ represent the star's proper motion vector and its diagonalized proper motion covariance matrix scaled by the re-normalized unit weight error (RUWE factor), respectively. Scaling the covariance matrix by the RUWE factor enabled us to weed out stars with unreliable astrometric solutions (mainly stars with RUWE~$ > 1.4$). The \texttt{ipd\_gof\_harmonic\_amplitude} denotes the amplitude of changes in the goodness-of-fit in image parameter determination, and together with \texttt{ipd\_frac\_multi\_peak}, it indicates the amount of image asymmetry, which implies possible binarity. The metric on how well the astrometric model matches the observations is expressed in the \texttt{astrometric\_excess\_noise\_sig} flag, which can indicate whether the object is astrometrically well-behaved. The last criterion on \texttt{visibility\_periods\_used} ensured that our selected variables were well-observed sources. The criteria in Eqs.~\ref{eq:PMcut}-\ref{eq:LastEq} reduced our initial sample from 6187 RR~Lyrae stars down to \sample\footnote{We note that our astrometric cuts yield similar results as using a classifier of spurious astrometry \citep{Rybizki2022} in $98.7$\% cases.}, which we will refer from here on as the study sample. 

\subsection{Estimating of dynamical properties and analysis setup} \label{subsec:SetupDyn} 

To find and examine the possible extreme velocity candidates among our study sample, we utilized the python package for Galactic dynamics, \texttt{galpy v1.6}\footnote{Available at \url{http://github.com/jobovy/galpy}.}\citep{Bovy2015}. In our setup, we assumed the distance between Sun and the Galactic center to be equal to R$_{\odot} = 8.2$\,kpc \citep{Bland-Hawthorn2016}, and we placed the Solar system at Z$_{\odot} = 20.8$\,pc \citep{Bennett2019}, thus slightly above the Galactic plane. For the Solar peculiar motion we used $\left(U_{\odot}, V_{\odot},W_{\odot} \right) = \left(11.1, 12.24, 7.25\right)$\kms~from \citet{Schonrich2010}, and we set the rotation velocity to $230$\kms \citep{Eilers2019}. 

With the purpose of examining the orbits and possible ejection mechanism of high-velocity RR~Lyrae stars, we took advantage of the composite MW potential implemented in the \texttt{galpy} package (\texttt{MWPotential2014}). The \texttt{MWPotential2014} is a combination of three gravitational potentials representing the Galactic disk \citep{Miyamoto1975}:
\begin{equation}\label{eq:MWdisk}
\Phi_d(R, z) = -\frac{\mathrm{G} M_\mathrm{d}}{\sqrt{R^2 + \left(a_\mathrm{d} + \sqrt{z^2 + b_\mathrm{d}^2}\right)^2}} \\,
\end{equation}
where G is the gravitational constant, $M_\mathrm{d}$ represents the mass of the disk, and $a_\mathrm{d}$ and $b_\mathrm{d}$ stand for scale length and height, respectively. 

The Galactic bulge in the \texttt{MWPotential2014} is implemented as a power-law density profile with an exponential cut-off potential:
\begin{equation}\label{eq:MWbulge}
\rho_{r} = \mathrm{G} M_\mathrm{b} \left ( \frac{r_1}{r} \right )^{\alpha} \mathrm{exp}\left ( -\left ( \frac{r}{r_\mathrm{c}} \right )^{2} \right ) \\,
\end{equation}
where $r_\mathrm{c}$ represents the cut-off radius, $r_1$ stands for a reference radius, and $\alpha$ serves as the inner power-law index. Finally, $M_\mathrm{b}$ represents the mass of the Galactic bulge. 

The spherical MW halo is defined as the Navarro-Frenk-White spherical dark matter halo \citep[NFW,][]{Navarro1997}: 
\begin{equation} \label{eq:MWhalo}
\rho_{r} = \frac{\mathrm{G} M_\mathrm{h}}{4\pi a^3} \frac{1}{\left ( r / a \right ) \left (1+ r / a \right )^2} \\,
\end{equation}
where $M_\mathrm{h}$ defines the mass of the Galactic halo, and the scale radius is represented by the variable $a$. For a detailed description of the physical properties of the potential on the individual MW substructures, see Tab.~1 in \citet{Bovy2015}. The original implementation of the \texttt{MWPotential2014} adopts a virial mass of the MW of $M_{\rm vir} = 0.8\cdot 10^{12}$\,M$_{\odot}$ which lies on the low-end of the virial mass estimates for the MW \citep[see Fig.~7 in][]{Callingham2019}. For comparison on how our results depend on the assumed MW mass, we explored two additional setups of the \texttt{MWPotential2014} with slightly larger virial mass; we will refer to them similarly to the initial potential as \texttt{MWPotential2014}$^{\rm M}$ ($25$ percent mass increase) and \texttt{MWPotential2014}$^{\rm H}$ ($50$ percent mass increase). 

Using \texttt{galpy}, we estimated the Galactocentric rectangular coordinates ($x$, $y$, and $z$) and velocities ($v_{x}$, $v_{y}$, and $v_{z}$). We used a multivariate Gaussian distribution where individual values for astrometric properties, distances, and systemic velocities are treated as a mean vector. For the full covariance matrix of the individual quantities, we utilized uncertainties and correlations provided in the \textit{Gaia}~EDR3 together with our estimates on distance and systemic velocity (where off-diagonal elements for distances and systemic velocities were set to zero). 

We estimated the full covariances between individual coordinates and velocities by running a Monte Carlo (MC) error sampling (for 100\,000 iterations). We also assessed the chance for a given star to be unbound ($P_{\rm unb}$) at the given escape velocity, $v_{\rm esc}$, the profile of the \texttt{MWPotential2014} and the star's Galactocentric radius. In this case, we use a definition of escape velocity from the \texttt{galpy} module, where it is defined as a velocity necessary to escape to infinity\footnote{Infinity (INF), in this case, is described as the radius at $10^{12}$.}:
\begin{equation} \label{eq:DecisionEQ}
v_{\rm esc}\left ( \text{R}_{\odot} \right ) = \sqrt{2\left | \Phi \left ( \text{R}_{\odot} \right ) - \Phi \left ( \text{INF} \right ) \right |} .
\end{equation}
where the potential at infinity is $\Phi \left ( \text{INF} \right )=0$ and $R_{\odot}$ represents the Solar radius. For each star, we estimated whether it exceeds $v_{\rm esc}$ at its Galactocentric radii in each MC iteration. For a given star to be considered \textit{unbound} in the assumed MW potential, we required it to exceed $v_{\rm esc}$ in at least half of the cases of the MC simulations: 
\begin{equation} \label{eq:Pubcon}
P_{\rm unb} > 0.5 \\.
\end{equation}
Each RR~Lyrae star that satisfied the above condition was further examined. We will refer to them from here on as the high-velocity RR~Lyrae stars (HV-RRL). We note that variable $P_{\rm unb}$ does not represent the probability that each star is bound or unbound considering our model.

\section{High velocity variables} \label{sec:WeShallSee}

Based on our setup from the previous Section~\ref{sec:RR-Setup}, we found in total \numberR~RR~Lyrae variables that fulfilled the condition in Eq.~\ref{eq:Pubcon}. Their complete list, with calculated properties and associated covariances, and \gaia~astrometric properties, can be found in Tab.~\ref{tab:HV-RR} and Tab.~\ref{tab:HV-RRastr} in the appendix. Besides the quality cuts in Section~\ref{subsec:AstroCut} all \numberR~candidates have RUWE$ < 1.4$, and none of them had been flagged as a \texttt{duplicated\_source} in \gaia~EDR3. In the following subsections, we will discuss their observed and calculated properties (displayed in Fig.~\ref{fig:HVRRProp}), and their RR~Lyrae classification. 

\subsection{Position in the color--magnitude diagram and variable classification} \label{subsec:CMD}

\begin{figure*}
\includegraphics[width=2\columnwidth]{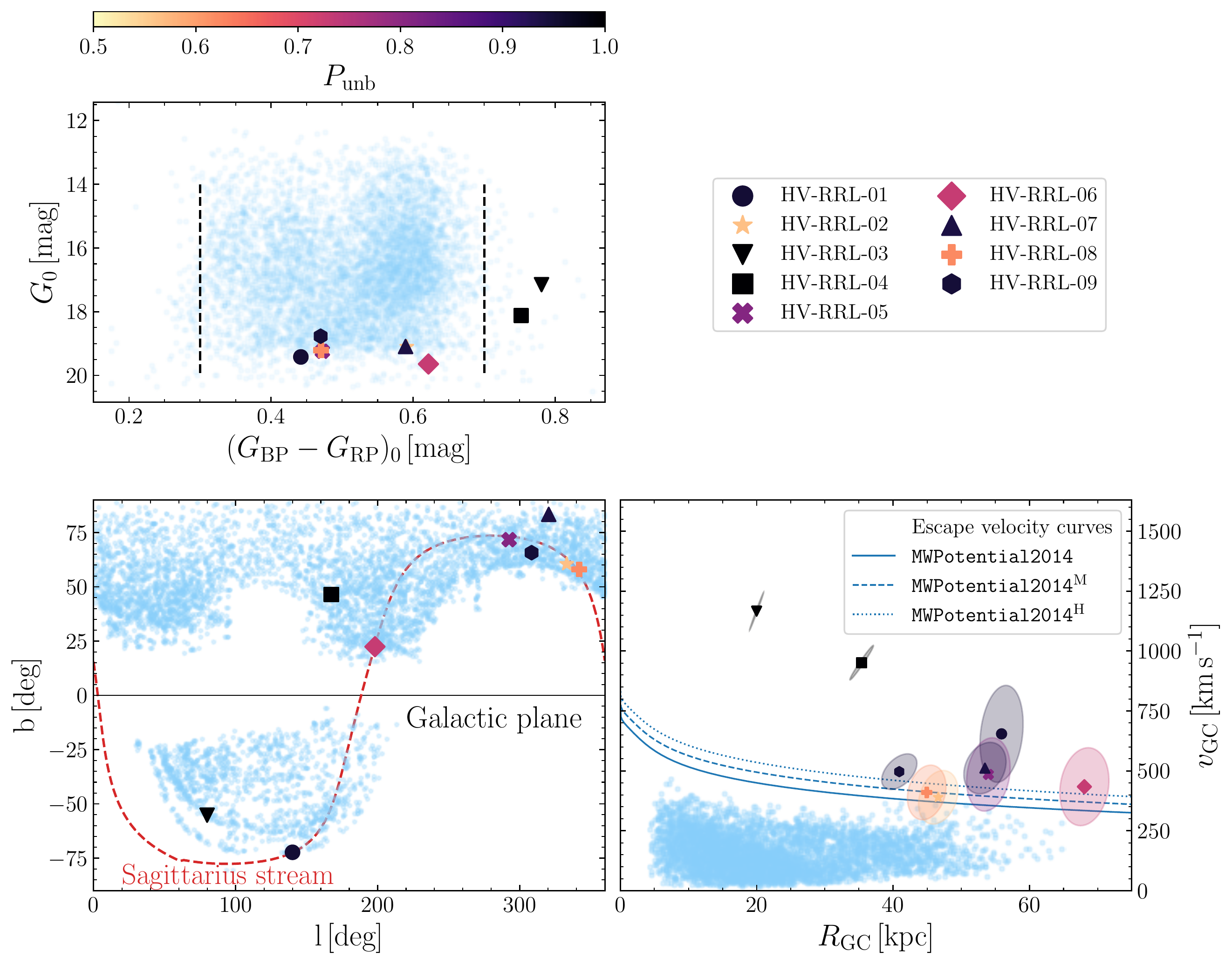}
\caption{Observed and estimated properties for our study sample of RR~Lyrae stars (light blue dots) and the candidates for high-velocity RR~Lyrae stars (HV-RRL, with individual markers color-coded on the basis of the $P_{\rm unb}$). The top left panel depicts the color--magnitude diagram based on the \gaia~passbands, the bottom left panel shows the spatial distribution in Galactic coordinates of our found HV-RRL candidates. The Galactic rest-frame velocity $v_{\rm GC}$ with respect to the Galactocentric distance $R_{\rm GC}$ is depicted in the bottom right panel. For each point fulfilling the condition in Eq.~\ref{eq:PMcut}, we plotted also its 1$\sigma$ covariance in the form of an error ellipse using the same color-coding as for the data point. In the bottom left panel, the Galactic plane is denoted by the solid black line, and the red dashed line outlines the Sagittarius stream. Individual blue lines in the bottom right panel represent escape velocity curves calculated using a different gravitational potential for the MW with the solid line representing the \texttt{MWPotential2014}, the dashed line denoting the \texttt{MWPotential2014}$^{\rm M}$, and the dotted line marking the \texttt{MWPotential2014}$^{\rm H}$.}
\label{fig:HVRRProp}
\end{figure*}

In the top left panel of Fig.~\ref{fig:HVRRProp} we display the color--magnitude diagram (CMD) for the study sample of RR~Lyrae variables, via the \gaia~photometry dereddened using the reddening maps from \citet{Schlafly2011} and the extinction coefficients from \citet{Casagrande2018}. In the CMD we see two populations separated by color, one at $(G_{\rm BP} - G_{\rm RP} )_{0} \approx 0.4$\,mag (characterizing first-overtone RR~Lyrae variables), and a second one at $(G_{\rm BP} - G_{\rm RP} )_{0} \approx 0.6$\,mag (representing fundamental-mode RR~Lyrae stars). 

We see that two stars (HV-RRL-03, HV-RRL-04), in the top left panel of Fig.~\ref{fig:HVRRProp}, with the highest value of $P_{\rm unb}$ ($P_{\rm unb} = 1.0$, in all three assumed MW potentials) are located beyond the assumed red edge of the IS for the distribution of the RR~Lyrae stars. Both of these stars were identified as RR~Lyrae stars in the Catalina sky survey \citep[CSS,][]{Drake2009} in their variable stars catalog \citet{Drake2014CatVari}. They were also found in the catalog of RR~Lyrae stars based on the Panoramic Survey Telescope and Rapid Response System \citep[PanSTARRS-1, PS1 catalog,][]{Chambers2016,Sesar2017a}. Unlike the CSS variable catalog, we reclassified HV-RRL-04 from a first-overtone pulsator to a fundamental-mode RR~Lyrae star, based on the re-analysis of its CSS photometric data. We established a different pulsation period in comparison to the CSS catalog \citep[for details see][]{Prudil2021Orphan}. The new classification and pulsation period are in line with the classification by the PanSTARRS-1 survey. It also agrees with the dereddened colors illustrated in the CMD, where fundamental pulsators, on average, lie closer to the red edge of the IS. This is in contrast with the first-overtone pulsators, which are generally located closer to the blue edge of the IS. To verify our classification for HV-RRL-03 and HV-RRL-04, we used a recently published catalog of variable stars in the Zwicky Transient Facility \citep[ZTF,][]{Masci2019ZTF,Bellm2019ZTF,Chen2020ZTFVAR}, a time-domain survey of the entire northern sky in the $g$ and $r$ passbands, respectively. The ZTF variable stars catalog contains photometry for both of the aforementioned stars, and it was utilized in our effort to confirm or refute our initial classification. 

The HV-RRL-03 (in ZTF: \texttt{SourceID}$=764936$) is classified as a BY~Dra type variable\footnote{BY~Draconis-type stars belong to the rotation class of dwarf variables with a spectral classification ranging from K to M with an occasional chromospheric activity in the form of starspots.}. Both our and ZTF period analyses yield a similar period of $P=0.61771$\,day and low amplitudes of light changes: Amp$(g)=0.233$\,mag, Amp$(r)=0.203$\,mag, Amp$^{V_{\rm CSS}}=0.17$\,mag. Under the assumption that HV-RRL-03 is an RR~Lyrae star, its low amplitude would be in agreement with its position at the red edge of the IS. If we assume its classification as a BY~Dra variable, we would expect it to have a redder color (lower effective temperature, $T_{\rm eff}$, approximately between $2500$ to $5200$\,K), while using \gaia~colors and a prescription for $T_{\rm eff}$ from \citet{Mucciarelli2020TempGaia}, we obtained temperatures between $5400$ to $5800$\,K for a range of metallicities \citep[expected for RR~Lyrae stars from $0$\,dex up to $-3$\,dex][]{Crestani2021}, and both for giants and dwarfs. This temperature range is also supported by the color-temperature relation for $(V-K_{s})_{0}$ from \citet{Ramirez2005} and \citet{Gonzalez2009}. From the latter we obtained a dereddened color for HV-RRL-03 $(V-K_{s})_{0}=1.41$\,mag\footnote{We note that this value lies close to the edge of applicability as defined both in \citet{Ramirez2005} and \citet{Gonzalez2009}.} using $V$-band measurements from the CSS survey in combination with $K_{s}$-band measurements from the Two Micron All Sky Survey \citep[2MASS,][]{Cutri2003}. We conclude that the $T_{\rm eff}$ of HV-RRL-03 lies around $\approx 5900$\,K. This range of effective temperatures is more in line with HV-RRL-03 being an RR~Lyrae star, although from photometry its classification remains uncertain. On the other hand, in \gaia~astrometry the parallax value $\varpi = 0.45 \pm 0.11$\,mas suggests that HV-RRL-03 lies at $d=2182^{+625}_{-488}$\,pc \citep{Bailer2021}, thus closer than our estimate based on period-metallicity-luminosity relation ($d = 19 \pm 1$\,kpc). Such a large difference in the distance would significantly change the calculated $v_{\rm GC}$ and $R_{\rm GC}$, and subsequently mark HV-RRL-03 as bound.

The star HV-RRL-04 (in ZTF: \texttt{SourceID}$=190951$) is classified in the ZTF catalog as an EA variable\footnote{This label refers to a class of evolved detached binaries, $\beta$~Persei type (Algol-type).} with a period $P=1.2928597$\,days\footnote{In our analysis we found $P=0.649351$\,day based on the CSS data.}, which we subsequently confirmed by re-analyzing the ZTF photometry \citep[using the \texttt{Period04} software,][]{Lenz2004Period04}. We corroborated the pulsation period and variability type from the ZTF survey based on the dispersion of the points in the phased light curves. Both stars, HV-RRL-03 and HV-RRL-04, are marked with an asterisk in Tab.~\ref{tab:HV-RR} and in Tab.~\ref{tab:HV-RRastr} since their classification as RR~Lyrae stars is highly uncertain (in the case of HV-RRL-03) or incorrect (in the case of HV-RRL-04), but for the remainder of this paper, we will treat them as RR~Lyrae stars. 

The remaining seven RR~Lyrae variables are located inside the IS, and their position in the CMD matches their assumed RR~Lyrae sub-type. Three RR~Lyrae stars from our sample (HV-RRL-02, HV-RRL-05, HV-RRL-09) have been identified by the ZTF survey as variable stars, and in all three objects, both pulsation types and periods agree between our sample and the ZTF survey. One RR~Lyrae variable, HV-RRL-08, has been identified as a variable star candidate by the ZTF survey with a period of $0.33348$\,days while in our analysis we found a pulsation period of $0.500229$\,days. In the frequency space, both pulsation periods lie just $~1$\,cd$^{-1}$\footnote{Counts-per-day} apart, implying that one or the other is a daily alias of the true pulsation period. A careful examination of the light curves phased using both pulsation periods, together with a high amplitude of for HV-RRL-08 (Amp$^{\rm CSS} = 0.980$\,mag), supports its classification as a fundamental-mode RR~Lyrae star with a pulsation period of $0.500229$\,days. The complete list of pulsation parameters, metallicities and classification into RR~Lyrae sub-types for the \numberR~RR~Lyrae variables can be found in Table~\ref{tab:Puls}. In addition, the phased light curves of the \numberR~high-velocity RR~Lyrae candidates are shown in Fig.\ref{fig:LightCurves} in the appendix. 

\begin{table} 
\caption{List of pulsation and chemical properties for high-velocity RR~Lyrae candidates found in our analysis. The first column lists our adopted object ID. Columns 2 and 3 provide the star's pulsation period $P$ and amplitude of the total light changes based on the CSS photometry. In the last two columns, we list the assumed RR~Lyrae sub-class and metallicity derived using the $\Delta$S method. Similarly to Table~\ref{tab:HV-RR} we marked variables with uncertain classification with an asterisk.}
\label{tab:Puls}
\begin{tabular}{lcclc}
\hline
ID & $P$ & Amp$^{V_{\rm CSS}}$ & Type & [Fe/H] \\
 & [day] & [mag] & & [dex] \\ \hline
HV-RRL-01 & 0.33118 & 0.34221 & RRc & $-0.90 \pm 0.08$ \\ 
HV-RRL-02 & 0.56564 & 0.91313 & RRab & $-0.80 \pm 0.13$ \\ 
HV-RRL-03$^{\ast}$ & 0.61772 & 0.17046 & RRab & $-1.36 \pm 0.19$ \\ 
HV-RRL-04$^{\ast}$ & 0.39219 & 0.26468 & RRc & $-1.92 \pm 0.19$ \\ 
HV-RRL-05 & 0.36655 & 0.44251 & RRc & $-1.78 \pm 0.19$ \\ 
HV-RRL-06 & 0.40890 & 0.74320 & RRab & $-1.82 \pm 0.19$ \\ 
HV-RRL-07 & 0.75194 & 0.45923 & RRab & $-1.64 \pm 0.19$ \\ 
HV-RRL-08 & 0.50023 & 0.97984 & RRab & $-1.06 \pm 0.19$ \\ 
HV-RRL-09 & 0.35240 & 0.42820 & RRc & $-1.55 \pm 0.51$ \\ 
\hline
\end{tabular}
\end{table}

\subsection{Kinematical and spatial properties of the high-velocity sample} \label{subsec:Kin}

The lower panels of Fig.~\ref{fig:HVRRProp} depict the kinematical properties for our study sample of RR~Lyrae variables. Pulsators identified potentially as unbound with the selected potential (see Sect.~\ref{sec:RR-Setup}) are color-coded by their $P_{\rm unb}$. The solid, dashed, and dotted lines denote escape velocity curves for three different gravitational potentials of the MW (see Sect.~\ref{sec:RR-Setup}). We see that besides the two HV-RRL stars with the highest $P_{\rm unb}$, and at the same time also probably misclassified as RR~Lyrae stars ($P_{\rm unb} =1.00$, for HV-RRL-03 and HV-RRL-04), the rest of our high-velocity sample lies beyond $40$\,kpc in Galactocentric distance. The large distance of these seven RR~Lyrae stars leads to potential issues of whether they can genuinely be considered unbound. Mainly due to their considerable error in $v_{\rm GC}$ and $R_{\rm GC}$ depicted with error ellipses in the right-hand panel of Fig.~\ref{fig:HVRRProp}, and because none of them appears to be unbound due to its sizeable systemic velocity, their unbound status is uncertain. These seven pulsators might as well be $\approx 1 - 2$\,$\sigma$ outliers, especially since their unbound status is based heavily on their large tangential velocities and not on $v_{\rm sys}$. Notably, the uncertainties in tangential velocities at such distances are dominated by uncertainties in proper motions, highlighting their probably spurious unbound origin.

Moreover, we varied the value for the rotation velocity \citep[previously set to $230$\kms,][]{Eilers2019} between $220$\kms~and $240$\kms to assess the influence of rotation velocity on $P_{\rm unb}$. We found that some of our HV-RRL do not fulfill the condition in Eq.~\ref{eq:Pubcon} when the rotation velocity is changed, particularly those with $P_{\rm unb}$ in the range between $0.5$ and $0.6$. In addition, looking at the number of identified HV-RRL candidates out of the entire sample (\numberR~out of \sample) casts a doubt on high-velocity nature specially looking at the yield of high-velocity stars in samples. For instance, \citet{Li2021High591} reported $591$ high-velocity star candidates out of $\approx10$ million stars, or case of \citet{Marchetti2019} where $125$ high-velocity stars were reported out of $\approx7$ million stars. Based on these numbers, generaly speaking, we would expect less than a single RR~Lyrae star from our sample to be unbound. In addition, from the statistical point of view, having a large dataset of~\sample, it is likely to have three or more $\sigma$ outliers from the general kinematical distribution, therefore, resulting in high $P_{\rm unb}$.

Regarding the possible origin of our HV-RRL candidates we integrated their orbits backward for $6$\,Gyr. During the integration, we observe each crossing of $z_{\rm GC}=0.0$\,kpc and mark the crossing radius:
\begin{equation}
r_{\rm cr} = \sqrt{x^{2}_{\rm GC} + y^{2}_{\rm GC}} \\. 
\end{equation}
This approach helped us to identify objects that could have been ejected from the Galactic disk (hyper runaway stars) or the Galactic center (hypervelocity stars). We estimated the minimum crossing radius ($r_{\rm min}$, calculated using the MC simulation of orbital integration) for individual HV-RRL and compared it to an arbitrary threshold marking the edge of MW disk \citep[usually assumed $25$\,kpc, see, e.g.,][]{Marchetti2019,Li2021High591,Marchetti2021}. Our analysis has shown that none of the found candidates for high-velocity RR~Lyrae stars appears to have originated in the two central MW stellar substructures. This is expected, since our sample of HV-RRL candidates is located at large distances and has highly tangential orbits.

The bottom right-hand panel of Fig.~\ref{fig:HVRRProp} depicts the spatial distribution of our HV-RRL candidates in Galactic coordinates. Here we see that seven out of \numberR~HV-RRL candidates are located along the Sagittarius stellar stream in the coordinate space. Three stars in the HV-RRL candidates dataset (HV-RRL-02, HV-RRL-07, HV-RRL-08) had been previously linked with the Sagittarius stream based on the spatial study of the Sagittarius stream by \citet{Hernitschek2017}. Based on this overlap, we decide to test whether some of the HV-RRL candidates can be associated with the Sagittarius dwarf galaxy (from here on referred to as Sgr) based on their orbit using the \texttt{galpy} package.

We implemented the aforementioned satellite as a Hernquist profile \citep{Hernquist1990} with a mass and scale radius from Table~1 in \citet[][and references therein]{Garrow2020}. We incorporated the dynamical friction affecting the orbit of the Sgr within the MW by utilizing the Chandrasekhar prescription \citep{Chandrasekhar1943} implemented in \texttt{galpy}. To determine whether any of the HV-RRL can be associated with Sgr, we integrated their orbits backward in time for $3$\,Gyr utilizing the MC error simulation (1000 iterations). In each step, we varied the orbital properties of Sgr\footnote{We used uncertainties on proper motions, distances and systemic velocities from \citet{McConnachie2020Dwarfs} and \citet{Vasiliev2020Sag}.} and the given HV-RRL, by their uncertainties by assuming they follow a Gaussian distribution. We also utilized the full covariance matrix as provided in the \gaia~EDR3 data products. In addition, in each alteration, we randomly selected one of the three potentials implemented for the MW (see Sect.~\ref{subsec:SetupDyn} for details). To assess the possible association of individual HV-RRL with the Sgr, in each MC realization, we estimated whether a given HV-RRL becomes associated with the Sgr by measuring the total distance between HV-RRLs and the Sgr. When the total distance decreased below two times the Sgr's scale radius we considered given HV-RRL linked with the Sgr. The resulting association probabilities were estimated based on the number of instances when we linked Sgr and a given HV-RRL divided by the number of iterations. We note that each HV-RRL is represented as a mass-less particle which we follow over the past $3$\,Gyr.

Considering only stars that exhibited a probability of being bound in more than ten percent of the instances, we found two stars that may be weakly linked with the Sgr. The stars HV-RRL-02 and HV-RRL-08 both show probabilities of $0.13$, of being linked to Sgr in the past $3$\,Gyr, marking a possibility that they have been ejected recently in one of its pericentric approaches \citep[see Fig.~9 in][]{Vasiliev2021Tango}. From the metallicity point of view (listed in Table~\ref{tab:Puls}), two HV-RRL stars fit reasonably well to the metallicity distribution of the Sgr stream and Sgr \citep{Hayes2020SagStrGrad}. Both of them had been previously marked as members of the Sgr stellar stream based on their spatial properties \citep{Hernitschek2017}. Therefore, two stars in our sample could be tentatively linked with the Sgr in our simplified simulation. A more sophisticated approach taking into account the disruption of Sgr, its mass-loss, and the LMC in the MW halo could constrain more realistically the link between these two RR~Lyrae stars and the Sgr.

\section{High velocity tail} \label{sec:HighTail}

In this section, we focus on those RR~Lyrae variables that possess high $v_{\rm GC}$ but remain bound to the MW. In the bottom right panel of Fig.~\ref{fig:HVRRProp} a few hundred stars from our study sample have $v_{\rm GC} > 250$\kms~and populate the high-velocity tail of the halo RR~Lyrae stars. The $v_{\rm GC}$ distribution of stars that occupy a region close to the escape velocity can be described by a power-law distribution function, $f$, as derived by \citet{Leonard1990}:
\begin{equation} \label{eq:distFC}
f\left ( v_{\rm GC}| v_{\rm esc}, k \right ) \propto \left ( v_{\rm GC} - v_{\rm esc} \right )^{k} , v_{\rm cut} < v_{\rm GC} < v_{\rm esc} \\,
\end{equation}
where $v_{\rm esc}$ is the escape velocity, and the free parameter $k$ is the exponent. The $v_{\rm cut}$ represents the lowest threshold for $v_{\rm GC}$ (usualy assumed between $250$\kms to $300$\,\kms) where the velocity distribution can still be described by Eq.~\ref{eq:distFC}. 

Equation~\ref{eq:distFC} assumes a velocity distribution that is populated up to the escape velocity and that does not contain any disk contaminants and unbound objects. The first assumption is often broken since the distribution is not always smooth due to substructures, and the velocity distribution is often truncated below the true escape velocity \citep{Smith2007,Grand2019,Koppelman2021}. Due to this effect, the impact of underestimating the escape velocity is around $7$ to $10$ percent below the actual escape velocity \citep{Grand2019,Koppelman2021}.

The last assumption on the purity of the sample is somewhat easier to ensure, since unbound objects have been indentified in Section~\ref{sec:WeShallSee}. The condition on excluding disk objects can be achieved using the Toomre diagram and probabilistic procedures described in \citet{Bensby2003,Bensby2014} or by the Toomre selection \citep[based on the velocity of the local standard of rest, used, e.g., in][]{Nissen2010,Bonaca2017,Koppelman2021}. In our work, we used the Toomre selection to select only halo RR~Lyrae stars that fulfilled following condition:
\begin{equation}
\left | \mathbf{v} - \mathbf{v}_{\rm LSR}\right | \geq 230\,\,\text{km\,s$^{-1}$} \\.
\end{equation}
where $\mathbf{v}_{\rm LSR}$ represents a velocity vector with respect to the local standard of rest $\mathbf{v}_{\rm LSR} = (0, 230, 0)$\kms, and $\mathbf{v} = (v_{x}, v_{y}, v_{z})$ is a velocity vector.

\subsection{Modeling approach}

We adopted the formalism prescribed in \citet{Koppelman2021}, which takes into account uncertainties in $v_{\rm GC}$ \citep[similarly to the study by][]{Necib2021I} by convolving the distribution function $f$ with a Gaussian function that is represented by the measured $v_{\rm GC}$, its uncertainities $\sigma_{v_{\rm GC}}$, and the true Galactocentric velocity $v{'}_{\rm GC}$:
\begin{equation}
C\left (v_{\rm GC}, \sigma_{v_{\rm GC}}, \boldsymbol{\theta}_{\rm M} \right ) = \int_{v_{\rm cut}}^{v_{\rm esc}} f\left ( v{'}_{\rm GC}| \boldsymbol{\theta}_{\rm M} \right ) g(v{'}_{\rm GC} | v_{\rm GC}, \sigma_{v_{\rm GC}}) \text{d}v \\.
\end{equation}
where $\boldsymbol{\theta}_{\rm M}$ represents the model parameters ($k, v_{\rm esc}, v_{\rm cut}$), and $g(v{'}_{\rm GC} | v_{\rm GC}, \sigma_{v_{\rm GC}})$ is defined as:
\begin{equation}
g(v{'}_{\rm GC} | v_{\rm GC}, \sigma_{v_{\rm GC}}) = \frac{1}{\sqrt{2\pi \sigma_{v_{\rm GC}}}} \exp\left ( - \frac{\left ( v_{\rm GC}-v{'}_{\rm GC} \right )^{2}}{2\sigma_{v_{\rm GC}}^{2}} \right )  \\.
\end{equation}
The likelihood of this approach is described as:
\begin{equation}
P\left (v_{\rm GC}, \sigma_{v_{\rm GC}} | \boldsymbol{\theta}_{\rm M} \right ) = \prod_{i=1}^{N} C\left (v^{i}_{\rm GC}, \sigma^{i}_{v_{\rm GC}}, \boldsymbol{\theta}_{\rm M} \right ) \\,
\end{equation}
with $N$ representing the number of stars fulfilling the condition $v_{\rm GC} > v_{\rm cut}$. In addition, since the $v_{\rm esc}$ varies with the Galactic radius \citep[see, e.g.,][]{Monari2018,Koppelman2021}, and due to the large range over which we determine $v_{\rm esc}$ we modified our approach to take into account changes in escape velocity over radii. We parametrized $v_{\rm esc}$ following prescription by \citet{Williams2017} and \citet{Deason2019}:
\begin{equation} \label{eq:repGAM}
v_{\rm esc} = v_{\rm esc}^{\rm 0} \left ( R / \text{R}_{\odot} \right )^{-\gamma / 2} \\,
\end{equation} 
where $v_{\rm esc}^{\rm 0}$ stand for escape velocity at the Solar radius. Parameter $\gamma$ represents gravitational potential slope and enters $\boldsymbol{\theta}_{\rm M} = (k, \gamma, v_{\rm esc}, v_{\rm cut})$ as a model parameter. Applying the Bayesian formalism to the likelihood function above yields:
\begin{equation} \label{eq:LHood}
P\left (\boldsymbol{\theta}_{\rm M} | v_{\rm GC}, \sigma_{v_{\rm GC}} \right ) \propto P(k)P(\gamma)P(v_{\rm esc}) P\left (v_{\rm GC}, \sigma_{v_{\rm GC}} | \boldsymbol{\theta}_{\rm M} \right ) \\,
\end{equation}
where $P(k)$, $ P(\gamma)$, and $P(v_{\rm esc})$ represent the priors on $k$, $\gamma$ and $v_{\rm esc}$. 

The prior assumption on the two quantities mentioned above significantly influences the resulting posterior distribution. It has been the subject of debate since the introduction of the distribution function describing the high-velocity tail. In their seminal work, \citet{Leonard1990} preferred a lower range for the exponent $k$ (between $0.5$ to $2.5$), which was later supported by the study of \citet{Deason2019} based on the Auriga simulation suite \citep{Grand2017}. Other studies like \citet{Smith2007}, \citet{Piffl2014}, and \citet{Monari2018} used a higher range for $k$ (between $2.3$ up to $4.7$). Some recent studies, like \citet{Koppelman2021} and \citet{Necib2021I, Necib2021II}, assumed an agnostic approach on the value of $k$ by using a large range for $k$ \citep[$1$ to $6$ in][]{Koppelman2021} and \citep[$1$ to $15$ in][]{Necib2021I}. 

In our analysis, we selected a similarly broad uniform prior on $k$ like in studies by \citet{Koppelman2021} and \citet{Necib2021I} with the following properties:
\begin{equation}
P(k) = 0.1 < k < 10 \\.
\end{equation}
The consensus on the prior for the escape velocity in the literature has seemingly converged to:
\begin{equation}
P(v_{\rm esc}) \propto 1/v_{\rm esc} \\,
\end{equation} 
and we will utilize it in our study as well. We also included a range of possible values for $v_{\rm esc}$ in the following form:
\begin{equation}
v_{\rm cut} < v_{\rm esc} < 1000~\text{kms}^{-1} \\.
\end{equation}
where in our analysis we used $v_{\rm cut} = 250$\kms. For the parameter $\gamma$ we addopted simple uniform prior:
\begin{equation}
P(\gamma) = 0 < \gamma < 1 \\.
\end{equation}

\subsection{Results for the Solar neighborhood} \label{subsec:MCMCrun}

In the following, we determined $k$ and $v_{\rm esc}$ for the region around the Solar position ($4 < R_{\rm GC} < 12$\,kpc) and with $v_{\rm GC}$ above $250$\kms~by maximizing the log-likelihood in Eq.~\ref{eq:LHood}. From our study sample, $234$ stars\footnote{We note that we used a comparable sample size to studies like \citet{Monari2018} and \citet{Deason2019}.} fulfilled these two conditions. We note that we removed all \numberR~HV-RRL from the subsequent analysis since they would tamper with the results, particularly when analysing radii outside this local sample.

We utilized the Markov Chain Monte Carlo (MCMC) sampler, implemented in \texttt{emcee} \citep{Foreman-Mackey2013} to find the posterior distribution of $k$ and $v_{\rm esc}$ using $50$ walkers, an initial $3000$ steps as a burn-in, and restarted the sampler for an additional $2000$ steps. Using the \texttt{emcee} implementation we estimated the maximum autocorrelation length and thin the chains by $\tau = 19$ which yield $12600$ posterior samples of the $k$ and $v_{\rm esc}$ posterior distributions. 

\begin{figure}
\includegraphics[width=\columnwidth]{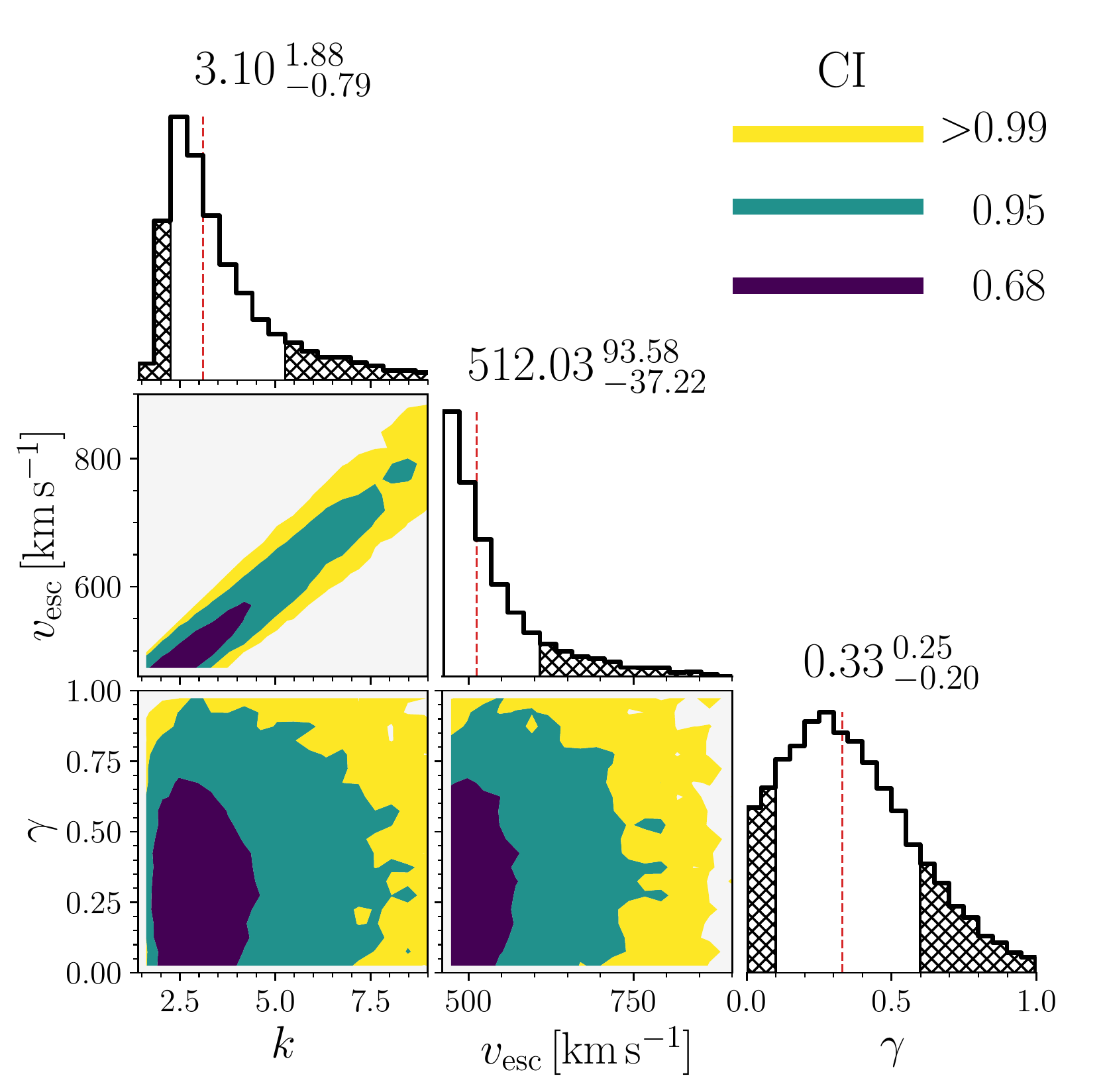}
\caption{The corner plot for the fitted parameters $k$, $v_{\rm esc}$ and $\gamma$, based on Eq.~\ref{eq:LHood}. The color contours denote the confidence intervals (CIs). The numbers above the histograms represent medians of distributions and their subscripts and superscripts stand for the 16th and 84th percentiles of the distribution.}
\label{fig:CornerSolar}
\end{figure}

In Fig.~\ref{fig:CornerSolar}, where the posterior distributions of $k$ and $v_{\rm esc}$ are displayed as corner plots, the degeneracy between both parameters is obvious. Since the distributions for both parameters have skewed distributions (non-Gaussians), we used their medians and percentiles ($15.9$ and $84.1$) to represent the resulting values of the fit. We found $v_{\rm esc} = 512^{+92}_{-38}$\kms and $k = 3.1^{+1.9}_{-0.8}$. The values for $k$ and $v_{\rm esc}$ agree very well with the values derived by \citet{Koppelman2021} \citet{Necib2021II}. On the other hand, the comparison with studies by, e.g., \citet{Deason2019} and \citet{Monari2018} shows that our estimate on $v_{\rm esc}$, lies slightly on the lower end, but still within the 16th and 84th of our percentile range. 

\subsection{Beyond the 12\,kpc radius} \label{subsec:MCMCrunBey}

The vast majority of the studies measuring the escape velocity of the MW focused on the Solar radius (from $4$\,kpc up to $12$\,kpc from the Galactic center) mainly due to the limitation on distance information (the lack thereof, or due to large uncertainties). Our dataset does not have such a limitation since our derived distances are not dependent on parallaxes, but on the period-metallicity-luminosity relation in the near-infrared $i$-passband. In addition, uncertainties on distances in our sample vary around $5$ to $6$ percent. Therefore, with a sufficient number of targets we can explore the high-velocity tail beyond the 12\,kpc radius.

Applying the same criteria on $v_{\rm GC}$ as in the previous subsection (using only variables with $v_{\rm GC} > 250$\kms) and using two ranges for $R_{\rm GC}$:
\begin{equation} \label{eq:VescCon1}
12 < R_{\rm GC} < 20\,\text{kpc}
\end{equation}
and
\begin{equation} \label{eq:VescCon2}
20 < R_{\rm GC} < 28\,\text{kpc} \\,
\end{equation}
we obtained $102$ and $97$ variables, respectively, that were used to determine $k$, $v_{\rm esc}$, and $\gamma$ beyond the Solar neighborhood. We proceeded in the same way as in Section \ref{subsec:MCMCrun}, with the same set of priors and the same setup for the MCMC sampling. Figure~\ref{fig:Corner12_20_28} shows results of our analysis for two regions beyond 12\,kpc. We see that with a decreasing number of stars that populate the high-velocity tail, the degeneracy between the parameters $k$ and $v_{\rm esc}$ starts to dominate the posterior distribution. 

In order to counter the degeneracy in both parameters, we explored the possibility of utilizing the posterior distribution of $k$ for the Solar neighborhood in regions further away from the Sun's position. This appears to be an appropriate path since $k$ only weakly varies with distance \citep[see Fig.~7 in][]{Koppelman2021} based on the \textit{Aurigaia} catalogues \citep{Grand2018} generated from the Auriga suite of cosmological magneto-hydrodynamical zoom simulations \citep{Grand2017}. Instead of fixing a value for $k$, like in \citet{Monari2018}, we used the entire posterior distribution of $k$ determined in the previous subsection prescribed by a kernel-density estimate (KDE)\footnote{We utilized the \texttt{scipy} \citep{scipy} implementation of 1D kernel-density estimate using Gaussian kernels.} and evaluated the probability density function at each sampled value of $k$.

In Figure~\ref{fig:Corner12_20_28_kde} we show the posterior distributions for both $R_{\rm GC}$ ranges obtained using the same MCMC setup as in Subsection~\ref{subsec:MCMCrun}. We see that the approximation for $k$ using a KDE improved the resulting posterior for the $v_{\rm esc}$ and in what follows we will use these estimates of escape velocity for regions beyond $12$\,kpc\footnote{We note that for regions beyond the Solar radius Eq.~\ref{eq:repGAM} was modified to represent parametrization of $v_{\rm esc}$ at larger radii, by replacing the R$_{\odot}$ with values of 16\,kpc and 24\,kpc that better represented the two regions (see Eq.~\ref{eq:VescCon1} and Eq.~\ref{eq:VescCon2}).}.

\section{Mass of the MW based on the RR~Lyrae velocity tail} \label{sec:MassOfMW}

The measured escape velocity at a given radius is connected with the gravitational potential of a galaxy, $\Phi$, and can be used to determine that galaxy's properties, e.g., its mass. In Section~\ref{sec:WeShallSee} we defined unbound stars as those exceeding the MW escape velocity estimated at infinite radius ($r=\infty$). In what follows, we will assume that the limiting radius for a star to be considered unbound is two times the virial radius $r_{\rm 200}$, which is a radius where the average galaxy density is $200$ times the critical density. This modification leads to a change in Equation~\ref{eq:DecisionEQ} to:
\begin{equation}
v_{\rm esc}\left (\text{R}_{\odot} \right ) = \sqrt{2\left | \Phi \left ( \text{R}_{\odot} \right ) - \Phi \left (2 r_{\rm 200} \right ) \right |} \\.
\end{equation}
To estimate the mass of the MW we use \texttt{MWPotential2014} as introduced in Section~\ref{subsec:SetupDyn}, and we define the NFW potential by the virial mass $M_{\rm 200}$ and concentration $c$. The escape velocity provides mainly insight into the mass distribution beyond the Solar radius, thus as a second constraint for the inner mass distribution. We utilize the circular velocity, which in our case is equal to v$_{\odot}=230$\kms \citep{Eilers2019}. We tested different values for v$_{\odot}$ (e.g., $220$\kms~and~$240$\kms), and their effect on $M_{\rm 200}$ and concluded that they have a small effect on the final $M_{\rm 200}$, which is still within one $\sigma$ of the estimated value.

To explore the range of possible parameters for our MW potential, we again employ the Bayesian formalism, as in the previous Section (see Sect.~\ref{sec:HighTail}):
\begin{equation} \label{eq:MassBay}
P\left (\boldsymbol{\theta}_{n} | v_{\rm obs}, \sigma_{\rm obs}^{+}, \sigma_{\rm obs}^{-} \right ) \propto P(\boldsymbol{\theta}_{n}) P\left (v_{\rm obs}, \sigma_{\rm obs}^{+}, \sigma_{\rm obs}^{-} | \boldsymbol{\theta}_{n} \right ) \\,
\end{equation}
In this case, we use Eq.~\ref{eq:MassBay} to estimate the mass and concentration based on the calculated escape velocities (at $8$, $16$, and $24$\,kpc,  which represent the centers of our radius ranges) and the circular velocity at the position of Sun. In Eq.~\ref{eq:MassBay}, $v_{\rm obs}$ stands for observed velocities (in our case escape and circular velocities) and $v_{\rm mod}$ represent the model velocities derived based on a set of parameters $\boldsymbol{\theta}_{n} = (M_{\rm 200}, c)$ for the MW potential. The upper and lower uncertainties ($\sigma^{+}_{\rm obs}$ and $\sigma^{-}_{\rm obs}$) are determined as the $15.9$th and $84.1$th percentiles of the posterior distributions for the escape velocities. For the circular velocity, we used a symmetric uncertainty of $10$\kms~similarly to \citet{Necib2021II}.

\begin{figure}
\includegraphics[width=\columnwidth]{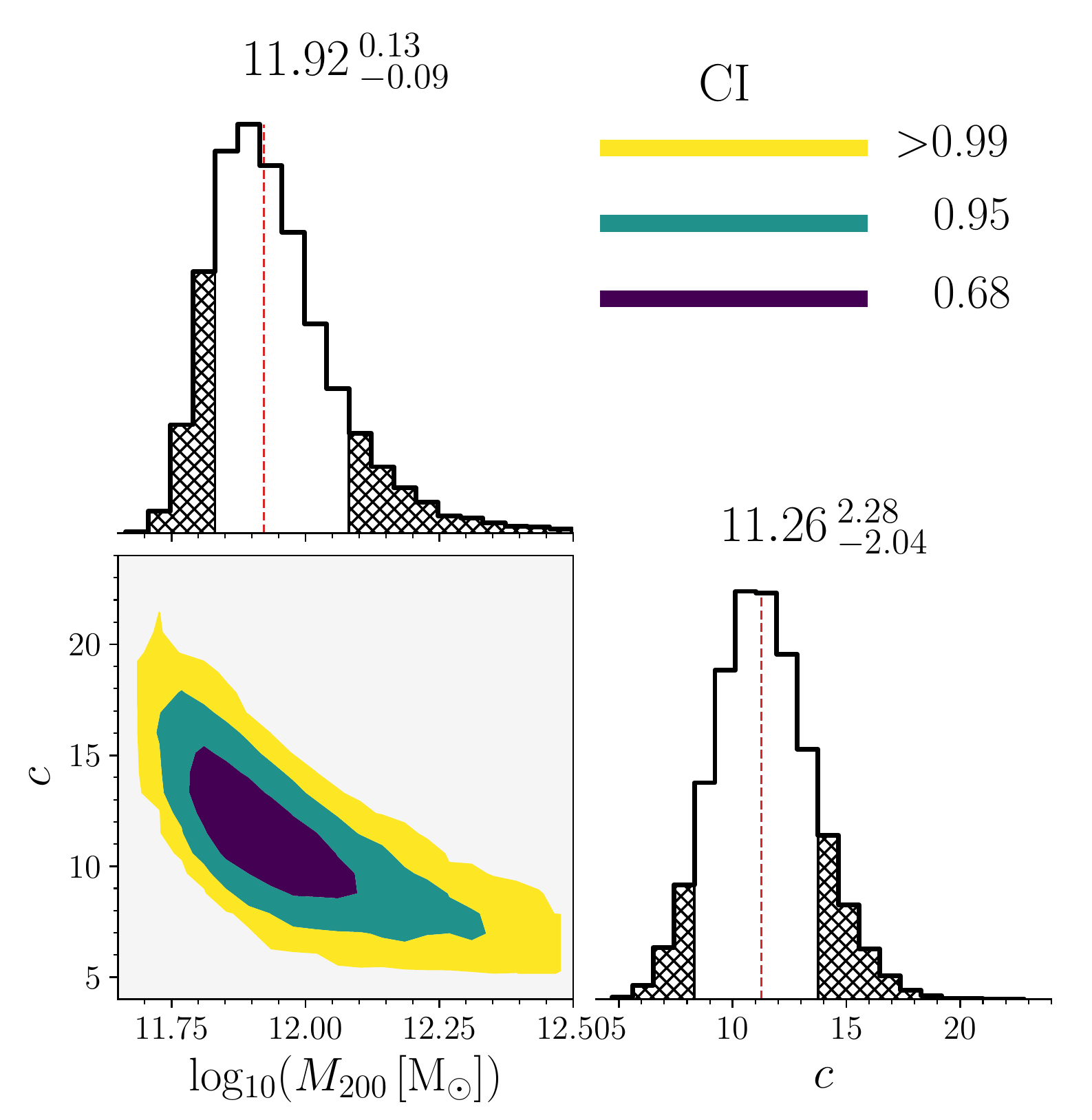}
\caption{The posterior distribution for sampled parameters $c$ and $\text{log}_{10}(M_{\rm 200})$ with color-coding representing the confidence intervals of our distribution.}
\label{fig:Mass_low}
\end{figure}

For the priors, $P(\boldsymbol{\theta}_{n})$, we used the following uniform ($\mathcal{U}$) priors on mass and concentration:
\begin{equation} \label{eq:VsysPriors}
\begin{gathered} 
P(\boldsymbol{\theta}_{n}) = \mathcal{U}(11.4 < \text{log}_{10}(M_{\rm 200}) < 12.5) \hspace{0.1cm} \cap \\ \hspace{-1.3cm} \mathcal{U}(1 < c < 30) .
\end{gathered}
\end{equation}
Here $\boldsymbol{\theta}_{n}$ stands for model parameters. Since the posterior distributions for our calculated escape velocities are highly asymmetric (as a result of the degeneracy between $k$ and $v_{\rm esc}$, and our set priors), we use the following prescription for log-likelihood that includes a treatment for asymmetric errors \citep[see sect.~6.3.1 in][for more details]{Barlow2019AsymErr}:
{\footnotesize
\begin{equation}\label{eq:LikeBarlow}
\begin{gathered} 
\hspace{-5.5cm} P\left (v_{\text{obs},i}, \sigma_{\text{obs},i}^{+}, \sigma_{\text{obs},i}^{-} | \boldsymbol{\theta}_{n} \right ) \propto \\ \exp{\left ( -\frac{1}{2} \left ( \frac{\left ( v_{\text{obs},i} - v_{\text{mod},i}\left ( \boldsymbol{\theta}_{n} \right ) \right ) \left ( \sigma^{+}_{\text{obs},i} + \sigma^{-}_{\text{obs},i} \right )}{2\sigma^{+}_{\text{obs},i}\sigma^{-}_{\text{obs},i} + \left ( \sigma^{+}_{\text{obs},i} - \sigma^{-}_{\text{obs},i} \right ) \left ( v_{\text{obs},i} - v_{\text{mod},i}\left ( \boldsymbol{\theta}_{n} \right ) \right )} \right )^2\right )}  .
\end{gathered}
\end{equation}}
We note that if the upper and lower uncertainties were equal, the likelihood in Eq.~\ref{eq:LikeBarlow} would turn into a regular likelihood. The posterior probability of our model is then:
\begin{equation}\label{eq:minimize}
\begin{gathered} 
\hspace{-4cm} P\left (\boldsymbol{\theta}_{n} | v_{\rm obs}, \sigma_{\rm obs}^{+}, \sigma_{\rm obs}^{-} \right ) \propto \\ \hspace{1.cm} P(M_{\rm 200}) P(c) \prod_{i=1}^{4} P\left (v_{\text{obs},i}, \sigma_{\text{obs},i}^{+}, \sigma_{\text{obs},i}^{-} | \boldsymbol{\theta}_{n} \right ) .
\end{gathered}
\end{equation}
Using the \texttt{emcee} module, we maximized the log-likelihood in Eq.~\ref{eq:minimize}. We used $150$ walkers with the initial 2000 steps as a burn-in and then restarting the sampler we ran the walkers for $3000$ additional steps. We thinned our chains using the autocorrelation $\tau=16$, and obtained $30000$ samples for the posterior distributions of concentration and mass. Figure~\ref{fig:Mass_low} displays the resulting distributions. We find $c = 11.3^{+2.3}_{-2.0}$ and a mass of $\text{log}_{10}(M_{\rm 200}) = 11.92^{+0.13}_{-0.09}$, which translates to $M_{\rm 200} = 0.83^{+0.29}_{-0.16} \cdot 10^{12}$\,M$_{\odot}$. Since our analysis covers Galactic radii up to $28$\,kpc we also estimated the mass within $20$\,kpc and $30$\,kpc. We found masses for the MW equal to $M_{\rm MW} \left(r < 20\,\text{kpc} \right) = 1.9^{+0.2}_{-0.1} \times 10^{11}$\,M$_{\odot}$ and $M_{\rm MW} \left(r < 30\,\text{kpc} \right) = 2.6^{+0.4}_{-0.2} \times 10^{11}$\,M$_{\odot}$.
 
Here it is important to emphasize that this estimate is based on three estimates of the escape velocity, and a single estimate of the circular velocity at the Solar position. If we consider only one single measurement of the escape velocity in the Solar neighborhood we would obtain a similar result for the MW mass ($M_{\rm 200} = 0.87^{+0.82}_{-0.33} \cdot 10^{12}$\,M$_{\odot}$), only with significantly larger uncertainties. Therefore, including other regions beyond $12$\,kpc is valuable to increase the precision of the resulting mass estimate.

Based on results from simulations \citep[see, e.g.,][]{Smith2007,Grand2019,Koppelman2021}, we can suspect that our calculated escape velocities are underestimated. If we then artificially boost our escape velocity estimates by 10\,percent \citep{Koppelman2021}, we obtain a mass estimate that seems to be more in line with other mass tracers; $\text{log}_{10}(M_{\rm 200}) = 12.10^{+0.12}_{-0.08}$, which translates to $M_{\rm 200} = 1.26^{+0.40}_{-0.22} \cdot 10^{12}$\,M$_{\odot}$. For the regions where we actually meassure the escape velocity (up to $30$\,kpc) we obtain the following masses for the MW: $M_{\rm MW} \left(r < 20\,\text{kpc} \right) = 2.1^{+0.2}_{-0.1} \times 10^{11}$\,M$_{\odot}$ and $M_{\rm MW} \left(r < 30\,\text{kpc} \right) = 3.0^{+0.4}_{-0.2} \times 10^{11}$\,M$_{\odot}$. Considering only the boosted $v_{\rm esc}$ and circular velocity for the Sun's position, we obtain a similar result for $M_{\rm 200} = 1.45^{+0.82}_{-0.46} \cdot 10^{12}$\,M$_{\odot}$, which again leads to a larger uncertainty on the mass.


\section{Conclusions} \label{sec:Summary}

In this work, we utilized our dataset of RR~Lyrae variables with complete 7-D information on their chemo-dynamical properties described in our previous work \citep{Prudil2021Orphan,Crestani2021,Fabrizio2021}. We expand our sample by using low-resolution spectra from the LAMOST survey, which yields a total of 6187 RR~Lyrae stars with complete information on their kinematics and chemistry. After cuts based on the quality of the astrometric solution, our study sample consists of 4847 variables.

We estimated kinematical and spatial properties like the Galactocentric velocity and radius for each object. We compared the calculated velocity with the escape velocity determined based on a simple axisymmetric MW potential. We found \numberR~RR~Lyrae stars with $P_{\rm unb}$ higher than $0.5$. Two of them were likely misclassified as RR~Lyrae stars, while the remaining seven objects appear to be genuine RR~Lyrae stars, based on their light curve shape and \gaia~colors. These variables appear to be on tangential orbits, which indicates that they do not originate in the Galactic disk and bulge.

Here it is important to emphasize that these seven HV-RRL candidates may be $\approx 1 -2$~$\sigma$ outliers in the $v_{\rm GC}$ and $R_{\rm GC}$ space, since their \textit{unbound} status is grounded on their high tangential velocities where uncertainties in proper motions dominate the error budget at large distances. Moreover, from the pure probabilistic perspective, one would not expect to find \numberR~unbound candidates out of \sample~of stars, in comparison with studies like \citet{Marchetti2019} and \citet{Li2021High591}, we would expect to find less than a single unbound RR~Lyrae stars. In addition, the unbound classification of some of the HV-RRL candidates can be influenced by modification of the rotation velocity, furthermore casting doubt on their unbound nature.

Seven out of \numberR~high velocity RR~Lyrae candidates follow the trail of the Sagittarius stellar stream. Three of these (HV-RRL-02, HV-RRL-07, and HV-RRL-08) were previously identified as members of the stream by \citet{Hernitschek2017} based on their spatial properties. We can tentatively link HV-RRL-02 and HV-RRL-08 to the Sagittarius dwarf spheroidal galaxy using their orbits. Since they both had already been associated with the Sagittarius stream by \citet{Hernitschek2017}, they could have been ejected quite recently.

Some high velocity candidates could be reclassified as bound if we consider a more massive MW halo. We explored this possibility by determining the MW escape velocity based on the high-velocity tail of our RR~Lyrae sample from which both disk contamination and our candidate unbound RR~Lyrae have been removed. Using a power-law distribution to describe the high-velocity tail \citep{Leonard1990}, we expanded our escape velocity analysis beyond the Solar circle ($4 < R_{\rm GC} < 12$\,kpc) in two regions reaching up to $28$\,kpc. Thus, we obtained $v_{\rm esc} = 512^{+94}_{-37}$\kms for the Solar neighborhood, which is lower than some of the more recent studies \citep[see, e.g.,][]{Piffl2014,Monari2018,Deason2019}, but more in line with some of the current ones like \citet{Koppelman2021} and \citet{Necib2021II}. It is important to emphasize that the large positive error on our escape velocity estimate is still within the 16th and 84th percentiles with previous measurements. The measurement beyond $12$\,kpc yields $v_{\rm esc} = 436^{+44}_{-22}$\kms and $v_{\rm esc} = 393^{+53}_{-26}$\kms at $16$\,kpc and $24$\,kpc, respectively. 

Together with the circular velocity at the Sun's position, all three escape velocity estimates entered into our analysis of the MW mass, where we worked with a MW potential that included the NFW halo, for which we varied the mass and concentration. The resulting mass was equal to $\text{log}_{10}(M_{\rm 200}) = 11.92^{+0.13}_{-0.09}$ ($M_{\rm 200} = 0.83^{+0.29}_{-0.16} \cdot 10^{12}$\,M$_{\odot}$, which falls on the low end of MW mass estimates \citep[e.g.,][]{Watkins2019,Callingham2019,Shen2022}, but agrees reasonably well with some of the recent estimates \citep{Eadie2016,Eadie2018,Eilers2019,Necib2021II}. Based on escape velocity estimates from simulations, we expect the measured escape velocity to be underestimated by approximately $10$ percent. Using this approximate correction for our escape velocities, we find the mass of the MW to be $\text{log}_{10}(M_{\rm 200}) = 12.10^{+0.12}_{-0.08}$ ($M_{\rm 200} = 1.26^{+0.40}_{-0.22} \cdot 10^{12}$\,M$_{\odot}$) which is in agreement with the previous mass estimates.

Mass estimates for the regions where we measure the escape velocity (up to $28$\,kpc) center around $M_{\rm MW} \left(r < 20\,\text{kpc} \right) = 1.9^{+0.2}_{-0.1} \times 10^{11}$\,M$_{\odot}$ and $M_{\rm MW} \left(r < 30\,\text{kpc} \right) = 2.6^{+0.4}_{-0.2} \times 10^{11}$\,M$_{\odot}$. Considering the possible $10$ percent bias in the escape velocity, estimates for the MW mass shift to $M_{\rm MW} \left(r < 20\,\text{kpc} \right) = 2.1^{+0.2}_{-0.1} \times 10^{11}$\,M$_{\odot}$ and $M_{\rm MW} \left(r < 30\,\text{kpc} \right) = 3.0^{+0.4}_{-0.2} \times 10^{11}$\,M$_{\odot}$. Particularly values for $M_{\rm MW} \left(r < 20\,\text{kpc} \right)$ allow for comparison with other tracers where, e.g., \citet{Malhan2018} found $M_{\rm MW} \left(r < 20\,\text{kpc} \right) = 2.5 \pm 0.2 \times 10^{11}$\,M$_{\odot}$ (using the GD-1 stream) and \citet{Bovy2016} estimated mass of $M_{\rm MW} \left(r < 20\,\text{kpc} \right) = 1.1 \pm 0.1 \times 10^{11}$\,M$_{\odot}$ based on the Pal~5 and GD-1 stellar streams. Our value falls between both aforementioned mass estimates and agrees well with the MW mass calculated by \citet{Watkins2019} $M_{\rm MW} \left(r < 20\,\text{kpc} \right) = 2.1^{+0.4}_{-0.3} \times 10^{11}$\,M$_{\odot}$.

An increase in the mass of the MW halo could potentially change the classification of the unbound stars HV-RRL-02 and HV-RRL-08 as being bound. On the other hand, stars like HV-RRL-05, HV-RRL-06, HV-RRL-07, and HV-RRL-09 would require an even higher MW mass to be considered bound \citep[up to $M_{\rm 200} = 1.4 \cdot 10^{12}$\,M$_{\odot}$ and higher, similarly to the mass estimates from][]{Hattori2018metPoor_LMC,Watkins2019}. HV-RRL-03 and HV-RRL-04 are likely misclassified RR~Lyrae variables, and thus are probably bound. In that case, only HV-RRL-01 is remaining for which the ejection mechanism is unclear, but at $d=53$\,kpc it stands out as unbound RR~Lyrae star with $v_{\rm GC} = 655$\kms.

Lastly with the ongoing and upcoming large spectroscopic surveys \citep[e.g., 4MOST, $WEAVE$, and SDSS-V,][]{4MOST2014,WEAVE2014,Kollmeier2017}, large populations of variable stars in the MW halo will be targeted that could serve as tracers of the high-velocity tail and subsequently improve as estimates especially across larger Galactic radii.

\begin{acknowledgements}

Z.P. acknowledges fruitful discussion with Helmer Koppelman and Andrea Kunder that significantly improved the paper. Z.P., A.J.K.H., B.L., and E.K.G. acknowledge support by the Deutsche Forschungsgemeinschaft (DFG, German Research Foundation) -- Project-ID 138713538 -- SFB 881 (``The Milky Way System", subprojects A03, A05, A11). Funding for the Sloan Digital Sky Survey IV has been provided by the Alfred P. Sloan Foundation, the U.S. Department of Energy Office of Science, and the Participating Institutions. SDSS-IV acknowledges support and resources from the Center for High Performance Computing  at the University of Utah. The SDSS website is www.sdss.org. SDSS-IV is managed by the Astrophysical Research Consortium for the Participating Institutions of the SDSS Collaboration including the Brazilian Participation Group, the Carnegie Institution for Science, Carnegie Mellon University, Center for Astrophysics | Harvard \& Smithsonian, the Chilean Participation Group, the French Participation Group, Instituto de Astrof\'isica de Canarias, The Johns Hopkins University, Kavli Institute for the Physics and Mathematics of the Universe (IPMU) / University of Tokyo, the Korean Participation Group, Lawrence Berkeley National Laboratory, Leibniz Institut f\"ur Astrophysik Potsdam (AIP), Max-Planck-Institut f\"ur Astronomie (MPIA Heidelberg), Max-Planck-Institut f\"ur Astrophysik (MPA Garching), Max-Planck-Institut f\"ur Extraterrestrische Physik (MPE), National Astronomical Observatories of China, New Mexico State University, New York University, University of Notre Dame, Observat\'ario Nacional / MCTI, The Ohio State University, Pennsylvania State University, Shanghai Astronomical Observatory, United Kingdom Participation Group, Universidad Nacional Aut\'onoma de M\'exico, University of Arizona, University of Colorado Boulder, University of Oxford, University of Portsmouth, University of Utah, University of Virginia, University of Washington, University of Wisconsin, Vanderbilt University, and Yale University. This work has made use of data from the European Space Agency (ESA) mission {\it Gaia} (\url{https://www.cosmos.esa.int/gaia}), processed by the {\it Gaia} Data Processing and Analysis Consortium (DPAC, \url{https://www.cosmos.esa.int/web/gaia/dpac/consortium}). Funding for the DPAC has been provided by national institutions, in particular the institutions participating in the {\it Gaia} Multilateral Agreement. The CSS survey is funded by the National Aeronautics and Space Administration under Grant No. NNG05GF22G issued through the Science Mission Directorate Near-Earth Objects Observations Program. The CRTS survey is supported by the U.S.~National Science Foundation under grants AST-0909182 and AST-1313422. Guoshoujing Telescope (the Large Sky Area Multi-Object Fiber Spectroscopic Telescope LAMOST) is a National Major Scientific Project built by the Chinese Academy of Sciences. Funding for the project has been provided by the National Development and Reform Commission. LAMOST is operated and managed by the National Astronomical Observatories, Chinese Academy of Sciences.

This research made use of the following Python packages: \texttt{Astropy} \citep{astropy2013,astropy2018}, \texttt{dustmaps} \citep{Green2018dust}, \texttt{emcee} \citep{Foreman-Mackey2013}, \texttt{galpy} \citep{Bovy2015}, \texttt{IPython} \citep{ipython}, \texttt{Matplotlib} \citep{matplotlib}, \texttt{NumPy} \citep{numpy}, \texttt{scikit-learn} \citep{Pedregosa2012}, and \texttt{SciPy} \citep{scipy}.
\end{acknowledgements}

\bibliographystyle{aa}
\bibliography{biby}

\begin{thebibliography}{143}
\expandafter\ifx\csname natexlab\endcsname\relax\def\natexlab#1{#1}\fi

\bibitem[{{Abadi} {et~al.}(2009){Abadi}, {Navarro}, \& {Steinmetz}}]{Abadi2009}
{Abadi}, M.~G., {Navarro}, J.~F., \& {Steinmetz}, M. 2009, \apjl, 691, L63

\bibitem[{{Abbas} {et~al.}(2014){Abbas}, {Grebel}, {Martin}, {Burgett},
  {Flewelling}, \& {Wainscoat}}]{Abbas2014}
{Abbas}, M.~A., {Grebel}, E.~K., {Martin}, N.~F., {et~al.} 2014, \mnras, 441,
  1230

\bibitem[{{Aguado} {et~al.}(2019){Aguado}, {Ahumada}, {Almeida}, {Anderson},
  {Andrews}, {Anguiano}, {Aquino Ort{\'\i}z}, {Arag{\'o}n-Salamanca},
  {Argudo-Fern{\'a}ndez}, {Aubert}, {Avila-Reese}, {Badenes}, {Barboza
  Rembold}, {Barger}, {Barrera-Ballesteros}, {Bates}, {Bautista}, {Beaton},
  {Beers}, {Belfiore}, {Bernardi}, {Bershady}, {Beutler}, {Bird}, {Bizyaev},
  {Blanc}, {Blanton}, {Blomqvist}, {Bolton}, {Boquien}, {Borissova}, {Bovy},
  {Brand t}, {Brinkmann}, {Brownstein}, {Bundy}, {Burgasser}, {Byler}, {Cano
  Diaz}, {Cappellari}, {Carrera}, {Cervantes Sodi}, {Chen}, {Cherinka}, {Choi},
  {Chung}, {Coffey}, {Comerford}, {Comparat}, {Covey}, {da Silva Ilha}, {da
  Costa}, {Dai}, {Damke}, {Darling}, {Davies}, {Dawson}, {de Sainte Agathe},
  {Deconto Machado}, {Del Moro}, {De Lee}, {Diamond-Stanic}, {Dom{\'\i}nguez
  S{\'a}nchez}, {Donor}, {Drory}, {du Mas des Bourboux}, {Duckworth}, {Dwelly},
  {Ebelke}, {Emsellem}, {Escoffier}, {Fern{\'a}ndez-Trincado}, {Feuillet},
  {Fischer}, {Fleming}, {Fraser-McKelvie}, {Freischlad}, {Frinchaboy}, {Fu},
  {Galbany}, {Garcia-Dias}, {Garc{\'\i}a-Hern{\'a}ndez}, {Garma Oehmichen},
  {Geimba Maia}, {Gil-Mar{\'\i}n}, {Grabowski}, {Gu}, {Guo}, {Ha},
  {Harrington}, {Hasselquist}, {Hayes}, {Hearty}, {Hernandez Toledo}, {Hicks},
  {Hogg}, {Holley-Bockelmann}, {Holtzman}, {Hsieh}, {Hunt}, {Hwang},
  {Ibarra-Medel}, {Jimenez Angel}, {Johnson}, {Jones}, {J{\"o}nsson},
  {Kinemuchi}, {Kollmeier}, {Krawczyk}, {Kreckel}, {Kruk}, {Lacerna}, {Lan},
  {Lane}, {Law}, {Lee}, {Li}, {Lian}, {Lin}, {Lin}, {Lintott}, {Long},
  {Longa-Pe{\~n}a}, {Mackereth}, {de la Macorra}, {Majewski}, {Malanushenko},
  {Manchado}, {Maraston}, {Mariappan}, {Marinelli}, {Marques-Chaves},
  {Masseron}, {Masters}, {McDermid}, {Medina Pe{\~n}a}, {Meneses-Goytia},
  {Merloni}, {Merrifield}, {Meszaros}, {Minniti}, {Minsley}, {Muna}, {Myers},
  {Nair}, {Correa do Nascimento}, {Newman}, {Nitschelm}, {Olmstead}, {Oravetz},
  {Oravetz}, {Ortega Minakata}, {Pace}, {Padilla}, {Palicio}, {Pan}, {Pan},
  {Parikh}, {Parker}, {Peirani}, {Penny}, {Percival}, {Perez-Fournon},
  {Peterken}, {Pinsonneault}, {Prakash}, {Raddick}, {Raichoor}, {Riffel},
  {Riffel}, {Rix}, {Robin}, {Roman-Lopes}, {Rose}, {Ross}, {Rossi}, {Rowlands},
  {Rubin}, {S{\'a}nchez}, {S{\'a}nchez-Gallego}, {Sayres}, {Schaefer},
  {Schiavon}, {Schimoia}, {Schlafly}, {Schlegel}, {Schneider}, {Schultheis},
  {Seo}, {Shamsi}, {Shao}, {Shen}, {Shetty}, {Simonian}, {Smethurst}, {Sobeck},
  {Souter}, {Spindler}, {Stark}, {Stassun}, {Steinmetz}, {Storchi-Bergmann},
  {Stringfellow}, {Su{\'a}rez}, {Sun}, {Taghizadeh-Popp}, {Talbot}, {Tayar},
  {Thakar}, {Thomas}, {Tissera}, {Tojeiro}, {Troup}, {Unda-Sanzana},
  {Valenzuela}, {Vargas-Maga{\~n}a}, {V{\'a}zquez-Mata}, {Wake}, {Weaver},
  {Weijmans}, {Westfall}, {Wild}, {Wilson}, {Woods}, {Yan}, {Yang}, {Zamora},
  {Zasowski}, {Zhang}, {Zheng}, {Zheng}, {Zhu}, {Zinn}, \& {Zou}}]{Aguado2019}
{Aguado}, D.~S., {Ahumada}, R., {Almeida}, A., {et~al.} 2019, \apjs, 240, 23

\bibitem[{{Asplund} {et~al.}(2009){Asplund}, {Grevesse}, {Sauval}, \&
  {Scott}}]{Asplund2009}
{Asplund}, M., {Grevesse}, N., {Sauval}, A.~J., \& {Scott}, P. 2009, \araa, 47,
  481

\bibitem[{{Astropy Collaboration} {et~al.}(2018){Astropy Collaboration},
  {Price-Whelan}, {Sip{\H{o}}cz}, {G{\"u}nther}, {Lim}, {Crawford}, {Conseil},
  {Shupe}, {Craig}, {Dencheva}, {Ginsburg}, {Vand erPlas}, {Bradley},
  {P{\'e}rez-Su{\'a}rez}, {de Val-Borro}, {Aldcroft}, {Cruz}, {Robitaille},
  {Tollerud}, {Ardelean}, {Babej}, {Bach}, {Bachetti}, {Bakanov}, {Bamford},
  {Barentsen}, {Barmby}, {Baumbach}, {Berry}, {Biscani}, {Boquien}, {Bostroem},
  {Bouma}, {Brammer}, {Bray}, {Breytenbach}, {Buddelmeijer}, {Burke},
  {Calderone}, {Cano Rodr{\'\i}guez}, {Cara}, {Cardoso}, {Cheedella}, {Copin},
  {Corrales}, {Crichton}, {D'Avella}, {Deil}, {Depagne}, {Dietrich}, {Donath},
  {Droettboom}, {Earl}, {Erben}, {Fabbro}, {Ferreira}, {Finethy}, {Fox},
  {Garrison}, {Gibbons}, {Goldstein}, {Gommers}, {Greco}, {Greenfield},
  {Groener}, {Grollier}, {Hagen}, {Hirst}, {Homeier}, {Horton}, {Hosseinzadeh},
  {Hu}, {Hunkeler}, {Ivezi{\'c}}, {Jain}, {Jenness}, {Kanarek}, {Kendrew},
  {Kern}, {Kerzendorf}, {Khvalko}, {King}, {Kirkby}, {Kulkarni}, {Kumar},
  {Lee}, {Lenz}, {Littlefair}, {Ma}, {Macleod}, {Mastropietro}, {McCully},
  {Montagnac}, {Morris}, {Mueller}, {Mumford}, {Muna}, {Murphy}, {Nelson},
  {Nguyen}, {Ninan}, {N{\"o}the}, {Ogaz}, {Oh}, {Parejko}, {Parley}, {Pascual},
  {Patil}, {Patil}, {Plunkett}, {Prochaska}, {Rastogi}, {Reddy Janga},
  {Sabater}, {Sakurikar}, {Seifert}, {Sherbert}, {Sherwood-Taylor}, {Shih},
  {Sick}, {Silbiger}, {Singanamalla}, {Singer}, {Sladen}, {Sooley},
  {Sornarajah}, {Streicher}, {Teuben}, {Thomas}, {Tremblay}, {Turner},
  {Terr{\'o}n}, {van Kerkwijk}, {de la Vega}, {Watkins}, {Weaver}, {Whitmore},
  {Woillez}, {Zabalza}, \& {Astropy Contributors}}]{astropy2018}
{Astropy Collaboration}, {Price-Whelan}, A.~M., {Sip{\H{o}}cz}, B.~M., {et~al.}
  2018, \aj, 156, 123

\bibitem[{{Astropy Collaboration} {et~al.}(2013){Astropy Collaboration},
  {Robitaille}, {Tollerud}, {Greenfield}, {Droettboom}, {Bray}, {Aldcroft},
  {Davis}, {Ginsburg}, {Price-Whelan}, {Kerzendorf}, {Conley}, {Crighton},
  {Barbary}, {Muna}, {Ferguson}, {Grollier}, {Parikh}, {Nair}, {Unther},
  {Deil}, {Woillez}, {Conseil}, {Kramer}, {Turner}, {Singer}, {Fox}, {Weaver},
  {Zabalza}, {Edwards}, {Azalee Bostroem}, {Burke}, {Casey}, {Crawford},
  {Dencheva}, {Ely}, {Jenness}, {Labrie}, {Lim}, {Pierfederici}, {Pontzen},
  {Ptak}, {Refsdal}, {Servillat}, \& {Streicher}}]{astropy2013}
{Astropy Collaboration}, {Robitaille}, T.~P., {Tollerud}, E.~J., {et~al.} 2013,
  \aap, 558, A33

\bibitem[{{Bailer-Jones} {et~al.}(2021){Bailer-Jones}, {Rybizki}, {Fouesneau},
  {Demleitner}, \& {Andrae}}]{Bailer2021}
{Bailer-Jones}, C.~A.~L., {Rybizki}, J., {Fouesneau}, M., {Demleitner}, M., \&
  {Andrae}, R. 2021, \aj, 161, 147

\bibitem[{{Barlow}(2019)}]{Barlow2019AsymErr}
{Barlow}, R.~J. 2019, arXiv e-prints, arXiv:1905.12362

\bibitem[{{Bellm} {et~al.}(2019){Bellm}, {Kulkarni}, {Graham}, {Dekany},
  {Smith}, {Riddle}, {Masci}, {Helou}, {Prince}, {Adams}, {Barbarino},
  {Barlow}, {Bauer}, {Beck}, {Belicki}, {Biswas}, {Blagorodnova}, {Bodewits},
  {Bolin}, {Brinnel}, {Brooke}, {Bue}, {Bulla}, {Burruss}, {Cenko}, {Chang},
  {Connolly}, {Coughlin}, {Cromer}, {Cunningham}, {De}, {Delacroix}, {Desai},
  {Duev}, {Eadie}, {Farnham}, {Feeney}, {Feindt}, {Flynn}, {Franckowiak},
  {Frederick}, {Fremling}, {Gal-Yam}, {Gezari}, {Giomi}, {Goldstein},
  {Golkhou}, {Goobar}, {Groom}, {Hacopians}, {Hale}, {Henning}, {Ho}, {Hover},
  {Howell}, {Hung}, {Huppenkothen}, {Imel}, {Ip}, {Ivezi{\'c}}, {Jackson},
  {Jones}, {Juric}, {Kasliwal}, {Kaspi}, {Kaye}, {Kelley}, {Kowalski},
  {Kramer}, {Kupfer}, {Landry}, {Laher}, {Lee}, {Lin}, {Lin}, {Lunnan},
  {Giomi}, {Mahabal}, {Mao}, {Miller}, {Monkewitz}, {Murphy}, {Ngeow},
  {Nordin}, {Nugent}, {Ofek}, {Patterson}, {Penprase}, {Porter}, {Rauch},
  {Rebbapragada}, {Reiley}, {Rigault}, {Rodriguez}, {van Roestel}, {Rusholme},
  {van Santen}, {Schulze}, {Shupe}, {Singer}, {Soumagnac}, {Stein}, {Surace},
  {Sollerman}, {Szkody}, {Taddia}, {Terek}, {Van Sistine}, {van Velzen},
  {Vestrand}, {Walters}, {Ward}, {Ye}, {Yu}, {Yan}, \&
  {Zolkower}}]{Bellm2019ZTF}
{Bellm}, E.~C., {Kulkarni}, S.~R., {Graham}, M.~J., {et~al.} 2019, \pasp, 131,
  018002

\bibitem[{{Belokurov} {et~al.}(2018){Belokurov}, {Erkal}, {Evans}, {Koposov},
  \& {Deason}}]{Belokurov2018}
{Belokurov}, V., {Erkal}, D., {Evans}, N.~W., {Koposov}, S.~E., \& {Deason},
  A.~J. 2018, \mnras, 478, 611

\bibitem[{{Bennett} \& {Bovy}(2019)}]{Bennett2019}
{Bennett}, M. \& {Bovy}, J. 2019, \mnras, 482, 1417

\bibitem[{{Bensby} {et~al.}(2003){Bensby}, {Feltzing}, \&
  {Lundstr{\"o}m}}]{Bensby2003}
{Bensby}, T., {Feltzing}, S., \& {Lundstr{\"o}m}, I. 2003, \aap, 410, 527

\bibitem[{{Bensby} {et~al.}(2014){Bensby}, {Feltzing}, \& {Oey}}]{Bensby2014}
{Bensby}, T., {Feltzing}, S., \& {Oey}, M.~S. 2014, \aap, 562, A71

\bibitem[{{Blaauw}(1961)}]{Blaauw1961}
{Blaauw}, A. 1961, \bain, 15, 265

\bibitem[{{Blanco-Cuaresma}(2019)}]{Blanco2019iSpec}
{Blanco-Cuaresma}, S. 2019, \mnras, 486, 2075

\bibitem[{{Blanco-Cuaresma} {et~al.}(2014){Blanco-Cuaresma}, {Soubiran},
  {Heiter}, \& {Jofr{\'e}}}]{Blanco2014}
{Blanco-Cuaresma}, S., {Soubiran}, C., {Heiter}, U., \& {Jofr{\'e}}, P. 2014,
  \aap, 569, A111

\bibitem[{{Bland-Hawthorn} \& {Gerhard}(2016)}]{Bland-Hawthorn2016}
{Bland-Hawthorn}, J. \& {Gerhard}, O. 2016, \araa, 54, 529

\bibitem[{{Bonaca} {et~al.}(2017){Bonaca}, {Conroy}, {Wetzel}, {Hopkins}, \&
  {Kere{\v{s}}}}]{Bonaca2017}
{Bonaca}, A., {Conroy}, C., {Wetzel}, A., {Hopkins}, P.~F., \& {Kere{\v{s}}},
  D. 2017, \apj, 845, 101

\bibitem[{{Bono} {et~al.}(2019){Bono}, {Iannicola}, {Braga}, {Ferraro},
  {Stetson}, {Magurno}, {Matsunaga}, {Beaton}, {Buonanno}, {Chaboyer},
  {Dall'Ora}, {Fabrizio}, {Fiorentino}, {Freedman}, {Gilligan}, {Madore},
  {Marconi}, {Marengo}, {Marinoni}, {Marrese}, {Martinez-Vazquez}, {Mateo},
  {Monelli}, {Neeley}, {Nonino}, {Sneden}, {Thevenin}, {Valenti}, \&
  {Walker}}]{Bono2019}
{Bono}, G., {Iannicola}, G., {Braga}, V.~F., {et~al.} 2019, \apj, 870, 115

\bibitem[{{Boubert} {et~al.}(2017){Boubert}, {Erkal}, {Evans}, \&
  {Izzard}}]{Boubert2017}
{Boubert}, D., {Erkal}, D., {Evans}, N.~W., \& {Izzard}, R.~G. 2017, \mnras,
  469, 2151

\bibitem[{{Bovy}(2015)}]{Bovy2015}
{Bovy}, J. 2015, \apjs, 216, 29

\bibitem[{{Bovy} {et~al.}(2016){Bovy}, {Bahmanyar}, {Fritz}, \&
  {Kallivayalil}}]{Bovy2016}
{Bovy}, J., {Bahmanyar}, A., {Fritz}, T.~K., \& {Kallivayalil}, N. 2016, \apj,
  833, 31

\bibitem[{{Braga} {et~al.}(2021){Braga}, {Crestani}, {Fabrizio}, {Bono},
  {Sneden}, {Preston}, {Storm}, {Kamann}, {Latour}, {Lala}, {Lemasle},
  {Prudil}, {Altavilla}, {Chaboyer}, {Dall'Ora}, {Ferraro}, {Gilligan},
  {Fiorentino}, {Iannicola}, {Inno}, {Kwak}, {Marengo}, {Marinoni}, {Marrese},
  {Mart{\'\i}nez-V{\'a}zquez}, {Monelli}, {Mullen}, {Matsunaga}, {Neeley},
  {Stetson}, {Valenti}, \& {Zoccali}}]{Braga2021}
{Braga}, V.~F., {Crestani}, J., {Fabrizio}, M., {et~al.} 2021, \apj, 919, 85

\bibitem[{{Bromley} {et~al.}(2006){Bromley}, {Kenyon}, {Geller}, {Barcikowski},
  {Brown}, \& {Kurtz}}]{Bromley2006}
{Bromley}, B.~C., {Kenyon}, S.~J., {Geller}, M.~J., {et~al.} 2006, \apj, 653,
  1194

\bibitem[{{Brown}(2015)}]{Brown2015Rev}
{Brown}, W.~R. 2015, \araa, 53, 15

\bibitem[{{Brown} {et~al.}(2015){Brown}, {Anderson}, {Gnedin}, {Bond},
  {Geller}, \& {Kenyon}}]{Brown2015}
{Brown}, W.~R., {Anderson}, J., {Gnedin}, O.~Y., {et~al.} 2015, \apj, 804, 49

\bibitem[{{Brown} {et~al.}(2014){Brown}, {Geller}, \& {Kenyon}}]{Brown2014}
{Brown}, W.~R., {Geller}, M.~J., \& {Kenyon}, S.~J. 2014, \apj, 787, 89

\bibitem[{{Brown} {et~al.}(2005){Brown}, {Geller}, {Kenyon}, \&
  {Kurtz}}]{Brown2005}
{Brown}, W.~R., {Geller}, M.~J., {Kenyon}, S.~J., \& {Kurtz}, M.~J. 2005,
  \apjl, 622, L33

\bibitem[{{Callingham} {et~al.}(2019){Callingham}, {Cautun}, {Deason}, {Frenk},
  {Wang}, {G{\'o}mez}, {Grand}, {Marinacci}, \& {Pakmor}}]{Callingham2019}
{Callingham}, T.~M., {Cautun}, M., {Deason}, A.~J., {et~al.} 2019, \mnras, 484,
  5453

\bibitem[{{Capuzzo-Dolcetta} \& {Fragione}(2015)}]{Capuzzo2015}
{Capuzzo-Dolcetta}, R. \& {Fragione}, G. 2015, \mnras, 454, 2677

\bibitem[{{Casagrande} \& {VandenBerg}(2018)}]{Casagrande2018}
{Casagrande}, L. \& {VandenBerg}, D.~A. 2018, \mnras, 479, L102

\bibitem[{Castelli \& Kurucz(2003)}]{Castelli2003}
Castelli, F. \& Kurucz, R. 2003, arXiv: Astrophysics
  [\eprint[arXiv]{astro-ph/0405087}]

\bibitem[{{Chambers} {et~al.}(2016){Chambers}, {Magnier}, {Metcalfe},
  {Flewelling}, {Huber}, {Waters}, {Denneau}, {Draper}, {Farrow}, {Finkbeiner},
  {Holmberg}, {Koppenhoefer}, {Price}, {Rest}, {Saglia}, {Schlafly}, {Smartt},
  {Sweeney}, {Wainscoat}, {Burgett}, {Chastel}, {Grav}, {Heasley}, {Hodapp},
  {Jedicke}, {Kaiser}, {Kudritzki}, {Luppino}, {Lupton}, {Monet}, {Morgan},
  {Onaka}, {Shiao}, {Stubbs}, {Tonry}, {White}, {Ba{\~n}ados}, {Bell},
  {Bender}, {Bernard}, {Boegner}, {Boffi}, {Botticella}, {Calamida},
  {Casertano}, {Chen}, {Chen}, {Cole}, {Deacon}, {Frenk}, {Fitzsimmons},
  {Gezari}, {Gibbs}, {Goessl}, {Goggia}, {Gourgue}, {Goldman}, {Grant},
  {Grebel}, {Hambly}, {Hasinger}, {Heavens}, {Heckman}, {Henderson}, {Henning},
  {Holman}, {Hopp}, {Ip}, {Isani}, {Jackson}, {Keyes}, {Koekemoer}, {Kotak},
  {Le}, {Liska}, {Long}, {Lucey}, {Liu}, {Martin}, {Masci}, {McLean}, {Mindel},
  {Misra}, {Morganson}, {Murphy}, {Obaika}, {Narayan}, {Nieto-Santisteban},
  {Norberg}, {Peacock}, {Pier}, {Postman}, {Primak}, {Rae}, {Rai}, {Riess},
  {Riffeser}, {Rix}, {R{\"o}ser}, {Russel}, {Rutz}, {Schilbach}, {Schultz},
  {Scolnic}, {Strolger}, {Szalay}, {Seitz}, {Small}, {Smith}, {Soderblom},
  {Taylor}, {Thomson}, {Taylor}, {Thakar}, {Thiel}, {Thilker}, {Unger},
  {Urata}, {Valenti}, {Wagner}, {Walder}, {Walter}, {Watters}, {Werner},
  {Wood-Vasey}, \& {Wyse}}]{Chambers2016}
{Chambers}, K.~C., {Magnier}, E.~A., {Metcalfe}, N., {et~al.} 2016, arXiv
  e-prints, arXiv:1612.05560

\bibitem[{{Chandrasekhar}(1943)}]{Chandrasekhar1943}
{Chandrasekhar}, S. 1943, \apj, 97, 255

\bibitem[{{Chen} {et~al.}(2020){Chen}, {Wang}, {Deng}, {de Grijs}, {Yang}, \&
  {Tian}}]{Chen2020ZTFVAR}
{Chen}, X., {Wang}, S., {Deng}, L., {et~al.} 2020, \apjs, 249, 18

\bibitem[{{Ciddor}(1996)}]{Ciddor1996}
{Ciddor}, P.~E. 1996, \ao, 35, 1566

\bibitem[{{Clementini} {et~al.}(2019){Clementini}, {Ripepi}, {Molinaro},
  {Garofalo}, {Muraveva}, {Rimoldini}, {Guy}, {Jevardat de Fombelle},
  {Nienartowicz}, {Marchal}, {Audard}, {Holl}, {Leccia}, {Marconi}, {Musella},
  {Mowlavi}, {Lecoeur-Taibi}, {Eyer}, {De Ridder}, {Regibo}, {Sarro},
  {Szabados}, {Evans}, \& {Riello}}]{Clementini2019}
{Clementini}, G., {Ripepi}, V., {Molinaro}, R., {et~al.} 2019, \aap, 622, A60

\bibitem[{{Crestani} {et~al.}(2021){Crestani}, {Fabrizio}, {Braga}, {Sneden},
  {Preston}, {Ferraro}, {Iannicola}, {Bono}, {Alves-Brito}, {Nonino},
  {D'Orazi}, {Inno}, {Monelli}, {Storm}, {Altavilla}, {Chaboyer}, {Dall'Ora},
  {Fiorentino}, {Gilligan}, {Grebel}, {Lala}, {Lemasle}, {Marengo}, {Marinoni},
  {Marrese}, {Mart{\'\i}nez-V{\'a}zquez}, {Matsunaga}, {Mullen}, {Neeley},
  {Prudil}, {da Silva}, {Stetson}, {Th{\'e}venin}, {Valenti}, {Walker}, \&
  {Zoccali}}]{Crestani2021}
{Crestani}, J., {Fabrizio}, M., {Braga}, V.~F., {et~al.} 2021, \apj, 908, 20

\bibitem[{{Cutri} {et~al.}(2003){Cutri}, {Skrutskie}, {van Dyk}, {Beichman},
  {Carpenter}, {Chester}, {Cambresy}, {Evans}, {Fowler}, {Gizis}, {Howard},
  {Huchra}, {Jarrett}, {Kopan}, {Kirkpatrick}, {Light}, {Marsh}, {McCallon},
  {Schneider}, {Stiening}, {Sykes}, {Weinberg}, {Wheaton}, {Wheelock}, \&
  {Zacarias}}]{Cutri2003}
{Cutri}, R.~M., {Skrutskie}, M.~F., {van Dyk}, S., {et~al.} 2003, VizieR Online
  Data Catalog, II/246

\bibitem[{{Dalton} {et~al.}(2014){Dalton}, {Trager}, {Abrams}, {Bonifacio},
  {L{\'o}pez Aguerri}, {Middleton}, {Benn}, {Dee}, {Say{\`e}de}, {Lewis},
  {Pragt}, {Pico}, {Walton}, {Rey}, {Allende Prieto}, {Pe{\~n}ate}, {Lhome},
  {Ag{\'o}cs}, {Alonso}, {Terrett}, {Brock}, {Gilbert}, {Ridings}, {Guinouard},
  {Verheijen}, {Tosh}, {Rogers}, {Steele}, {Stuik}, {Tromp}, {Jasko}, {Kragt},
  {Lesman}, {Mottram}, {Bates}, {Gribbin}, {Fernando Rodriguez}, {Delgado},
  {Martin}, {Cano}, {Navarro}, {Irwin}, {Lewis}, {Gonzalez Solares},
  {O'Mahony}, {Bianco}, {Zurita}, {ter Horst}, {Molinari}, {Lodi}, {Guerra},
  {Vallenari}, \& {Baruffolo}}]{WEAVE2014}
{Dalton}, G., {Trager}, S., {Abrams}, D.~C., {et~al.} 2014, in \procspie, Vol.
  9147, Ground-based and Airborne Instrumentation for Astronomy V, 91470L

\bibitem[{{de Jong} {et~al.}(2014){de Jong}, {Barden}, {Bellido-Tirado},
  {Brynnel}, {Chiappini}, {Depagne}, {Haynes}, {Johl}, {Phillips}, {Schnurr},
  {Schwope}, {Walcher}, {Bauer}, {Cescutti}, {Cioni}, {Dionies}, {Enke},
  {Haynes}, {Kelz}, {Kitaura}, {Lamer}, {Minchev}, {M{\"u}ller}, {Nuza},
  {Olaya}, {Piffl}, {Popow}, {Saviauk}, {Steinmetz}, {Ural}, {Valentini},
  {Winkler}, {Wisotzki}, {Ansorge}, {Banerji}, {Gonzalez Solares}, {Irwin},
  {Kennicutt}, {King}, {McMahon}, {Koposov}, {Parry}, {Sun}, {Walton},
  {Finger}, {Iwert}, {Krumpe}, {Lizon}, {Mainieri}, {Amans}, {Bonifacio},
  {Cohen}, {Fran{\c c}ois}, {Jagourel}, {Mignot}, {Royer}, {Sartoretti},
  {Bender}, {Hess}, {Lang-Bardl}, {Muschielok}, {Schlichter}, {B{\"o}hringer},
  {Boller}, {Bongiorno}, {Brusa}, {Dwelly}, {Merloni}, {Nandra}, {Salvato},
  {Pragt}, {Navarro}, {Gerlofsma}, {Roelfsema}, {Dalton}, {Middleton}, {Tosh},
  {Boeche}, {Caffau}, {Christlieb}, {Grebel}, {Hansen}, {Koch}, {Ludwig},
  {Mandel}, {Quirrenbach}, {Sbordone}, {Seifert}, {Thimm}, {Helmi}, {trager},
  {Bensby}, {Feltzing}, {Ruchti}, {Edvardsson}, {Korn}, {Lind}, {Boland},
  {Colless}, {Frost}, {Gilbert}, {Gillingham}, {Lawrence}, {Legg}, {Saunders},
  {Sheinis}, {Driver}, {Robotham}, {Bacon}, {Caillier}, {Kosmalski}, {Laurent},
  \& {Richard}}]{4MOST2014}
{de Jong}, R.~S., {Barden}, S., {Bellido-Tirado}, O., {et~al.} 2014, in
  \procspie, Vol. 9147, Ground-based and Airborne Instrumentation for Astronomy
  V, 91470M

\bibitem[{{Deason} {et~al.}(2019){Deason}, {Fattahi}, {Belokurov}, {Evans},
  {Grand}, {Marinacci}, \& {Pakmor}}]{Deason2019}
{Deason}, A.~J., {Fattahi}, A., {Belokurov}, V., {et~al.} 2019, \mnras, 485,
  3514

\bibitem[{{Deng} {et~al.}(2012){Deng}, {Newberg}, {Liu}, {Carlin}, {Beers},
  {Chen}, {Chen}, {Christlieb}, {Grillmair}, {Guhathakurta}, {Han}, {Hou},
  {Lee}, {L{\'e}pine}, {Li}, {Liu}, {Pan}, {Sellwood}, {Wang}, {Wang}, {Yang},
  {Yanny}, {Zhang}, {Zhang}, {Zheng}, \& {Zhu}}]{Deng2012}
{Deng}, L.-C., {Newberg}, H.~J., {Liu}, C., {et~al.} 2012, Research in
  Astronomy and Astrophysics, 12, 735

\bibitem[{{Drake} {et~al.}(2013{\natexlab{a}}){Drake}, {Catelan}, {Djorgovski},
  {Torrealba}, {Graham}, {Belokurov}, {Koposov}, {Mahabal}, {Prieto},
  {Donalek}, {Williams}, {Larson}, {Christensen}, \& {Beshore}}]{Drake2013}
{Drake}, A.~J., {Catelan}, M., {Djorgovski}, S.~G., {et~al.}
  2013{\natexlab{a}}, \apj, 763, 32

\bibitem[{{Drake} {et~al.}(2013{\natexlab{b}}){Drake}, {Catelan}, {Djorgovski},
  {Torrealba}, {Graham}, {Mahabal}, {Prieto}, {Donalek}, {Williams}, {Larson},
  {Christensen}, \& {Beshore}}]{Drake2013stream}
{Drake}, A.~J., {Catelan}, M., {Djorgovski}, S.~G., {et~al.}
  2013{\natexlab{b}}, \apj, 765, 154

\bibitem[{{Drake} {et~al.}(2009){Drake}, {Djorgovski}, {Mahabal}, {Beshore},
  {Larson}, {Graham}, {Williams}, {Christensen}, {Catelan}, {Boattini},
  {Gibbs}, {Hill}, \& {Kowalski}}]{Drake2009}
{Drake}, A.~J., {Djorgovski}, S.~G., {Mahabal}, A., {et~al.} 2009, \apj, 696,
  870

\bibitem[{{Drake} {et~al.}(2014){Drake}, {Graham}, {Djorgovski}, {Catelan},
  {Mahabal}, {Torrealba}, {Garc{\'{\i}}a-{\'A}lvarez}, {Donalek}, {Prieto},
  {Williams}, {Larson}, {Christen sen}, {Belokurov}, {Koposov}, {Beshore},
  {Boattini}, {Gibbs}, {Hill}, {Kowalski}, {Johnson}, \&
  {Shelly}}]{Drake2014CatVari}
{Drake}, A.~J., {Graham}, M.~J., {Djorgovski}, S.~G., {et~al.} 2014, \apjs,
  213, 9

\bibitem[{{Du} {et~al.}(2018){Du}, {Li}, {Newberg}, {Chen}, {Shi}, {Wu}, \&
  {Ma}}]{Du2018}
{Du}, C., {Li}, H., {Newberg}, H.~J., {et~al.} 2018, \apjl, 869, L31

\bibitem[{{Eadie} {et~al.}(2018){Eadie}, {Keller}, \& {Harris}}]{Eadie2018}
{Eadie}, G., {Keller}, B., \& {Harris}, W.~E. 2018, \apj, 865, 72

\bibitem[{{Eadie} \& {Harris}(2016)}]{Eadie2016}
{Eadie}, G.~M. \& {Harris}, W.~E. 2016, \apj, 829, 108

\bibitem[{{Edelmann} {et~al.}(2005){Edelmann}, {Napiwotzki}, {Heber},
  {Christlieb}, \& {Reimers}}]{Edelmann2005}
{Edelmann}, H., {Napiwotzki}, R., {Heber}, U., {Christlieb}, N., \& {Reimers},
  D. 2005, \apjl, 634, L181

\bibitem[{{Eilers} {et~al.}(2019){Eilers}, {Hogg}, {Rix}, \&
  {Ness}}]{Eilers2019}
{Eilers}, A.-C., {Hogg}, D.~W., {Rix}, H.-W., \& {Ness}, M.~K. 2019, \apj, 871,
  120

\bibitem[{{Erkal} {et~al.}(2019{\natexlab{a}}){Erkal}, {Belokurov}, {Laporte},
  {Koposov}, {Li}, {Grillmair}, {Kallivayalil}, {Price-Whelan}, {Evans},
  {Hawkins}, {Hendel}, {Mateu}, {Navarro}, {del Pino}, {Slater}, {Sohn}, \&
  {Orphan Aspen Treasury Collaboration}}]{Erkal2019}
{Erkal}, D., {Belokurov}, V., {Laporte}, C.~F.~P., {et~al.} 2019{\natexlab{a}},
  \mnras, 487, 2685

\bibitem[{{Erkal} {et~al.}(2019{\natexlab{b}}){Erkal}, {Boubert}, {Gualandris},
  {Evans}, \& {Antonini}}]{Erkal2019HypLMC}
{Erkal}, D., {Boubert}, D., {Gualandris}, A., {Evans}, N.~W., \& {Antonini}, F.
  2019{\natexlab{b}}, \mnras, 483, 2007

\bibitem[{{Evans} {et~al.}(2020){Evans}, {Renzo}, \& {Rossi}}]{Evans2020}
{Evans}, F.~A., {Renzo}, M., \& {Rossi}, E.~M. 2020, \mnras, 497, 5344

\bibitem[{{Fabricius} {et~al.}(2021){Fabricius}, {Luri}, {Arenou}, {Babusiaux},
  {Helmi}, {Muraveva}, {Reyl{\'e}}, {Spoto}, {Vallenari}, {Antoja}, {Balbinot},
  {Barache}, {Bauchet}, {Bragaglia}, {Busonero}, {Cantat-Gaudin}, {Carrasco},
  {Diakit{\'e}}, {Fabrizio}, {Figueras}, {Garcia-Gutierrez}, {Garofalo},
  {Jordi}, {Kervella}, {Khanna}, {Leclerc}, {Licata}, {Lambert}, {Marrese},
  {Masip}, {Ramos}, {Robichon}, {Robin}, {Romero-G{\'o}mez}, {Rubele}, \&
  {Weiler}}]{Fabricius2020EDR3}
{Fabricius}, C., {Luri}, X., {Arenou}, F., {et~al.} 2021, \aap, 649, A5

\bibitem[{{Fabrizio} {et~al.}(2021){Fabrizio}, {Braga}, {Crestani}, {Bono},
  {Ferraro}, {Fiorentino}, {Iannicola}, {Preston}, {Sneden}, {Th{\'e}venin},
  {Altavilla}, {Chaboyer}, {Dall'Ora}, {da Silva}, {Grebel}, {Gilligan},
  {Lala}, {Lemasle}, {Magurno}, {Marengo}, {Marinoni}, {Marrese},
  {Mart{\'\i}nez-V{\'a}zquez}, {Matsunaga}, {Monelli}, {Mullen}, {Neeley},
  {Nonino}, {Prudil}, {Salaris}, {Stetson}, {Valenti}, \&
  {Zoccali}}]{Fabrizio2021}
{Fabrizio}, M., {Braga}, V.~F., {Crestani}, J., {et~al.} 2021, \apj, 919, 118

\bibitem[{{For} {et~al.}(2011){For}, {Sneden}, \& {Preston}}]{For2011chem}
{For}, B.-Q., {Sneden}, C., \& {Preston}, G.~W. 2011, \apjs, 197, 29

\bibitem[{{Foreman-Mackey} {et~al.}(2013){Foreman-Mackey}, {Hogg}, {Lang}, \&
  {Goodman}}]{Foreman-Mackey2013}
{Foreman-Mackey}, D., {Hogg}, D.~W., {Lang}, D., \& {Goodman}, J. 2013, \pasp,
  125, 306

\bibitem[{{Fragione} \& {Capuzzo-Dolcetta}(2016)}]{Fragione2016}
{Fragione}, G. \& {Capuzzo-Dolcetta}, R. 2016, \mnras, 458, 2596

\bibitem[{{Gaia Collaboration} {et~al.}(2021{\natexlab{a}}){Gaia
  Collaboration}, {Brown}, {Vallenari}, {Prusti}, {de Bruijne}, {Babusiaux},
  {Biermann}, {Creevey}, {Evans}, {Eyer}, {Hutton}, {Jansen}, {Jordi},
  {Klioner}, {Lammers}, {Lindegren}, {Luri}, {Mignard}, {Panem}, {Pourbaix},
  {Randich}, {Sartoretti}, {Soubiran}, {Walton}, {Arenou}, {Bailer-Jones},
  {Bastian}, {Cropper}, {Drimmel}, {Katz}, {Lattanzi}, {van Leeuwen}, {Bakker},
  {Cacciari}, {Casta{\~n}eda}, {De Angeli}, {Ducourant}, {Fabricius},
  {Fouesneau}, {Fr{\'e}mat}, {Guerra}, {Guerrier}, {Guiraud}, {Jean-Antoine
  Piccolo}, {Masana}, {Messineo}, {Mowlavi}, {Nicolas}, {Nienartowicz},
  {Pailler}, {Panuzzo}, {Riclet}, {Roux}, {Seabroke}, {Sordo}, {Tanga},
  {Th{\'e}venin}, {Gracia-Abril}, {Portell}, {Teyssier}, {Altmann}, {Andrae},
  {Bellas-Velidis}, {Benson}, {Berthier}, {Blomme}, {Brugaletta}, {Burgess},
  {Busso}, {Carry}, {Cellino}, {Cheek}, {Clementini}, {Damerdji}, {Davidson},
  {Delchambre}, {Dell'Oro}, {Fern{\'a}ndez-Hern{\'a}ndez}, {Galluccio},
  {Garc{\'\i}a-Lario}, {Garcia-Reinaldos}, {Gonz{\'a}lez-N{\'u}{\~n}ez},
  {Gosset}, {Haigron}, {Halbwachs}, {Hambly}, {Harrison}, {Hatzidimitriou},
  {Heiter}, {Hern{\'a}ndez}, {Hestroffer}, {Hodgkin}, {Holl}, {Jan{\ss}en},
  {Jevardat de Fombelle}, {Jordan}, {Krone-Martins}, {Lanzafame},
  {L{\"o}ffler}, {Lorca}, {Manteiga}, {Marchal}, {Marrese}, {Moitinho}, {Mora},
  {Muinonen}, {Osborne}, {Pancino}, {Pauwels}, {Petit}, {Recio-Blanco},
  {Richards}, {Riello}, {Rimoldini}, {Robin}, {Roegiers}, {Rybizki}, {Sarro},
  {Siopis}, {Smith}, {Sozzetti}, {Ulla}, {Utrilla}, {van Leeuwen}, {van
  Reeven}, {Abbas}, {Abreu Aramburu}, {Accart}, {Aerts}, {Aguado}, {Ajaj},
  {Altavilla}, {{\'A}lvarez}, {{\'A}lvarez Cid-Fuentes}, {Alves}, {Anderson},
  {Anglada Varela}, {Antoja}, {Audard}, {Baines}, {Baker},
  {Balaguer-N{\'u}{\~n}ez}, {Balbinot}, {Balog}, {Barache}, {Barbato},
  {Barros}, {Barstow}, {Bartolom{\'e}}, {Bassilana}, {Bauchet},
  {Baudesson-Stella}, {Becciani}, {Bellazzini}, {Bernet}, {Bertone}, {Bianchi},
  {Blanco-Cuaresma}, {Boch}, {Bombrun}, {Bossini}, {Bouquillon}, {Bragaglia},
  {Bramante}, {Breedt}, {Bressan}, {Brouillet}, {Bucciarelli}, {Burlacu},
  {Busonero}, {Butkevich}, {Buzzi}, {Caffau}, {Cancelliere}, {C{\'a}novas},
  {Cantat-Gaudin}, {Carballo}, {Carlucci}, {Carnerero}, {Carrasco},
  {Casamiquela}, {Castellani}, {Castro-Ginard}, {Castro Sampol}, {Chaoul},
  {Charlot}, {Chemin}, {Chiavassa}, {Cioni}, {Comoretto}, {Cooper}, {Cornez},
  {Cowell}, {Crifo}, {Crosta}, {Crowley}, {Dafonte}, {Dapergolas}, {David},
  {David}, {de Laverny}, {De Luise}, {De March}, {De Ridder}, {de Souza}, {de
  Teodoro}, {de Torres}, {del Peloso}, {del Pozo}, {Delbo}, {Delgado},
  {Delgado}, {Delisle}, {Di Matteo}, {Diakite}, {Diener}, {Distefano},
  {Dolding}, {Eappachen}, {Edvardsson}, {Enke}, {Esquej}, {Fabre}, {Fabrizio},
  {Faigler}, {Fedorets}, {Fernique}, {Fienga}, {Figueras}, {Fouron},
  {Fragkoudi}, {Fraile}, {Franke}, {Gai}, {Garabato}, {Garcia-Gutierrez},
  {Garc{\'\i}a-Torres}, {Garofalo}, {Gavras}, {Gerlach}, {Geyer}, {Giacobbe},
  {Gilmore}, {Girona}, {Giuffrida}, {Gomel}, {Gomez}, {Gonzalez-Santamaria},
  {Gonz{\'a}lez-Vidal}, {Granvik}, {Guti{\'e}rrez-S{\'a}nchez}, {Guy},
  {Hauser}, {Haywood}, {Helmi}, {Hidalgo}, {Hilger}, {H{\l}adczuk}, {Hobbs},
  {Holland}, {Huckle}, {Jasniewicz}, {Jonker}, {Juaristi Campillo}, {Julbe},
  {Karbevska}, {Kervella}, {Khanna}, {Kochoska}, {Kontizas}, {Kordopatis},
  {Korn}, {Kostrzewa-Rutkowska}, {Kruszy{\'n}ska}, {Lambert}, {Lanza}, {Lasne},
  {Le Campion}, {Le Fustec}, {Lebreton}, {Lebzelter}, {Leccia}, {Leclerc},
  {Lecoeur-Taibi}, {Liao}, {Licata}, {Lindstr{\o}m}, {Lister}, {Livanou},
  {Lobel}, {Madrero Pardo}, {Managau}, {Mann}, {Marchant}, {Marconi}, {Marcos
  Santos}, {Marinoni}, {Marocco}, {Marshall}, {Martin Polo},
  {Mart{\'\i}n-Fleitas}, {Masip}, {Massari}, {Mastrobuono-Battisti}, {Mazeh},
  {McMillan}, {Messina}, {Michalik}, {Millar}, {Mints}, {Molina}, {Molinaro},
  {Moln{\'a}r}, {Montegriffo}, {Mor}, {Morbidelli}, {Morel}, {Morris},
  {Mulone}, {Munoz}, {Muraveva}, {Murphy}, {Musella}, {Noval}, {Ord{\'e}novic},
  {Orr{\`u}}, {Osinde}, {Pagani}, {Pagano}, {Palaversa}, {Palicio}, {Panahi},
  {Pawlak}, {Pe{\~n}alosa Esteller}, {Penttil{\"a}}, {Piersimoni}, {Pineau},
  {Plachy}, {Plum}, {Poggio}, {Poretti}, {Poujoulet}, {Pr{\v{s}}a}, {Pulone},
  {Racero}, {Ragaini}, {Rainer}, {Raiteri}, {Rambaux}, {Ramos}, {Ramos-Lerate},
  {Re Fiorentin}, {Regibo}, {Reyl{\'e}}, {Ripepi}, {Riva}, {Rixon}, {Robichon},
  {Robin}, {Roelens}, {Rohrbasser}, {Romero-G{\'o}mez}, {Rowell}, {Royer},
  {Rybicki}, {Sadowski}, {Sagrist{\`a} Sell{\'e}s}, {Sahlmann}, {Salgado},
  {Salguero}, {Samaras}, {Sanchez Gimenez}, {Sanna}, {Santove{\~n}a},
  {Sarasso}, {Schultheis}, {Sciacca}, {Segol}, {Segovia}, {S{\'e}gransan},
  {Semeux}, {Shahaf}, {Siddiqui}, {Siebert}, {Siltala}, {Slezak}, {Smart},
  {Solano}, {Solitro}, {Souami}, {Souchay}, {Spagna}, {Spoto}, {Steele},
  {Steidelm{\"u}ller}, {Stephenson}, {S{\"u}veges}, {Szabados}, {Szegedi-Elek},
  {Taris}, {Tauran}, {Taylor}, {Teixeira}, {Thuillot}, {Tonello}, {Torra},
  {Torra}, {Turon}, {Unger}, {Vaillant}, {van Dillen}, {Vanel}, {Vecchiato},
  {Viala}, {Vicente}, {Voutsinas}, {Weiler}, {Wevers}, {Wyrzykowski}, {Yoldas},
  {Yvard}, {Zhao}, {Zorec}, {Zucker}, {Zurbach}, \&
  {Zwitter}}]{GaiaEDR3Summary2020}
{Gaia Collaboration}, {Brown}, A.~G.~A., {Vallenari}, A., {et~al.}
  2021{\natexlab{a}}, \aap, 649, A1

\bibitem[{{Gaia Collaboration} {et~al.}(2016){Gaia Collaboration}, {Prusti},
  {de Bruijne}, {Brown}, {Vallenari}, {Babusiaux}, {Bailer-Jones}, {Bastian},
  {Biermann}, {Evans}, {Eyer}, {Jansen}, {Jordi}, {Klioner}, {Lammers},
  {Lindegren}, {Luri}, {Mignard}, {Milligan}, {Panem}, {Poinsignon},
  {Pourbaix}, {Randich}, {Sarri}, {Sartoretti}, {Siddiqui}, {Soubiran},
  {Valette}, {van Leeuwen}, {Walton}, {Aerts}, {Arenou}, {Cropper}, {Drimmel},
  {H{\o}g}, {Katz}, {Lattanzi}, {O'Mullane}, {Grebel}, {Holland}, {Huc},
  {Passot}, {Bramante}, {Cacciari}, {Casta{\~n}eda}, {Chaoul}, {Cheek}, {De
  Angeli}, {Fabricius}, {Guerra}, {Hern{\'a}ndez}, {Jean-Antoine-Piccolo},
  {Masana}, {Messineo}, {Mowlavi}, {Nienartowicz}, {Ord{\'o}{\~n}ez-Blanco},
  {Panuzzo}, {Portell}, {Richards}, {Riello}, {Seabroke}, {Tanga},
  {Th{\'e}venin}, {Torra}, {Els}, {Gracia-Abril}, {Comoretto},
  {Garcia-Reinaldos}, {Lock}, {Mercier}, {Altmann}, {Andrae}, {Astraatmadja},
  {Bellas-Velidis}, {Benson}, {Berthier}, {Blomme}, {Busso}, {Carry},
  {Cellino}, {Clementini}, {Cowell}, {Creevey}, {Cuypers}, {Davidson}, {De
  Ridder}, {de Torres}, {Delchambre}, {Dell'Oro}, {Ducourant}, {Fr{\'e}mat},
  {Garc{\'\i}a-Torres}, {Gosset}, {Halbwachs}, {Hambly}, {Harrison}, {Hauser},
  {Hestroffer}, {Hodgkin}, {Huckle}, {Hutton}, {Jasniewicz}, {Jordan},
  {Kontizas}, {Korn}, {Lanzafame}, {Manteiga}, {Moitinho}, {Muinonen},
  {Osinde}, {Pancino}, {Pauwels}, {Petit}, {Recio-Blanco}, {Robin}, {Sarro},
  {Siopis}, {Smith}, {Smith}, {Sozzetti}, {Thuillot}, {van Reeven}, {Viala},
  {Abbas}, {Abreu Aramburu}, {Accart}, {Aguado}, {Allan}, {Allasia},
  {Altavilla}, {{\'A}lvarez}, {Alves}, {Anderson}, {Andrei}, {Anglada Varela},
  {Antiche}, {Antoja}, {Ant{\'o}n}, {Arcay}, {Atzei}, {Ayache}, {Bach},
  {Baker}, {Balaguer-N{\'u}{\~n}ez}, {Barache}, {Barata}, {Barbier}, {Barblan},
  {Baroni}, {Barrado y Navascu{\'e}s}, {Barros}, {Barstow}, {Becciani},
  {Bellazzini}, {Bellei}, {Bello Garc{\'\i}a}, {Belokurov}, {Bendjoya},
  {Berihuete}, {Bianchi}, {Bienaym{\'e}}, {Billebaud}, {Blagorodnova},
  {Blanco-Cuaresma}, {Boch}, {Bombrun}, {Borrachero}, {Bouquillon}, {Bourda},
  {Bouy}, {Bragaglia}, {Breddels}, {Brouillet}, {Br{\"u}semeister},
  {Bucciarelli}, {Budnik}, {Burgess}, {Burgon}, {Burlacu}, {Busonero}, {Buzzi},
  {Caffau}, {Cambras}, {Campbell}, {Cancelliere}, {Cantat-Gaudin}, {Carlucci},
  {Carrasco}, {Castellani}, {Charlot}, {Charnas}, {Charvet}, {Chassat},
  {Chiavassa}, {Clotet}, {Cocozza}, {Collins}, {Collins}, {Costigan}, {Crifo},
  {Cross}, {Crosta}, {Crowley}, {Dafonte}, {Damerdji}, {Dapergolas}, {David},
  {David}, {De Cat}, {de Felice}, {de Laverny}, {De Luise}, {De March}, {de
  Martino}, {de Souza}, {Debosscher}, {del Pozo}, {Delbo}, {Delgado},
  {Delgado}, {di Marco}, {Di Matteo}, {Diakite}, {Distefano}, {Dolding}, {Dos
  Anjos}, {Drazinos}, {Dur{\'a}n}, {Dzigan}, {Ecale}, {Edvardsson}, {Enke},
  {Erdmann}, {Escolar}, {Espina}, {Evans}, {Eynard Bontemps}, {Fabre},
  {Fabrizio}, {Faigler}, {Falc{\~a}o}, {Farr{\`a}s Casas}, {Faye}, {Federici},
  {Fedorets}, {Fern{\'a}ndez-Hern{\'a}ndez}, {Fernique}, {Fienga}, {Figueras},
  {Filippi}, {Findeisen}, {Fonti}, {Fouesneau}, {Fraile}, {Fraser}, {Fuchs},
  {Furnell}, {Gai}, {Galleti}, {Galluccio}, {Garabato}, {Garc{\'\i}a-Sedano},
  {Gar{\'e}}, {Garofalo}, {Garralda}, {Gavras}, {Gerssen}, {Geyer}, {Gilmore},
  {Girona}, {Giuffrida}, {Gomes}, {Gonz{\'a}lez-Marcos},
  {Gonz{\'a}lez-N{\'u}{\~n}ez}, {Gonz{\'a}lez-Vidal}, {Granvik}, {Guerrier},
  {Guillout}, {Guiraud}, {G{\'u}rpide}, {Guti{\'e}rrez-S{\'a}nchez}, {Guy},
  {Haigron}, {Hatzidimitriou}, {Haywood}, {Heiter}, {Helmi}, {Hobbs},
  {Hofmann}, {Holl}, {Holland }, {Hunt}, {Hypki}, {Icardi}, {Irwin}, {Jevardat
  de Fombelle}, {Jofr{\'e}}, {Jonker}, {Jorissen}, {Julbe}, {Karampelas},
  {Kochoska}, {Kohley}, {Kolenberg}, {Kontizas}, {Koposov}, {Kordopatis},
  {Koubsky}, {Kowalczyk}, {Krone-Martins}, {Kudryashova}, {Kull}, {Bachchan},
  {Lacoste-Seris}, {Lanza}, {Lavigne}, {Le Poncin-Lafitte}, {Lebreton},
  {Lebzelter}, {Leccia}, {Leclerc}, {Lecoeur-Taibi}, {Lemaitre}, {Lenhardt},
  {Leroux}, {Liao}, {Licata}, {Lindstr{\o}m}, {Lister}, {Livanou}, {Lobel},
  {L{\"o}ffler}, {L{\'o}pez}, {Lopez-Lozano}, {Lorenz}, {Loureiro},
  {MacDonald}, {Magalh{\~a}es Fernandes}, {Managau}, {Mann}, {Mantelet},
  {Marchal}, {Marchant}, {Marconi}, {Marie}, {Marinoni}, {Marrese},
  {Marschalk{\'o}}, {Marshall}, {Mart{\'\i}n-Fleitas}, {Martino}, {Mary},
  {Matijevi{\v{c}}}, {Mazeh}, {McMillan}, {Messina}, {Mestre}, {Michalik},
  {Millar}, {Miranda}, {Molina}, {Molinaro}, {Molinaro}, {Moln{\'a}r},
  {Moniez}, {Montegriffo}, {Monteiro}, {Mor}, {Mora}, {Morbidelli}, {Morel},
  {Morgenthaler}, {Morley}, {Morris}, {Mulone}, {Muraveva}, {Musella},
  {Narbonne}, {Nelemans}, {Nicastro}, {Noval}, {Ord{\'e}novic},
  {Ordieres-Mer{\'e}}, {Osborne}, {Pagani}, {Pagano}, {Pailler}, {Palacin},
  {Palaversa}, {Parsons}, {Paulsen}, {Pecoraro}, {Pedrosa}, {Pentik{\"a}inen},
  {Pereira}, {Pichon}, {Piersimoni}, {Pineau}, {Plachy}, {Plum}, {Poujoulet},
  {Pr{\v{s}}a}, {Pulone}, {Ragaini}, {Rago}, {Rambaux}, {Ramos-Lerate},
  {Ranalli}, {Rauw}, {Read}, {Regibo}, {Renk}, {Reyl{\'e}}, {Ribeiro},
  {Rimoldini}, {Ripepi}, {Riva}, {Rixon}, {Roelens}, {Romero-G{\'o}mez},
  {Rowell}, {Royer}, {Rudolph}, {Ruiz-Dern}, {Sadowski}, {Sagrist{\`a}
  Sell{\'e}s}, {Sahlmann}, {Salgado}, {Salguero}, {Sarasso}, {Savietto},
  {Schnorhk}, {Schultheis}, {Sciacca}, {Segol}, {Segovia}, {Segransan},
  {Serpell}, {Shih}, {Smareglia}, {Smart}, {Smith}, {Solano}, {Solitro},
  {Sordo}, {Soria Nieto}, {Souchay}, {Spagna}, {Spoto}, {Stampa}, {Steele},
  {Steidelm{\"u}ller}, {Stephenson}, {Stoev}, {Suess}, {S{\"u}veges}, {Surdej},
  {Szabados}, {Szegedi-Elek}, {Tapiador}, {Taris}, {Tauran}, {Taylor},
  {Teixeira}, {Terrett}, {Tingley}, {Trager}, {Turon}, {Ulla}, {Utrilla},
  {Valentini}, {van Elteren}, {Van Hemelryck}, {van Leeuwen}, {Varadi},
  {Vecchiato}, {Veljanoski}, {Via}, {Vicente}, {Vogt}, {Voss}, {Votruba},
  {Voutsinas}, {Walmsley}, {Weiler}, {Weingrill}, {Werner}, {Wevers},
  {Whitehead}, {Wyrzykowski}, {Yoldas}, {{\v{Z}}erjal}, {Zucker}, {Zurbach},
  {Zwitter}, {Alecu}, {Allen}, {Allende Prieto}, {Amorim},
  {Anglada-Escud{\'e}}, {Arsenijevic}, {Azaz}, {Balm}, {Beck}, {Bernstein},
  {Bigot}, {Bijaoui}, {Blasco}, {Bonfigli}, {Bono}, {Boudreault}, {Bressan},
  {Brown}, {Brunet}, {Bunclark}, {Buonanno}, {Butkevich}, {Carret}, {Carrion},
  {Chemin}, {Ch{\'e}reau}, {Corcione}, {Darmigny}, {de Boer}, {de Teodoro}, {de
  Zeeuw}, {Delle Luche}, {Domingues}, {Dubath}, {Fodor}, {Fr{\'e}zouls},
  {Fries}, {Fustes}, {Fyfe}, {Gallardo}, {Gallegos}, {Gardiol}, {Gebran},
  {Gomboc}, {G{\'o}mez}, {Grux}, {Gueguen}, {Heyrovsky}, {Hoar}, {Iannicola},
  {Isasi Parache}, {Janotto}, {Joliet}, {Jonckheere}, {Keil}, {Kim},
  {Klagyivik}, {Klar}, {Knude}, {Kochukhov}, {Kolka}, {Kos}, {Kutka}, {Lainey},
  {LeBouquin}, {Liu}, {Loreggia}, {Makarov}, {Marseille}, {Martayan},
  {Martinez-Rubi}, {Massart}, {Meynadier}, {Mignot}, {Munari}, {Nguyen},
  {Nordlander}, {Ocvirk}, {O'Flaherty}, {Olias Sanz}, {Ortiz}, {Osorio},
  {Oszkiewicz}, {Ouzounis}, {Palmer}, {Park}, {Pasquato}, {Peltzer}, {Peralta},
  {P{\'e}turaud}, {Pieniluoma}, {Pigozzi}, {Poels}, {Prat}, {Prod'homme},
  {Raison}, {Rebordao}, {Risquez}, {Rocca-Volmerange}, {Rosen}, {Ruiz-Fuertes},
  {Russo}, {Sembay}, {Serraller Vizcaino}, {Short}, {Siebert}, {Silva},
  {Sinachopoulos}, {Slezak}, {Soffel}, {Sosnowska}, {Strai{\v{z}}ys}, {ter
  Linden}, {Terrell}, {Theil}, {Tiede}, {Troisi}, {Tsalmantza}, {Tur},
  {Vaccari}, {Vachier}, {Valles}, {Van Hamme}, {Veltz}, {Virtanen}, {Wallut},
  {Wichmann}, {Wilkinson}, {Ziaeepour}, \& {Zschocke}}]{Gaia2016}
{Gaia Collaboration}, {Prusti}, T., {de Bruijne}, J.~H.~J., {et~al.} 2016,
  \aap, 595, A1

\bibitem[{{Gaia Collaboration} {et~al.}(2021{\natexlab{b}}){Gaia
  Collaboration}, {Smart}, {Sarro}, {Rybizki}, {Reyl{\'e}}, {Robin}, {Hambly},
  {Abbas}, {Barstow}, {de Bruijne}, {Bucciarelli}, {Carrasco}, {Cooper},
  {Hodgkin}, {Masana}, {Michalik}, {Sahlmann}, {Sozzetti}, {Brown},
  {Vallenari}, {Prusti}, {Babusiaux}, {Biermann}, {Creevey}, {Evans}, {Eyer},
  {Hutton}, {Jansen}, {Jordi}, {Klioner}, {Lammers}, {Lindegren}, {Luri},
  {Mignard}, {Panem}, {Pourbaix}, {Randich}, {Sartoretti}, {Soubiran},
  {Walton}, {Arenou}, {Bailer-Jones}, {Bastian}, {Cropper}, {Drimmel}, {Katz},
  {Lattanzi}, {van Leeuwen}, {Bakker}, {Casta{\~n}eda}, {De Angeli},
  {Ducourant}, {Fabricius}, {Fouesneau}, {Fr{\'e}mat}, {Guerra}, {Guerrier},
  {Guiraud}, {Jean-Antoine Piccolo}, {Messineo}, {Mowlavi}, {Nicolas},
  {Nienartowicz}, {Pailler}, {Panuzzo}, {Riclet}, {Roux}, {Seabroke}, {Sordo},
  {Tanga}, {Th{\'e}venin}, {Gracia-Abril}, {Portell}, {Teyssier}, {Altmann},
  {Andrae}, {Bellas-Velidis}, {Benson}, {Berthier}, {Blomme}, {Brugaletta},
  {Burgess}, {Busso}, {Carry}, {Cellino}, {Cheek}, {Clementini}, {Damerdji},
  {Davidson}, {Delchambre}, {Dell'Oro}, {Fern{\'a}ndez-Hern{\'a}ndez},
  {Galluccio}, {Garc{\'\i}a-Lario}, {Garcia-Reinaldos},
  {Gonz{\'a}lez-N{\'u}{\~n}ez}, {Gosset}, {Haigron}, {Halbwachs}, {Harrison},
  {Hatzidimitriou}, {Heiter}, {Hern{\'a}ndez}, {Hestroffer}, {Holl},
  {Jan{\ss}en}, {Jevardat de Fombelle}, {Jordan}, {Krone-Martins}, {Lanzafame},
  {L{\"o}ffler}, {Lorca}, {Manteiga}, {Marchal}, {Marrese}, {Moitinho}, {Mora},
  {Muinonen}, {Osborne}, {Pancino}, {Pauwels}, {Recio-Blanco}, {Richards},
  {Riello}, {Rimoldini}, {Roegiers}, {Siopis}, {Smith}, {Ulla}, {Utrilla}, {van
  Leeuwen}, {van Reeven}, {Abreu Aramburu}, {Accart}, {Aerts}, {Aguado},
  {Ajaj}, {Altavilla}, {{\'A}lvarez}, {{\'A}lvarez Cid-Fuentes}, {Alves},
  {Anderson}, {Anglada Varela}, {Antoja}, {Audard}, {Baines}, {Baker},
  {Balaguer-N{\'u}{\~n}ez}, {Balbinot}, {Balog}, {Barache}, {Barbato},
  {Barros}, {Bartolom{\'e}}, {Bassilana}, {Bauchet}, {Baudesson-Stella},
  {Becciani}, {Bellazzini}, {Bernet}, {Bertone}, {Bianchi}, {Blanco-Cuaresma},
  {Boch}, {Bombrun}, {Bossini}, {Bouquillon}, {Bragaglia}, {Bramante},
  {Breedt}, {Bressan}, {Brouillet}, {Burlacu}, {Busonero}, {Butkevich},
  {Buzzi}, {Caffau}, {Cancelliere}, {C{\'a}novas}, {Cantat-Gaudin}, {Carballo},
  {Carlucci}, {Carnerero}, {Casamiquela}, {Castellani}, {Castro-Ginard},
  {Castro Sampol}, {Chaoul}, {Charlot}, {Chemin}, {Chiavassa}, {Cioni},
  {Comoretto}, {Cornez}, {Cowell}, {Crifo}, {Crosta}, {Crowley}, {Dafonte},
  {Dapergolas}, {David}, {David}, {de Laverny}, {De Luise}, {De March}, {De
  Ridder}, {de Souza}, {de Teodoro}, {de Torres}, {del Peloso}, {del Pozo},
  {Delgado}, {Delgado}, {Delisle}, {Di Matteo}, {Diakite}, {Diener},
  {Distefano}, {Dolding}, {Eappachen}, {Edvardsson}, {Enke}, {Esquej}, {Fabre},
  {Fabrizio}, {Faigler}, {Fedorets}, {Fernique}, {Fienga}, {Figueras},
  {Fouron}, {Fragkoudi}, {Fraile}, {Franke}, {Gai}, {Garabato},
  {Garcia-Gutierrez}, {Garc{\'\i}a-Torres}, {Garofalo}, {Gavras}, {Gerlach},
  {Geyer}, {Giacobbe}, {Gilmore}, {Girona}, {Giuffrida}, {Gomel}, {Gomez},
  {Gonzalez-Santamaria}, {Gonz{\'a}lez-Vidal}, {Granvik},
  {Guti{\'e}rrez-S{\'a}nchez}, {Guy}, {Hauser}, {Haywood}, {Helmi}, {Hidalgo},
  {Hilger}, {H{\l}adczuk}, {Hobbs}, {Holland}, {Huckle}, {Jasniewicz},
  {Jonker}, {Juaristi Campillo}, {Julbe}, {Karbevska}, {Kervella}, {Khanna},
  {Kochoska}, {Kontizas}, {Kordopatis}, {Korn}, {Kostrzewa-Rutkowska},
  {Kruszy{\'n}ska}, {Lambert}, {Lanza}, {Lasne}, {Le Campion}, {Le Fustec},
  {Lebreton}, {Lebzelter}, {Leccia}, {Leclerc}, {Lecoeur-Taibi}, {Liao},
  {Licata}, {Lindstr{\o}m}, {Lister}, {Livanou}, {Lobel}, {Madrero Pardo},
  {Managau}, {Mann}, {Marchant}, {Marconi}, {Marcos Santos}, {Marinoni},
  {Marocco}, {Marshall}, {Martin Polo}, {Mart{\'\i}n-Fleitas}, {Masip},
  {Massari}, {Mastrobuono-Battisti}, {Mazeh}, {McMillan}, {Messina}, {Millar},
  {Mints}, {Molina}, {Molinaro}, {Moln{\'a}r}, {Montegriffo}, {Mor},
  {Morbidelli}, {Morel}, {Morris}, {Mulone}, {Munoz}, {Muraveva}, {Murphy},
  {Musella}, {Noval}, {Ord{\'e}novic}, {Orr{\`u}}, {Osinde}, {Pagani},
  {Pagano}, {Palaversa}, {Palicio}, {Panahi}, {Pawlak}, {Pe{\~n}alosa
  Esteller}, {Penttil{\"a}}, {Piersimoni}, {Pineau}, {Plachy}, {Plum},
  {Poggio}, {Poretti}, {Poujoulet}, {Pr{\v{s}}a}, {Pulone}, {Racero},
  {Ragaini}, {Rainer}, {Raiteri}, {Rambaux}, {Ramos}, {Ramos-Lerate}, {Re
  Fiorentin}, {Regibo}, {Ripepi}, {Riva}, {Rixon}, {Robichon}, {Robin},
  {Roelens}, {Rohrbasser}, {Romero-G{\'o}mez}, {Rowell}, {Royer}, {Rybicki},
  {Sadowski}, {Sagrist{\`a} Sell{\'e}s}, {Salgado}, {Salguero}, {Samaras},
  {Sanchez Gimenez}, {Sanna}, {Santove{\~n}a}, {Sarasso}, {Schultheis},
  {Sciacca}, {Segol}, {Segovia}, {S{\'e}gransan}, {Semeux}, {Shahaf},
  {Siddiqui}, {Siebert}, {Siltala}, {Slezak}, {Solano}, {Solitro}, {Souami},
  {Souchay}, {Spagna}, {Spoto}, {Steele}, {Steidelm{\"u}ller}, {Stephenson},
  {S{\"u}veges}, {Szabados}, {Szegedi-Elek}, {Taris}, {Tauran}, {Taylor},
  {Teixeira}, {Thuillot}, {Tonello}, {Torra}, {Torra}, {Turon}, {Unger},
  {Vaillant}, {van Dillen}, {Vanel}, {Vecchiato}, {Viala}, {Vicente},
  {Voutsinas}, {Weiler}, {Wevers}, {Wyrzykowski}, {Yoldas}, {Yvard}, {Zhao},
  {Zorec}, {Zucker}, {Zurbach}, \& {Zwitter}}]{Smart2020CloseEDR3}
{Gaia Collaboration}, {Smart}, R.~L., {Sarro}, L.~M., {et~al.}
  2021{\natexlab{b}}, \aap, 649, A6

\bibitem[{{Garrow} {et~al.}(2020){Garrow}, {Webb}, \& {Bovy}}]{Garrow2020}
{Garrow}, T., {Webb}, J.~J., \& {Bovy}, J. 2020, \mnras, 499, 804

\bibitem[{{Gonz{\'a}lez Hern{\'a}ndez} \& {Bonifacio}(2009)}]{Gonzalez2009}
{Gonz{\'a}lez Hern{\'a}ndez}, J.~I. \& {Bonifacio}, P. 2009, \aap, 497, 497

\bibitem[{{Grand} {et~al.}(2019){Grand}, {Deason}, {White}, {Simpson},
  {G{\'o}mez}, {Marinacci}, \& {Pakmor}}]{Grand2019}
{Grand}, R. J.~J., {Deason}, A.~J., {White}, S. D.~M., {et~al.} 2019, \mnras,
  487, L72

\bibitem[{{Grand} {et~al.}(2017){Grand}, {G{\'o}mez}, {Marinacci}, {Pakmor},
  {Springel}, {Campbell}, {Frenk}, {Jenkins}, \& {White}}]{Grand2017}
{Grand}, R. J.~J., {G{\'o}mez}, F.~A., {Marinacci}, F., {et~al.} 2017, \mnras,
  467, 179

\bibitem[{{Grand} {et~al.}(2018){Grand}, {Helly}, {Fattahi}, {Cautun}, {Cole},
  {Cooper}, {Deason}, {Frenk}, {G{\'o}mez}, {Hunt}, {Marinacci}, {Pakmor},
  {Simpson}, {Springel}, \& {Xu}}]{Grand2018}
{Grand}, R. J.~J., {Helly}, J., {Fattahi}, A., {et~al.} 2018, \mnras, 481, 1726

\bibitem[{{Green}(2018)}]{Green2018dust}
{Green}, G. 2018, The Journal of Open Source Software, 3, 695

\bibitem[{{Gvaramadze} {et~al.}(2009){Gvaramadze}, {Gualandris}, \& {Portegies
  Zwart}}]{Gvaramadze2009}
{Gvaramadze}, V.~V., {Gualandris}, A., \& {Portegies Zwart}, S. 2009, \mnras,
  396, 570

\bibitem[{{Hansen} {et~al.}(2011){Hansen}, {Nordstr{\"o}m}, {Bonifacio},
  {Spite}, {Andersen}, {Beers}, {Cayrel}, {Spite}, {Molaro}, {Barbuy},
  {Depagne}, {Fran{\c{c}}ois}, {Hill}, {Plez}, \& {Sivarani}}]{Hansen2011}
{Hansen}, C.~J., {Nordstr{\"o}m}, B., {Bonifacio}, P., {et~al.} 2011, \aap,
  527, A65

\bibitem[{{Hansen} {et~al.}(2016){Hansen}, {Rich}, {Koch}, {Xu}, {Kunder}, \&
  {Ludwig}}]{Hansen2016}
{Hansen}, C.~J., {Rich}, R.~M., {Koch}, A., {et~al.} 2016, \aap, 590, A39

\bibitem[{Harris {et~al.}(2020)Harris, Millman, van~der Walt, Gommers,
  Virtanen, Cournapeau, Wieser, Taylor, Berg, Smith, Kern, Picus, Hoyer, van
  Kerkwijk, Brett, Haldane, Fernández~del Río, Wiebe, Peterson,
  Gérard-Marchant, Sheppard, Reddy, Weckesser, Abbasi, Gohlke, \&
  Oliphant}]{numpy}
Harris, C.~R., Millman, K.~J., van~der Walt, S.~J., {et~al.} 2020, Nature, 585,
  357–362

\bibitem[{{Hattori} {et~al.}(2018{\natexlab{a}}){Hattori}, {Valluri}, {Bell},
  \& {Roederer}}]{Hattori2018metPoor_LMC}
{Hattori}, K., {Valluri}, M., {Bell}, E.~F., \& {Roederer}, I.~U.
  2018{\natexlab{a}}, \apj, 866, 121

\bibitem[{{Hattori} {et~al.}(2018{\natexlab{b}}){Hattori}, {Valluri}, \&
  {Castro}}]{Hattori2018}
{Hattori}, K., {Valluri}, M., \& {Castro}, N. 2018{\natexlab{b}}, \apj, 869, 33

\bibitem[{{Hayes} {et~al.}(2020){Hayes}, {Majewski}, {Hasselquist}, {Anguiano},
  {Shetrone}, {Law}, {Schiavon}, {Cunha}, {Smith}, {Beaton}, {Price-Whelan},
  {Allende Prieto}, {Battaglia}, {Bizyaev}, {Brownstein}, {Cohen},
  {Frinchaboy}, {Garc{\'\i}a-Hern{\'a}ndez}, {Lacerna}, {Lane},
  {M{\'e}sz{\'a}ros}, {Bidin}, {M{\~{u}}noz}, {Nidever}, {Oravetz}, {Oravetz},
  {Pan}, {Roman-Lopes}, {Sobeck}, \& {Stringfellow}}]{Hayes2020SagStrGrad}
{Hayes}, C.~R., {Majewski}, S.~R., {Hasselquist}, S., {et~al.} 2020, \apj, 889,
  63

\bibitem[{{Helmi} {et~al.}(2018){Helmi}, {Babusiaux}, {Koppelman}, {Massari},
  {Veljanoski}, \& {Brown}}]{Helmi2018Nature}
{Helmi}, A., {Babusiaux}, C., {Koppelman}, H.~H., {et~al.} 2018, \nat, 563, 85

\bibitem[{{Hernitschek} {et~al.}(2017){Hernitschek}, {Sesar}, {Rix},
  {Belokurov}, {Martinez-Delgado}, {Martin}, {Kaiser}, {Hodapp}, {Chambers},
  {Wainscoat}, {Magnier}, {Kudritzki}, {Metcalfe}, \&
  {Draper}}]{Hernitschek2017}
{Hernitschek}, N., {Sesar}, B., {Rix}, H.-W., {et~al.} 2017, \apj, 850, 96

\bibitem[{{Hernquist}(1990)}]{Hernquist1990}
{Hernquist}, L. 1990, \apj, 356, 359

\bibitem[{{Hills}(1988)}]{Hills1988}
{Hills}, J.~G. 1988, \nat, 331, 687

\bibitem[{{Huang} {et~al.}(2021){Huang}, {Li}, {Zhang}, {Li}, {Sun}, {Chang},
  {Dong}, \& {Liu}}]{Huang2021}
{Huang}, Y., {Li}, Q., {Zhang}, H., {et~al.} 2021, \apjl, 907, L42

\bibitem[{Hunter(2007)}]{matplotlib}
Hunter, J.~D. 2007, Computing in Science \& Engineering, 9, 90

\bibitem[{{Ibata} {et~al.}(2019){Ibata}, {Malhan}, \& {Martin}}]{Ibata2019}
{Ibata}, R.~A., {Malhan}, K., \& {Martin}, N.~F. 2019, \apj, 872, 152

\bibitem[{{Koch} {et~al.}(2012){Koch}, {L{\'e}pine}, \&
  {{\c{C}}al{\i}{\textcommabelow s}kan}}]{Koch2012}
{Koch}, A., {L{\'e}pine}, S., \& {{\c{C}}al{\i}{\textcommabelow s}kan},
  {\c{S}}. 2012, in European Physical Journal Web of Conferences, Vol.~19,
  European Physical Journal Web of Conferences, 03002

\bibitem[{{Kollmeier} {et~al.}(2009){Kollmeier}, {Gould}, {Knapp}, \&
  {Beers}}]{Kollmeier2009}
{Kollmeier}, J.~A., {Gould}, A., {Knapp}, G., \& {Beers}, T.~C. 2009, \apj,
  697, 1543

\bibitem[{{Kollmeier} {et~al.}(2017){Kollmeier}, {Zasowski}, {Rix}, {Johns},
  {Anderson}, {Drory}, {Johnson}, {Pogge}, {Bird}, {Blanc}, {Brownstein},
  {Crane}, {De Lee}, {Klaene}, {Kreckel}, {MacDonald}, {Merloni}, {Ness},
  {O'Brien}, {Sanchez-Gallego}, {Sayres}, {Shen}, {Thakar}, {Tkachenko},
  {Aerts}, {Blanton}, {Eisenstein}, {Holtzman}, {Maoz}, {Nandra}, {Rockosi},
  {Weinberg}, {Bovy}, {Casey}, {Chaname}, {Clerc}, {Conroy}, {Eracleous},
  {G{\"a}nsicke}, {Hekker}, {Horne}, {Kauffmann}, {McQuinn}, {Pellegrini},
  {Schinnerer}, {Schlafly}, {Schwope}, {Seibert}, {Teske}, \& {van
  Saders}}]{Kollmeier2017}
{Kollmeier}, J.~A., {Zasowski}, G., {Rix}, H.-W., {et~al.} 2017, arXiv
  e-prints, arXiv:1711.03234

\bibitem[{{Koposov} {et~al.}(2019){Koposov}, {Belokurov}, {Li}, {Mateu},
  {Erkal}, {Grillmair}, {Hendel}, {Price-Whelan}, {Laporte}, {Hawkins}, {Sohn},
  {del Pino}, {Evans}, {Slater}, {Kallivayalil}, {Navarro}, \& {Orphan Aspen
  Treasury Collaboration}}]{Koposov2019Orphan}
{Koposov}, S.~E., {Belokurov}, V., {Li}, T.~S., {et~al.} 2019, \mnras, 485,
  4726

\bibitem[{{Koposov} {et~al.}(2020){Koposov}, {Boubert}, {Li}, {Erkal}, {Da
  Costa}, {Zucker}, {Ji}, {Kuehn}, {Lewis}, {Mackey}, {Simpson}, {Shipp},
  {Wan}, {Belokurov}, {Bland-Hawthorn}, {Martell}, {Nordlander}, {Pace}, {De
  Silva}, {Wang}, \& {S5 Collaboration}}]{Koposov2020HighVel}
{Koposov}, S.~E., {Boubert}, D., {Li}, T.~S., {et~al.} 2020, \mnras, 491, 2465

\bibitem[{{Koppelman} \& {Helmi}(2021)}]{Koppelman2021}
{Koppelman}, H.~H. \& {Helmi}, A. 2021, \aap, 649, A136

\bibitem[{{Kunder} {et~al.}(2020){Kunder}, {P{\'e}rez-Villegas}, {Rich},
  {Ogata}, {Murari}, {Boren}, {Johnson}, {Nataf}, {Walker}, {Bono}, {Koch},
  {Propris}, {Storm}, \& {Wojno}}]{Kunder2020}
{Kunder}, A., {P{\'e}rez-Villegas}, A., {Rich}, R.~M., {et~al.} 2020, \aj, 159,
  270

\bibitem[{{Kunder} {et~al.}(2015){Kunder}, {Rich}, {Hawkins}, {Poleski},
  {Storm}, {Johnson}, {Shen}, {Li}, {Cordero}, {Nataf}, {Bono}, {Walker},
  {Koch}, {De Propris}, {Udalski}, {Szyma{\'n}ski}, {Soszy{\'n}ski},
  {Pietrzy{\'n}ski}, {Ulaczyk}, {Wyrzykowski}, {Pietrukowicz}, {Skowron},
  {Koz{\l}owski}, \& {Mr{\'o}z}}]{Kunder2015}
{Kunder}, A., {Rich}, R.~M., {Hawkins}, K., {et~al.} 2015, \apjl, 808, L12

\bibitem[{{Lenz} \& {Breger}(2004)}]{Lenz2004Period04}
{Lenz}, P. \& {Breger}, M. 2004, in IAU Symposium, Vol. 224, The A-Star Puzzle,
  ed. J.~{Zverko}, J.~{Ziznovsky}, S.~J. {Adelman}, \& W.~W. {Weiss}, 786--790

\bibitem[{{Leonard}(1991)}]{Leonard1991}
{Leonard}, P. J.~T. 1991, \aj, 101, 562

\bibitem[{{Leonard} \& {Tremaine}(1990)}]{Leonard1990}
{Leonard}, P. J.~T. \& {Tremaine}, S. 1990, \apj, 353, 486

\bibitem[{{L{\'e}pine} {et~al.}(2011){L{\'e}pine}, {Koch}, {Rich}, \&
  {Kuijken}}]{Lepine2011}
{L{\'e}pine}, S., {Koch}, A., {Rich}, R.~M., \& {Kuijken}, K. 2011, \apj, 741,
  100

\bibitem[{{Li} {et~al.}(2021){Li}, {Luo}, {Lu}, {Zhang}, {Li}, {Wang}, {Zuo},
  {Xiang}, {Ting}, {Marchetti}, {Li}, {Wang}, {Zhang}, {Hattori}, {Zhao},
  {Zhang}, \& {Zhao}}]{Li2021High591}
{Li}, Y.-B., {Luo}, A.~L., {Lu}, Y.-J., {et~al.} 2021, \apjs, 252, 3

\bibitem[{{Lindegren} {et~al.}(2021){Lindegren}, {Klioner}, {Hern{\'a}ndez},
  {Bombrun}, {Ramos-Lerate}, {Steidelm{\"u}ller}, {Bastian}, {Biermann}, {de
  Torres}, {Gerlach}, {Geyer}, {Hilger}, {Hobbs}, {Lammers}, {McMillan},
  {Stephenson}, {Casta{\~n}eda}, {Davidson}, {Fabricius}, {Gracia-Abril},
  {Portell}, {Rowell}, {Teyssier}, {Torra}, {Bartolom{\'e}}, {Clotet},
  {Garralda}, {Gonz{\'a}lez-Vidal}, {Torra}, {Abbas}, {Altmann}, {Anglada
  Varela}, {Balaguer-N{\'u}{\~n}ez}, {Balog}, {Barache}, {Becciani}, {Bernet},
  {Bertone}, {Bianchi}, {Bouquillon}, {Brown}, {Bucciarelli}, {Busonero},
  {Butkevich}, {Buzzi}, {Cancelliere}, {Carlucci}, {Charlot}, {Cioni},
  {Crosta}, {Crowley}, {del Peloso}, {del Pozo}, {Drimmel}, {Esquej}, {Fienga},
  {Fraile}, {Gai}, {Garcia-Reinaldos}, {Guerra}, {Hambly}, {Hauser},
  {Jan{\ss}en}, {Jordan}, {Kostrzewa-Rutkowska}, {Lattanzi}, {Liao}, {Licata},
  {Lister}, {L{\"o}ffler}, {Marchant}, {Masip}, {Mignard}, {Mints}, {Molina},
  {Mora}, {Morbidelli}, {Murphy}, {Pagani}, {Panuzzo}, {Pe{\~n}alosa Esteller},
  {Poggio}, {Re Fiorentin}, {Riva}, {Sagrist{\`a} Sell{\'e}s}, {Sanchez
  Gimenez}, {Sarasso}, {Sciacca}, {Siddiqui}, {Smart}, {Souami}, {Spagna},
  {Steele}, {Taris}, {Utrilla}, {van Reeven}, \&
  {Vecchiato}}]{Lindegren2020GaiaAstrometry}
{Lindegren}, L., {Klioner}, S.~A., {Hern{\'a}ndez}, J., {et~al.} 2021, \aap,
  649, A2

\bibitem[{{Liu} {et~al.}(2014){Liu}, {Yuan}, {Huo}, {Deng}, {Hou}, {Zhao},
  {Zhao}, {Shi}, {Luo}, {Xiang}, {Zhang}, {Huang}, \& {Zhang}}]{Liu2014}
{Liu}, X.~W., {Yuan}, H.~B., {Huo}, Z.~Y., {et~al.} 2014, in Setting the scene
  for Gaia and LAMOST, ed. S.~{Feltzing}, G.~{Zhao}, N.~A. {Walton}, \&
  P.~{Whitelock}, Vol. 298, 310--321

\bibitem[{{Malhan} \& {Ibata}(2018)}]{Malhan2018}
{Malhan}, K. \& {Ibata}, R.~A. 2018, \mnras, 477, 4063

\bibitem[{{Marchetti}(2021)}]{Marchetti2021}
{Marchetti}, T. 2021, \mnras, 503, 1374

\bibitem[{{Marchetti} {et~al.}(2019){Marchetti}, {Rossi}, \&
  {Brown}}]{Marchetti2019}
{Marchetti}, T., {Rossi}, E.~M., \& {Brown}, A.~G.~A. 2019, \mnras, 490, 157

\bibitem[{{Masci} {et~al.}(2019){Masci}, {Laher}, {Rusholme}, {Shupe}, {Groom},
  {Surace}, {Jackson}, {Monkewitz}, {Beck}, {Flynn}, {Terek}, {Landry},
  {Hacopians}, {Desai}, {Howell}, {Brooke}, {Imel}, {Wachter}, {Ye}, {Lin},
  {Cenko}, {Cunningham}, {Rebbapragada}, {Bue}, {Miller}, {Mahabal}, {Bellm},
  {Patterson}, {Juri{\'c}}, {Golkhou}, {Ofek}, {Walters}, {Graham}, {Kasliwal},
  {Dekany}, {Kupfer}, {Burdge}, {Cannella}, {Barlow}, {Van Sistine}, {Giomi},
  {Fremling}, {Blagorodnova}, {Levitan}, {Riddle}, {Smith}, {Helou}, {Prince},
  \& {Kulkarni}}]{Masci2019ZTF}
{Masci}, F.~J., {Laher}, R.~R., {Rusholme}, B., {et~al.} 2019, \pasp, 131,
  018003

\bibitem[{{McConnachie} \& {Venn}(2020)}]{McConnachie2020Dwarfs}
{McConnachie}, A.~W. \& {Venn}, K.~A. 2020, Research Notes of the American
  Astronomical Society, 4, 229

\bibitem[{{Miyamoto} \& {Nagai}(1975)}]{Miyamoto1975}
{Miyamoto}, M. \& {Nagai}, R. 1975, \pasj, 27, 533

\bibitem[{{Monari} {et~al.}(2018){Monari}, {Famaey}, {Carrillo}, {Piffl},
  {Steinmetz}, {Wyse}, {Anders}, {Chiappini}, \& {Jan{\ss}en}}]{Monari2018}
{Monari}, G., {Famaey}, B., {Carrillo}, I., {et~al.} 2018, \aap, 616, L9

\bibitem[{{Mucciarelli} \& {Bellazzini}(2020)}]{Mucciarelli2020TempGaia}
{Mucciarelli}, A. \& {Bellazzini}, M. 2020, Research Notes of the American
  Astronomical Society, 4, 52

\bibitem[{{Navarro} {et~al.}(1997){Navarro}, {Frenk}, \& {White}}]{Navarro1997}
{Navarro}, J.~F., {Frenk}, C.~S., \& {White}, S. D.~M. 1997, \apj, 490, 493

\bibitem[{{Necib} \& {Lin}(2021{\natexlab{a}})}]{Necib2021I}
{Necib}, L. \& {Lin}, T. 2021{\natexlab{a}}, arXiv e-prints, arXiv:2102.01704

\bibitem[{{Necib} \& {Lin}(2021{\natexlab{b}})}]{Necib2021II}
{Necib}, L. \& {Lin}, T. 2021{\natexlab{b}}, arXiv e-prints, arXiv:2102.02211

\bibitem[{{Nissen} \& {Schuster}(2010)}]{Nissen2010}
{Nissen}, P.~E. \& {Schuster}, W.~J. 2010, \aap, 511, L10

\bibitem[{{Pearson} {et~al.}(2017){Pearson}, {Price-Whelan}, \&
  {Johnston}}]{Pearson2017}
{Pearson}, S., {Price-Whelan}, A.~M., \& {Johnston}, K.~V. 2017, Nature
  Astronomy, 1, 633

\bibitem[{Pedregosa {et~al.}(2011)Pedregosa, Varoquaux, Gramfort, Michel,
  Thirion, Grisel, Blondel, Prettenhofer, Weiss, Dubourg, Vanderplas, Passos,
  Cournapeau, Brucher, Perrot, \& Duchesnay}]{Pedregosa2012}
Pedregosa, F., Varoquaux, G., Gramfort, A., {et~al.} 2011, Journal of Machine
  Learning Research, 12, 2825

\bibitem[{{Perets} \& {{\v{S}}ubr}(2012)}]{Perets2012}
{Perets}, H.~B. \& {{\v{S}}ubr}, L. 2012, \apj, 751, 133

\bibitem[{P\'erez \& Granger(2007)}]{ipython}
P\'erez, F. \& Granger, B.~E. 2007, Computing in Science and Engineering, 9, 21

\bibitem[{{Piffl} {et~al.}(2014){Piffl}, {Scannapieco}, {Binney}, {Steinmetz},
  {Scholz}, {Williams}, {de Jong}, {Kordopatis}, {Matijevi{\v{c}}},
  {Bienaym{\'e}}, {Bland-Hawthorn}, {Boeche}, {Freeman}, {Gibson}, {Gilmore},
  {Grebel}, {Helmi}, {Munari}, {Navarro}, {Parker}, {Reid}, {Seabroke},
  {Watson}, {Wyse}, \& {Zwitter}}]{Piffl2014}
{Piffl}, T., {Scannapieco}, C., {Binney}, J., {et~al.} 2014, \aap, 562, A91

\bibitem[{{Portegies Zwart}(2000)}]{Portegies2000}
{Portegies Zwart}, S.~F. 2000, \apj, 544, 437

\bibitem[{{Poveda} {et~al.}(1967){Poveda}, {Ruiz}, \& {Allen}}]{Poveda1967}
{Poveda}, A., {Ruiz}, J., \& {Allen}, C. 1967, Boletin de los Observatorios
  Tonantzintla y Tacubaya, 4, 86

\bibitem[{{Price-Whelan} {et~al.}(2019){Price-Whelan}, {Mateu}, {Iorio},
  {Pearson}, {Bonaca}, \& {Belokurov}}]{Price-Whelan2019}
{Price-Whelan}, A.~M., {Mateu}, C., {Iorio}, G., {et~al.} 2019, \aj, 158, 223

\bibitem[{{Prudil} {et~al.}(2019){Prudil}, {D{\'e}k{\'a}ny}, {Grebel},
  {Catelan}, {Skarka}, \& {Smolec}}]{Prudil2019Kin}
{Prudil}, Z., {D{\'e}k{\'a}ny}, I., {Grebel}, E.~K., {et~al.} 2019, \mnras,
  487, 3270

\bibitem[{{Prudil} {et~al.}(2020){Prudil}, {D{\'e}k{\'a}ny}, {Grebel}, \&
  {Kunder}}]{Prudil2020Disk}
{Prudil}, Z., {D{\'e}k{\'a}ny}, I., {Grebel}, E.~K., \& {Kunder}, A. 2020,
  \mnras, 492, 3408

\bibitem[{{Prudil} {et~al.}(2021){Prudil}, {Hanke}, {Lemasle}, {Crestani},
  {Braga}, {Fabrizio}, {Koch-Hansen}, {Bono}, {Grebel}, {Matsunaga}, {Marengo},
  {da Silva}, {Dall'Ora}, {Mart{\'\i}nez-V{\'a}zquez}, {Altavilla}, {Lala},
  {Chaboyer}, {Ferraro}, {Fiorentino}, {Gilligan}, {Nonino}, \&
  {Th{\'e}venin}}]{Prudil2021Orphan}
{Prudil}, Z., {Hanke}, M., {Lemasle}, B., {et~al.} 2021, \aap, 648, A78

\bibitem[{{Ram{\'\i}rez} \& {Mel{\'e}ndez}(2005)}]{Ramirez2005}
{Ram{\'\i}rez}, I. \& {Mel{\'e}ndez}, J. 2005, \apj, 626, 465

\bibitem[{{Rybizki} {et~al.}(2022){Rybizki}, {Green}, {Rix}, {El-Badry},
  {Demleitner}, {Zari}, {Udalski}, {Smart}, \& {Gould}}]{Rybizki2022}
{Rybizki}, J., {Green}, G.~M., {Rix}, H.-W., {et~al.} 2022, \mnras, 510, 2597

\bibitem[{{Savino} {et~al.}(2020){Savino}, {Koch}, {Prudil}, {Kunder}, \&
  {Smolec}}]{Savino2020}
{Savino}, A., {Koch}, A., {Prudil}, Z., {Kunder}, A., \& {Smolec}, R. 2020,
  \aap, 641, A96

\bibitem[{{Schlafly} \& {Finkbeiner}(2011)}]{Schlafly2011}
{Schlafly}, E.~F. \& {Finkbeiner}, D.~P. 2011, \apj, 737, 103

\bibitem[{{Sch{\"o}nrich} {et~al.}(2010){Sch{\"o}nrich}, {Binney}, \&
  {Dehnen}}]{Schonrich2010}
{Sch{\"o}nrich}, R., {Binney}, J., \& {Dehnen}, W. 2010, \mnras, 403, 1829

\bibitem[{{Sesar} {et~al.}(2017){Sesar}, {Hernitschek}, {Mitrovi{\'c}},
  {Ivezi{\'c}}, {Rix}, {Cohen}, {Bernard}, {Grebel}, {Martin}, {Schlafly},
  {Burgett}, {Draper}, {Flewelling}, {Kaiser}, {Kudritzki}, {Magnier},
  {Metcalfe}, {Tonry}, \& {Waters}}]{Sesar2017a}
{Sesar}, B., {Hernitschek}, N., {Mitrovi{\'c}}, S., {et~al.} 2017, \aj, 153,
  204

\bibitem[{{Shen} {et~al.}(2022){Shen}, {Eadie}, {Murray}, {Zaritsky},
  {Speagle}, {Ting}, {Conroy}, {Cargile}, {Johnson}, {Naidu}, \&
  {Han}}]{Shen2022}
{Shen}, J., {Eadie}, G.~M., {Murray}, N., {et~al.} 2022, \apj, 925, 1

\bibitem[{{Sherwin} {et~al.}(2008){Sherwin}, {Loeb}, \&
  {O'Leary}}]{Sherwin2008}
{Sherwin}, B.~D., {Loeb}, A., \& {O'Leary}, R.~M. 2008, \mnras, 386, 1179

\bibitem[{{Smith} {et~al.}(2007){Smith}, {Ruchti}, {Helmi}, {Wyse},
  {Fulbright}, {Freeman}, {Navarro}, {Seabroke}, {Steinmetz}, {Williams},
  {Bienaym{\'e}}, {Binney}, {Bland-Hawthorn}, {Dehnen}, {Gibson}, {Gilmore},
  {Grebel}, {Munari}, {Parker}, {Scholz}, {Siebert}, {Watson}, \&
  {Zwitter}}]{Smith2007}
{Smith}, M.~C., {Ruchti}, G.~R., {Helmi}, A., {et~al.} 2007, \mnras, 379, 755

\bibitem[{{Sneden} {et~al.}(2017){Sneden}, {Preston}, {Chadid}, \&
  {Adam{\'o}w}}]{Sneden2017}
{Sneden}, C., {Preston}, G.~W., {Chadid}, M., \& {Adam{\'o}w}, M. 2017, \apj,
  848, 68

\bibitem[{{Sneden}(1973)}]{Sneden1973}
{Sneden}, C.~A. 1973, PhD thesis, THE UNIVERSITY OF TEXAS AT AUSTIN.

\bibitem[{{Sohn} {et~al.}(2018){Sohn}, {Watkins}, {Fardal}, {van der Marel},
  {Deason}, {Besla}, \& {Bellini}}]{Sohn2018}
{Sohn}, S.~T., {Watkins}, L.~L., {Fardal}, M.~A., {et~al.} 2018, \apj, 862, 52

\bibitem[{{Vasiliev} \& {Belokurov}(2020)}]{Vasiliev2020Sag}
{Vasiliev}, E. \& {Belokurov}, V. 2020, \mnras, 497, 4162

\bibitem[{{Vasiliev} {et~al.}(2021){Vasiliev}, {Belokurov}, \&
  {Erkal}}]{Vasiliev2021Tango}
{Vasiliev}, E., {Belokurov}, V., \& {Erkal}, D. 2021, \mnras, 501, 2279

\bibitem[{{Vickers} {et~al.}(2015){Vickers}, {Smith}, \&
  {Grebel}}]{Vickers2015}
{Vickers}, J.~J., {Smith}, M.~C., \& {Grebel}, E.~K. 2015, \aj, 150, 77

\bibitem[{Virtanen {et~al.}(2020)Virtanen, Gommers, Oliphant, Haberland, Reddy,
  Cournapeau, Burovski, Peterson, Weckesser, Bright, {van der Walt}, Brett,
  Wilson, Millman, Mayorov, Nelson, Jones, Kern, Larson, Carey, Polat, Feng,
  Moore, {VanderPlas}, Laxalde, Perktold, Cimrman, Henriksen, Quintero, Harris,
  Archibald, Ribeiro, Pedregosa, {van Mulbregt}, \& {SciPy 1.0
  Contributors}}]{scipy}
Virtanen, P., Gommers, R., Oliphant, T.~E., {et~al.} 2020, Nature Methods, 17,
  261

\bibitem[{{Watkins} {et~al.}(2010){Watkins}, {Evans}, \& {An}}]{Watkins2010}
{Watkins}, L.~L., {Evans}, N.~W., \& {An}, J.~H. 2010, \mnras, 406, 264

\bibitem[{{Watkins} {et~al.}(2019){Watkins}, {van der Marel}, {Sohn}, \&
  {Evans}}]{Watkins2019}
{Watkins}, L.~L., {van der Marel}, R.~P., {Sohn}, S.~T., \& {Evans}, N.~W.
  2019, \apj, 873, 118

\bibitem[{{Williams} {et~al.}(2017){Williams}, {Belokurov}, {Casey}, \&
  {Evans}}]{Williams2017}
{Williams}, A.~A., {Belokurov}, V., {Casey}, A.~R., \& {Evans}, N.~W. 2017,
  \mnras, 468, 2359

\bibitem[{{York} {et~al.}(2000){York}, {Adelman}, {Anderson}, {Anderson},
  {Annis}, {Bahcall}, {Bakken}, {Barkhouser}, {Bastian}, {Berman}, {Boroski},
  {Bracker}, {Briegel}, {Briggs}, {Brinkmann}, {Brunner}, {Burles}, {Carey},
  {Carr}, {Castander}, {Chen}, {Colestock}, {Connolly}, {Crocker}, {Csabai},
  {Czarapata}, {Davis}, {Doi}, {Dombeck}, {Eisenstein}, {Ellman}, {Elms},
  {Evans}, {Fan}, {Federwitz}, {Fiscelli}, {Friedman}, {Frieman}, {Fukugita},
  {Gillespie}, {Gunn}, {Gurbani}, {de Haas}, {Haldeman}, {Harris}, {Hayes},
  {Heckman}, {Hennessy}, {Hindsley}, {Holm}, {Holmgren}, {Huang}, {Hull},
  {Husby}, {Ichikawa}, {Ichikawa}, {Ivezi{\'c}}, {Kent}, {Kim}, {Kinney},
  {Klaene}, {Kleinman}, {Kleinman}, {Knapp}, {Korienek}, {Kron}, {Kunszt},
  {Lamb}, {Lee}, {Leger}, {Limmongkol}, {Lindenmeyer}, {Long}, {Loomis},
  {Loveday}, {Lucinio}, {Lupton}, {MacKinnon}, {Mannery}, {Mantsch}, {Margon},
  {McGehee}, {McKay}, {Meiksin}, {Merelli}, {Monet}, {Munn}, {Narayanan},
  {Nash}, {Neilsen}, {Neswold}, {Newberg}, {Nichol}, {Nicinski}, {Nonino},
  {Okada}, {Okamura}, {Ostriker}, {Owen}, {Pauls}, {Peoples}, {Peterson},
  {Petravick}, {Pier}, {Pope}, {Pordes}, {Prosapio}, {Rechenmacher}, {Quinn},
  {Richards}, {Richmond}, {Rivetta}, {Rockosi}, {Ruthmansdorfer}, {Sand ford},
  {Schlegel}, {Schneider}, {Sekiguchi}, {Sergey}, {Shimasaku}, {Siegmund},
  {Smee}, {Smith}, {Snedden}, {Stone}, {Stoughton}, {Strauss}, {Stubbs},
  {SubbaRao}, {Szalay}, {Szapudi}, {Szokoly}, {Thakar}, {Tremonti}, {Tucker},
  {Uomoto}, {Vanden Berk}, {Vogeley}, {Waddell}, {Wang}, {Watanabe},
  {Weinberg}, {Yanny}, {Yasuda}, \& {SDSS Collaboration}}]{York2000}
{York}, D.~G., {Adelman}, J., {Anderson}, John~E., J., {et~al.} 2000, \aj, 120,
  1579

\bibitem[{{Yu} \& {Tremaine}(2003)}]{Yu2003}
{Yu}, Q. \& {Tremaine}, S. 2003, \apj, 599, 1129

\bibitem[{{Zhao} {et~al.}(2012){Zhao}, {Zhao}, {Chu}, {Jing}, \&
  {Deng}}]{Zhao2012Lamost}
{Zhao}, G., {Zhao}, Y.-H., {Chu}, Y.-Q., {Jing}, Y.-P., \& {Deng}, L.-C. 2012,
  Research in Astronomy and Astrophysics, 12, 723

\end{thebibliography}

\begin{appendix} 

\section{Additional figures}

\begin{figure*}
\includegraphics[width=2\columnwidth]{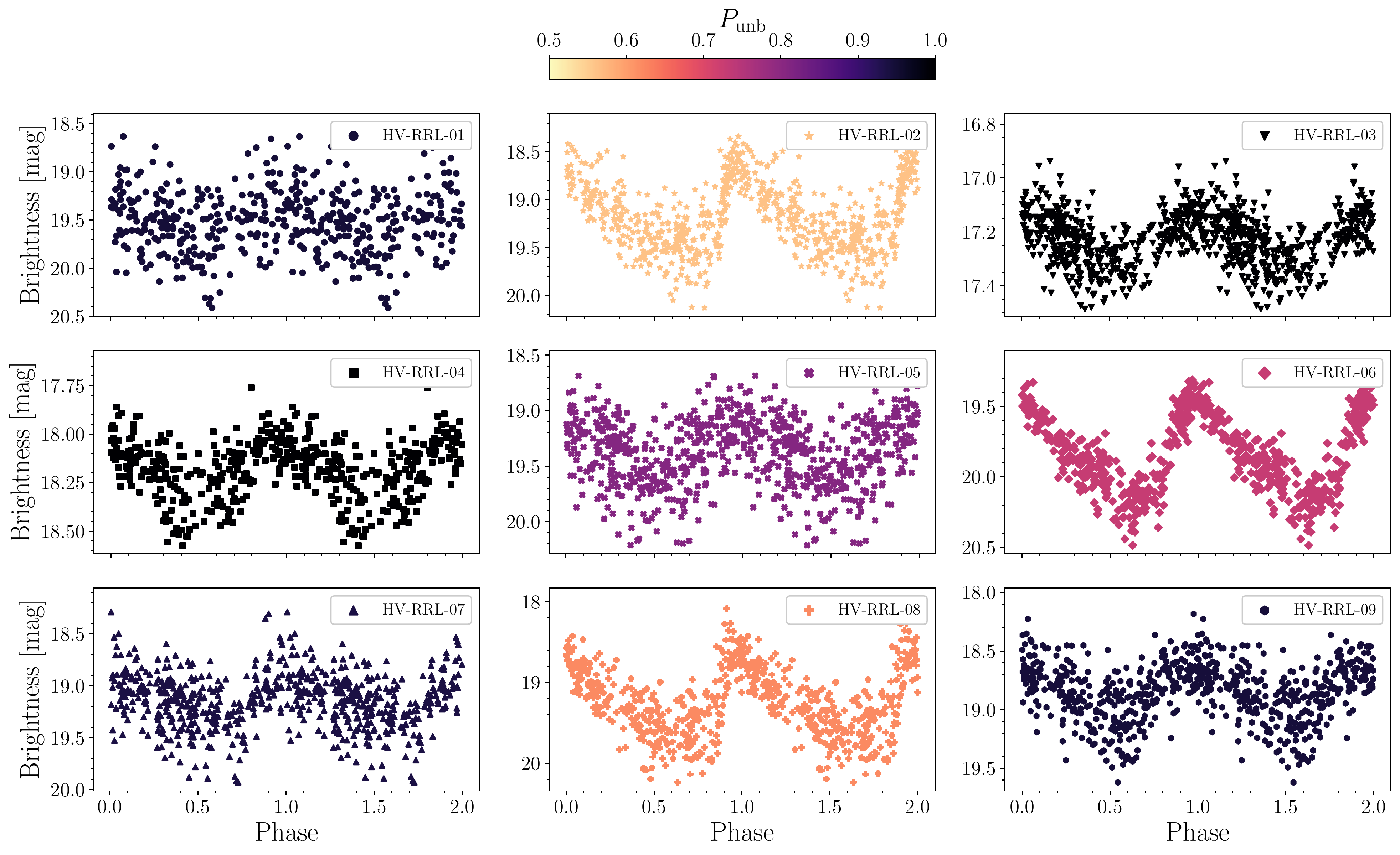}
\caption{Phased light curves for the high velocity RR~Lyrae stars in our sample from the CSS survey. The color-coding of the individual phased light curve represents the $P_{\rm unb}$ condition of them being unbound.}
\label{fig:LightCurves}
\end{figure*}

\begin{figure*}
\includegraphics[width=\columnwidth]{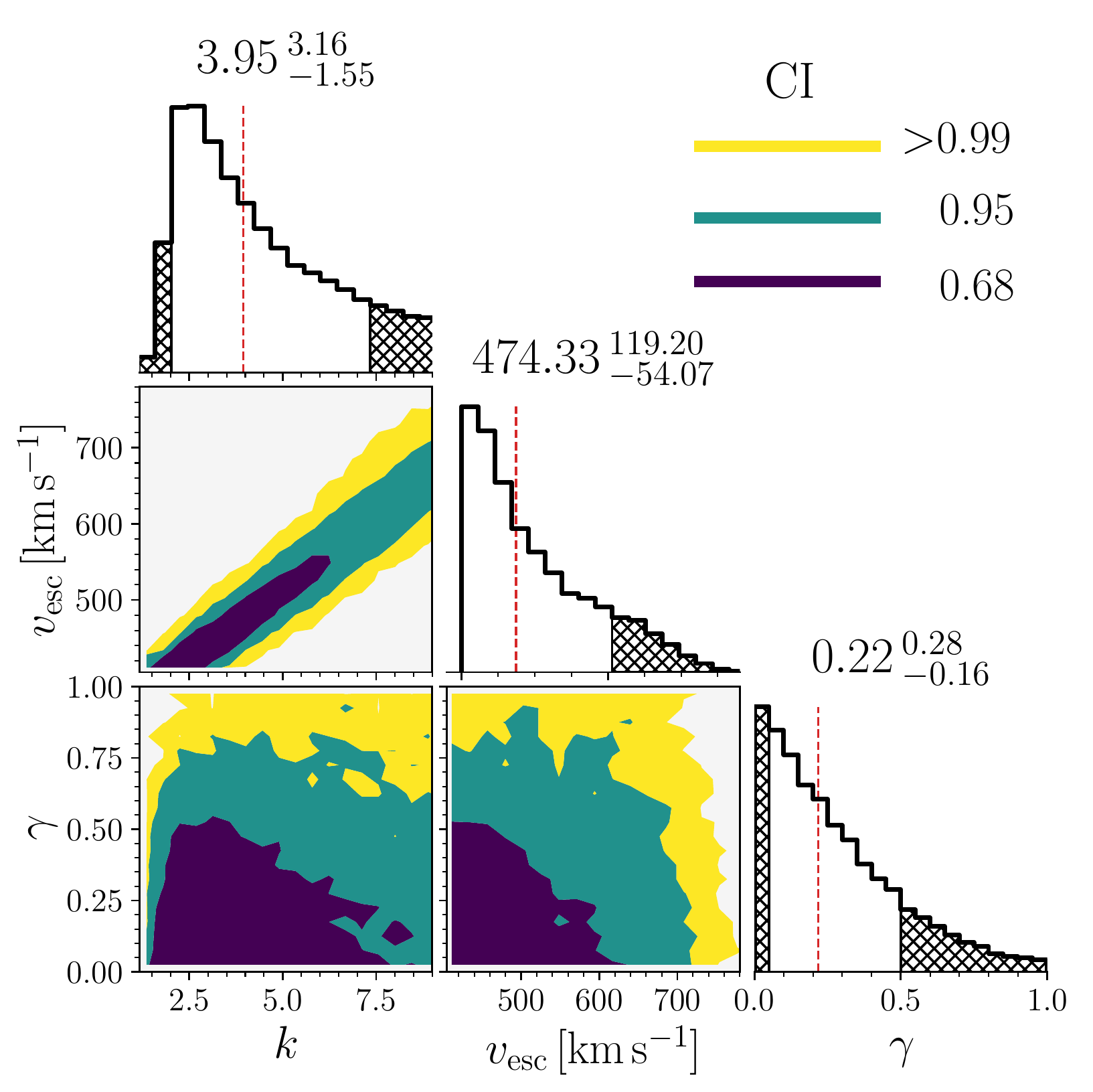}
\includegraphics[width=\columnwidth]{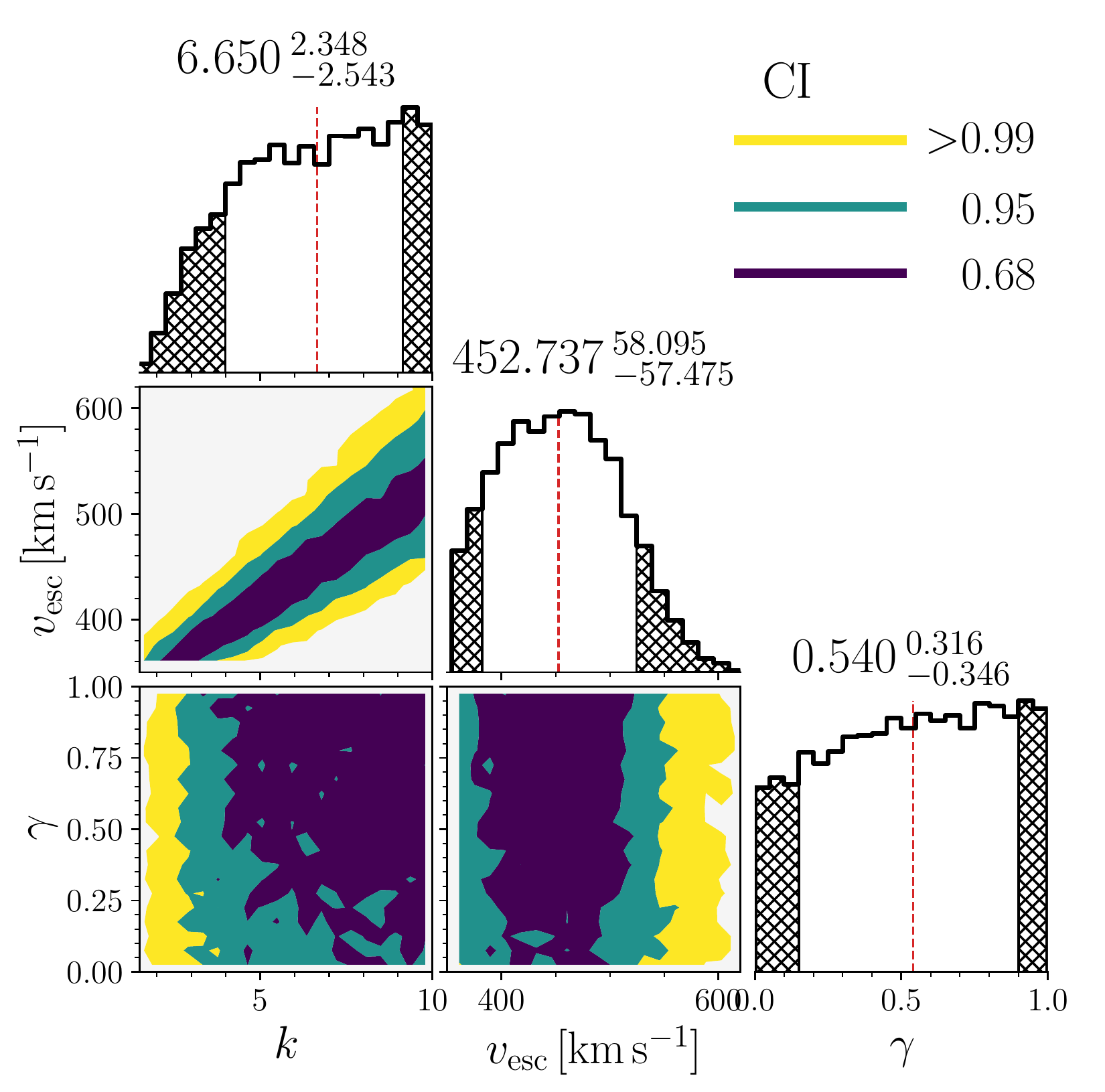}
\caption{Similarly to Fig.~\ref{fig:CornerSolar}, here we displayed posterior distribution of $k$, $v_{\rm esc}$ and $\gamma$, for $12 < R_{\rm GC} < 20\,\text{kpc}$ (left panel) and for $20 < R_{\rm GC} < 28\,\text{kpc}$ (right panel).}
\label{fig:Corner12_20_28}
\end{figure*}

\begin{figure*}
\includegraphics[width=\columnwidth]{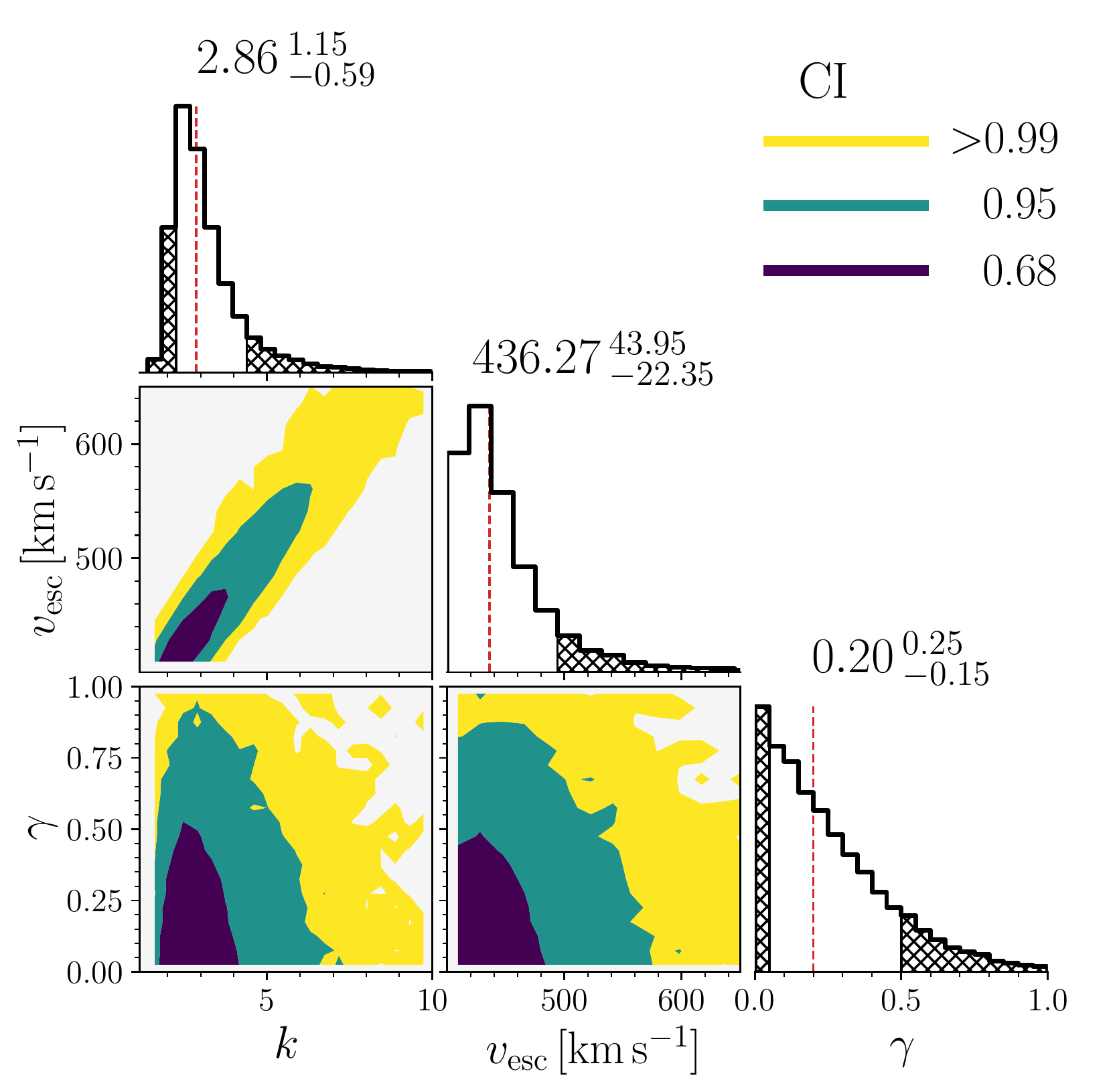}
\includegraphics[width=\columnwidth]{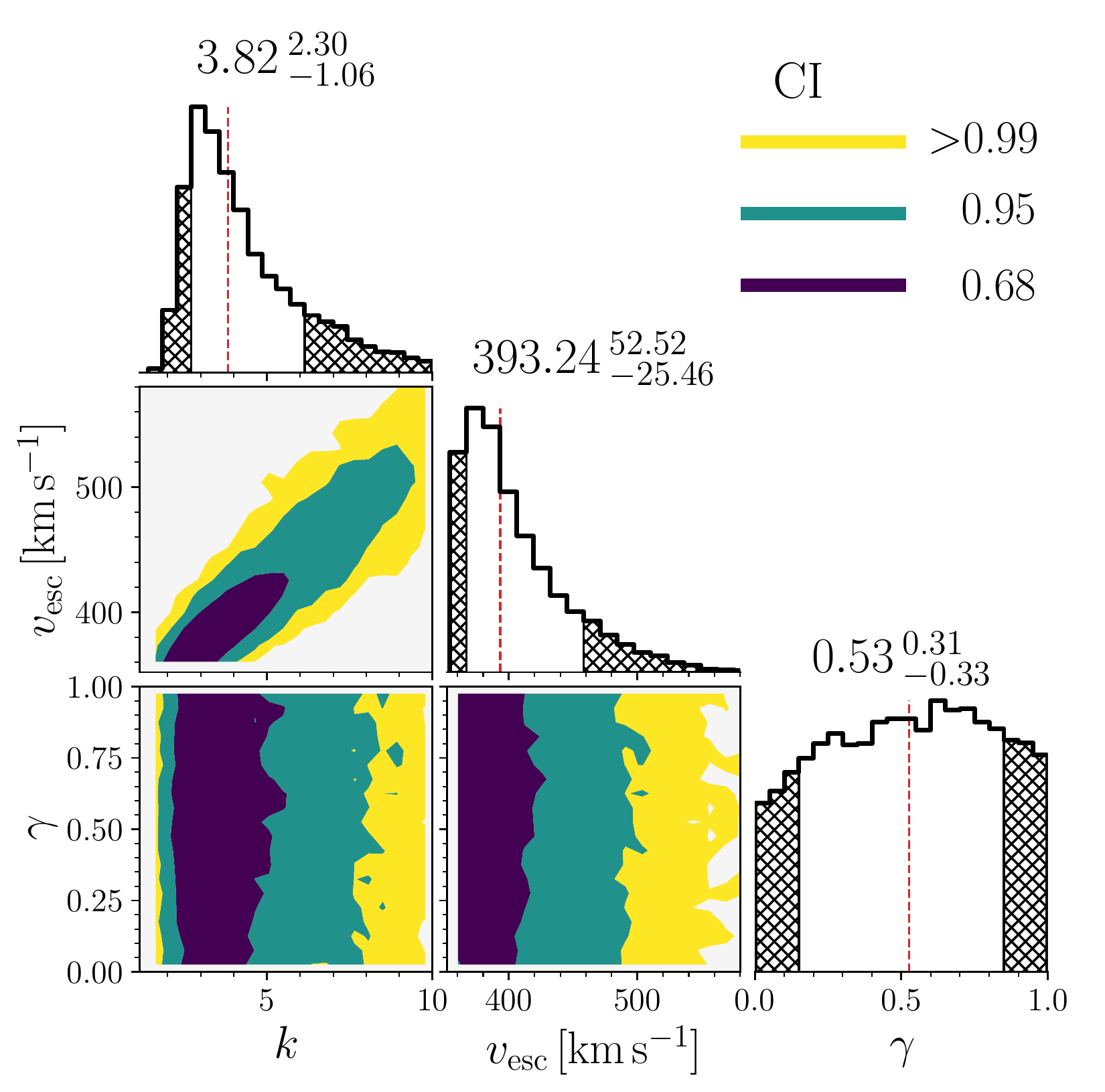}
\caption{Same as Fig.~\ref{fig:Corner12_20_28}. Both panels display the posterior distribution of $k$, $v_{\rm esc}$ and $\gamma$, with the left panel showing results for $12 < R_{\rm GC} < 20\,\text{kpc}$ and the right plot showing results for $20 < R_{\rm GC} < 28\,\text{kpc}$. To derive these distributions, we used a kernel-density estimate (Sect.~\ref{subsec:MCMCrun}) as a prior for the $k$ parameter.}
\label{fig:Corner12_20_28_kde}
\end{figure*}

\section{Additional tables}

\setlength{\tabcolsep}{3pt}    
\begin{landscape}
\begin{table}
\caption{List of likely unbound RR Lyrae variables in our sample with $P_{\rm unb} > 0.5$. The first two columns represent our adopted object ID and \gaia~EDR3 ID. Columns 3, 4, 5, and 6 list the heliocentric distances and systemic velocities together with their assumed uncertainties, as estimated in \citet{Prudil2021Orphan}. Column 8 contains estimated $P_{\rm unb}$ of the RR~Lyrae stars being unbound fulfilling the condition in Eq. 13. The calculated Galactocentric velocities and their covariances are listed in columns 9, 10, 11, 12, and 13 starting with the angular velocity $v_{y}$ followed by the quadrature of the radial and vertical velocity, $\sqrt{v_{x}^2 + v_{z}^2}$ both accompanied by their uncertainties and correlations. The last five columns (14, 15, 16, 17, and 18) represent the total Galactocentric velocity $v_{\rm GC}$ and distance $R_{\rm GC}$ followed be their covariances.}
\label{tab:HV-RR}
\begin{tabular}{lcccccccccccccccc}
\hline
ID & \textit{Gaia}-EDR3 ID & $d$ & $\sigma_{d}$ & $v_{\rm sys}$ & $\sigma_{v_{\rm sys}}$ & $P_{\rm unb}$ & $V$ & $\sigma_{V}$ & $\sqrt{U^2 + W^2}$ & $\sigma_{\sqrt{U^2 + W^2}}$ & $\rho^{V}_{\sqrt{U^2 + W^2}}$ & $v_{\rm GC}$ & $\sigma_{v_{\rm GC}}$ & $R_{\rm GC}$ & $\sigma_{R_{\rm GC}}$ & $\rho^{v_{\rm GC}}_{R_{\rm GC}}$ \\
 & & [kpc] & [kpc] & [km\,s$^{-1}$] & [km\,s$^{-1}$] & & [km\,s$^{-1}$] & [km\,s$^{-1}$] & [km\,s$^{-1}$] & [km\,s$^{-1}$] & & [km\,s$^{-1}$] & [km\,s$^{-1}$] & [kpc] & [kpc] & \\ \hline
HV-RRL-01 & 2471437298673558400 & 53 & 2 & -146.4 & 37.0 & 0.95 & -284 & 114 & 582 & 179 & -0.49 & 655 & 189 & 55.9 & 2.9 & 0.22 \\ 
HV-RRL-02 & 3664033198603611520 & 49 & 2 & 21.9 & 31.9 & 0.56 & -141 & 71 & 357 & 95 & -0.57 & 388 & 105 & 46.6 & 2.7 & 0.24 \\ 
HV-RRL-03$^{\ast}$ & 2644870582050682240 & 19 & 1 & -10.8 & 17.3 & 1.00 & -759 & 56 & 883 & 51 & -0.91 & 1165 & 73 & 20.0 & 1.0 & 0.99 \\ 
HV-RRL-04$^{\ast}$ & 1018414089653402496 & 29 & 1 & 6.9 & 46.9 & 1.00 & -927 & 66 & 206 & 46 & -0.12 & 950 & 66 & 35.4 & 1.6 & 0.98 \\ 
HV-RRL-05 & 3903031502808975616 & 54 & 3 & 57.5 & 32.0 & 0.81 & 69 & 93 & 470 & 141 & 0.47 & 483 & 143 & 54.0 & 3.0 & 0.17 \\ 
HV-RRL-06 & 674694939356598528 & 60 & 3 & 85.5 & 4.9 & 0.73 & 20 & 87 & 421 & 156 & 0.21 & 432 & 153 & 68.1 & 3.4 & 0.16 \\ 
HV-RRL-07 & 3942539665718239360 & 53 & 2 & 79.7 & 46.9 & 0.94 & 49 & 82 & 502 & 100 & 0.21 & 511 & 101 & 53.5 & 2.9 & 0.25 \\ 
HV-RRL-08 & 3661050979472106880 & 48 & 2 & -24.5 & 13.8 & 0.62 & -79 & 76 & 394 & 107 & -0.17 & 409 & 107 & 44.9 & 2.7 & 0.23 \\ 
HV-RRL-09 & 3692326965681261184 & 42 & 2 & 185.3 & 7.1 & 0.95 & -212 & 51 & 445 & 66 & -0.32 & 495 & 69 & 40.9 & 2.4 & 0.46 \\ 
\hline
\end{tabular}
\end{table}
\end{landscape}

\setlength{\tabcolsep}{3pt}    
\begin{landscape}
\begin{table}
\caption{Table of unbound RR Lyrae candidates with $P_{\rm unb} > 0.5$ and their astrometric and photometric properties from \gaia~catalog. The first column represents our object ID with an asterisk marking variables uncertain classification as RR~Lyrae stars. Columns 2, 3, 4, 5, and 6 lists \gaia~EDR3 ID, equatorial and Galactic coordinates. The proper motions in right ascension and declination, together with their errors, are listed in columns 7, 8, 9, and 10. Columns 11 and 12 contain parallax and its uncertainty. The last three columns (13, 14 and 15) represent re-normalized unit weight error (RUWE) and \gaia~photometry in $G$-band and color $G_{\rm BP} - G_{\rm RP}$.}
\label{tab:HV-RRastr}
\begin{tabular}{lcccccccccccccccc}
\hline
ID & \textit{Gaia}-EDR3 ID & $\alpha$ & $\delta$ & $l$ & $b$ & $\mu_{\alpha^{\ast}}$ & $\sigma_{\mu_{\alpha^{\ast}}}$ & $\mu_{\delta}$ & $\sigma_{\mu_{\delta}}$ & $\varpi$ & $\sigma_{\varpi}$ & RUWE & $G$ & $G_{\rm BP} - G_{\rm RP}$ \\ 
 & & [deg] & [deg] & [deg] & [deg] & [mas\,yr$^{-1}$] & [mas\,yr$^{-1}$] & [mas\,yr$^{-1}$] & [mas\,yr$^{-1}$] & [mas] & [mas] & & [mag] & [mag] \\ \hline
HV-RRL-01 & 2471437298673558400 & 18.0789 & -10.1312 & 140.0668 & -72.3048 & -0.707 & 0.506 & -2.994 & 0.670 & 0.257 & 0.424 & 1.048 & 19.52 & 0.49 \\ 
HV-RRL-02 & 3664033198603611520 & 207.1866 & 1.0410 & 333.1053 & 60.5110 & -2.157 & 0.440 & -0.359 & 0.234 & 0.244 & 0.345 & 0.908 & 19.19 & 0.63 \\ 
HV-RRL-03$^{\ast}$ & 2644870582050682240 & 349.9803 & -0.1865 & 79.9517 & -55.2301 & 11.478 & 0.106 & -9.324 & 0.111 & 0.446 & 0.112 & 1.080 & 17.26 & 0.83 \\ 
HV-RRL-04$^{\ast}$ & 1018414089653402496 & 144.0048 & 50.2509 & 167.2901 & 46.4708 & -0.917 & 0.114 & -8.469 & 0.088 & -0.201 & 0.112 & 1.004 & 18.16 & 0.77 \\ 
HV-RRL-05 & 3903031502808975616 & 189.5793 & 9.1371 & 292.5429 & 71.7437 & -1.833 & 0.396 & 0.433 & 0.528 & 0.494 & 0.387 & 1.036 & 19.30 & 0.50 \\ 
HV-RRL-06 & 674694939356598528 & 117.3741 & 22.6240 & 198.0853 & 22.4142 & -1.243 & 0.547 & -1.053 & 0.305 & -0.144 & 0.448 & 0.987 & 19.81 & 0.70 \\ 
HV-RRL-07 & 3942539665718239360 & 194.9736 & 20.8361 & 320.4377 & 83.4185 & -1.990 & 0.326 & 0.451 & 0.383 & -0.102 & 0.278 & 0.934 & 19.19 & 0.64 \\ 
HV-RRL-08 & 3661050979472106880 & 212.2624 & 1.1819 & 341.8310 & 58.0671 & -2.163 & 0.435 & -0.041 & 0.373 & -0.244 & 0.377 & 1.046 & 19.33 & 0.53 \\ 
HV-RRL-09 & 3692326965681261184 & 195.0834 & 2.9494 & 308.3415 & 65.7276 & -2.964 & 0.314 & -0.415 & 0.239 & -0.045 & 0.226 & 1.040 & 18.84 & 0.50 \\ 
\hline
\end{tabular}
\end{table}
\end{landscape}

\end{appendix}


\end{document}